\documentclass[11pt]{article}
\pdfoutput=1

\usepackage{jcapmod}
\newcommand{\red}[1]{\textcolor{red}{#1}}

\def\be{\begin{equation}}
\def\ee{\end{equation}}

\def\bea{\begin{eqnarray}}
\def\eea{\end{eqnarray}}

\begin{document}

\pagenumbering{roman}
\begin{titlepage}
\baselineskip=15.5pt \thispagestyle{empty}

\rightline{DESY 19-094}

\bigskip\

\vspace{1cm}
\begin{center}
{\fontsize{18}{24}\selectfont  \bfseries  {The Effective Field Theory of Large Scale Structure\\ 
at Three Loops \\ [12pt] }} 
\end{center}
\begin{center}
{\fontsize{12}{18}\selectfont Thomas Konstandin,$^{1}$ Rafael A.~Porto$^{1,2}$ and Henrique Rubira$^{1,3}$} 
\end{center}

\begin{center}

\textsl{$^1$  \small Deutsches Elektronen-Synchrotron DESY,\\ Notkestra$\beta$e 85, D-22607 Hamburg, Germany}
\vskip 8pt

\textsl{$^2$  \small The Abdus Salam International Center for Theoretical Physics,\\ Strada Costiera, 11, Trieste 34151, Italy}
\vskip 8pt

\textsl{$^3$  \small II. Institute of Theoretical Physics, University of Hamburg, D-22761 Hamburg, Germany}
\end{center}

\vspace{1.2cm}
\hrule \vspace{0.3cm}
\noindent {\bf Abstract}\\[0.1cm]
We study the power spectrum of dark matter density fluctuations in the framework of the Effective Field Theory of Large Scale Structures (EFTofLSS) up to three loop orders. We show that the extra coefficients in the EFT are sufficient to match numerical simulations with percent accuracy when a generic renormalization prescription is implemented (allowing for running of the individual counter-terms). We show that the level of accuracy increases with respect to the two loop results, up to $k \simeq 0.4\, h$~Mpc$^{-1}$ at redshift $z=0$, although the overall improvement is somewhat marginal. At the same time, we argue there is evidence that the behavior of the loop expansion in the EFTofLSS is typical of an asymptotic series, already on the brink of its maximum predictive power (at $z=0$). Hence, the inclusion of higher orders will likely deteriorate the matching to data, even at moderate values~of~$k$. Part of the reason for this behavior is due to large contributions to the (renormalized) power spectrum at three loop order from mildly non-linear scales, even after the UV counter-terms are included. In conclusion, the EFTofLSS to three loop orders provides the best approximation to the (deterministic part of the) power spectrum in the weakly non-linear~regime at $z=0$, and higher loops are not expected to improve our level of accuracy. 
\vskip10pt
\hrule
\vskip10pt

\end{titlepage}

\thispagestyle{empty}
\tableofcontents

\clearpage
\pagenumbering{arabic}
\setcounter{page}{1}

\clearpage
\section{Introduction}
\label{sec:introduction}

Ambitious observational programs are underway to make very precise measurements of the evolution of large scale
structures (LSS), e.g. \cite{Abbott:2005bi,Ivezic:2008fe,Laureijs:2011gra,Levi:2013gra}.  One of the main goals is to constrain the nature of dark energy, but at the same time future surveys will also probe the origin of the initial seed for structure formation, which --- very plausibly --- emanated from an early phase of accelerated expansion.  After the outstanding results from the Planck collaboration \cite{Aghanim:2018eyx,Akrami:2019izv}, current bounds on primordial non-Gaussianity are still far from reaching a well-motivated physical threshold \cite{threshold1,threshold2}, which may be difficult to achieve if restricted to observations of the cosmic microwave background (CMB) alone. The study of LSS  will then provide, not only powerful constraints on the properties of dark energy, but also the next leading probe for early universe cosmology. Regardless of one's motivation, an accurate analytic understanding will indisputably maximize the discovery potential for these remarkable experiments. As a consequence, the combination of a voluminous amount of new data has reinvigorated the constant effort to make accurate theoretical predictions in cosmology. This includes the fully non-linear region of structure formation, where simulations have reached an exquisite level, as well as the weakly non-linear regime, where the Effective Field Theory of LSS (EFT of LSS) --- both in Euler \cite{Baumann:2010tm,Carrasco:2012cv,Carrasco:2013mua, Carrasco:2013sva,Angulo:2014tfa,Foreman:2015lca,Baldauf:2015aha,Baldauf:2015zga,Baldauf:2015tla,error} and Lagrangian space \cite{left,left2,left3,Zaldarriaga:2015jrj} (see \cite{review} for a review) ---  has pushed forward the frontiers of `precision cosmology'. \vskip 4pt

In the EFTofLSS, as in any other EFT \cite{review}, the imprint of the short-distance physics in long-distance dynamics is encapsulated in a series of (symmetry-motivated) `Wilson coefficients'. In practice, these extra parameters play two roles. On the one hand, they can be chosen to remove the cutoff dependence introduced in an effective approach, simultaneously fixing the errors incurred by pushing the perturbative description beyond its realm of validity. On the other hand, the remaining finite part of the Wilson coefficients can incorporate the true knowledge from the non-perturbative regime in the long-distance dynamics. This information can be obtained either from observation or by comparison with a description which is assumed to be valid at short(er) scales. Hence, given a sought-after level of precision, the EFT provides an accurate analytic description of the dynamics up to a finite number of matching coefficients. The EFT formalism has several advantages over numerical methods attempting to cover the entirety of the parameter space over all scales. Not only the EFT approach provides an analytic description of the problem in a `universal' framework with systematic power-counting, but also because the EFT formalism is naturally suited to scan over different cosmologies and initial conditions to a high level of accuracy. Therefore, unlike standard perturbation theory (SPT) beyond leading order \cite{Bernardeau:2001qr} (where the perturbative (or loop) expansion is ill-defined), the EFT approach makes cosmological perturbative expansions a controlled theoretical framework.\vskip 4pt 

In a body of recent work, initiated in \cite{Carrasco:2013mua,Carrasco:2013sva}, the EFTofLSS was studied up to two loops, and shown to describe the evolution of dark matter density perturbations in the mildly non-linear regime with great success \cite{Angulo:2014tfa,Foreman:2015lca,error,Leofin}. In this paper, we continue the quest for accuracy by studying the power spectrum within the EFT approach to three loop orders. The SPT calculation at three loops was carried out in \cite{Blas:2013aba}, yet without cutting off the region of integration where the perturbative expansion is not expected to be valid. We point the reader to \cite{Blas:2013aba} for details on the diagrammatic and computational tools needed to achieve the third order in perturbation theory. In this paper we implement instead the EFTofLSS, introducing a series of counter-terms to remove the unwanted contributions from modes outside the realm of validity of SPT, and at the same time to properly incorporate the true non-linear information. As expected, precision increases with respect to the two loop results. As we shall see, this occurs with high precision up to $k \simeq 0.4\,h$\,Mpc$^{-1}$ at redshift $z=0$.\vskip 4pt

While this is a remarkable achievement of the EFTofLSS, we argue here that --- even corrected by the EFT methodology --- there is strong indication that the perturbative series is an asymptotic expansion reaching its maximum predictive power at weakly non-linear scales. (Similar speculations have been made in \cite{Sahni:1995rr, Pajer:2017ulp}.) As a consequence, we do not expect higher loop orders to have a positive impact, but rather to produce a departure from the true answer at moderate values of $k$. As a circumstantial evidence for this behavior, we find that the results up to two loop orders can match the numerical data with great precision up to high scales, $k \simeq 0.7\,h$ Mpc$^{-1}$, while at three loop orders --- and after including the necessary counter-terms --- the $\chi^2$ of the best fit presents a sharp increase for $k \gtrsim 0.55\, h$\, Mpc$^{-1}$. We argue the reason for the mismatch is not due to short-distance contributions, which we may be overlooking, but rather to large effects from mildly non-linear scales. 
Hence, our findings suggest that the EFTofLSS to three loop order provides the most accurate analytic model for the (deterministic part of the) dark matter power spectrum in the weakly non-linear regime at $z=0$. The study of higher $n$-point function would be required to analyze the behavior of the perturbative expansion more thoroughly.\vskip 4pt

This paper is organized as follows. In sec. 2 we discuss the general renormalization scheme and matching procedure we will utilize to extract the value of the extra parameters of the effective theory up to a given loop order. In sec. 3 we move on to the development of the EFTofLSS to three loops, and in particular the counter-terms that will be needed to properly remove the unwanted UV part of the SPT computation and incorporate the true information from non-linear scales. In that respect, we perform an `ultraviolet (UV) test' similar to what was proposed in \cite{Foreman:2015lca} to ensure we incorporate all of the necessary counter-terms that handled the UV-sensitivity of the SPT results. We incorporate further details in appendix A. In sec. 4 we match the EFTofLSS up to two and three loop orders to numerical simulations up to a given fitting scale $k_{\rm fit}$. We include the IR-resummation of \cite{Senatore:2014via,Cataneo:2016suz, Senatore:2017pbn}, which removes large oscillations in the power spectrum. We find a high level of accuracy up to scales of order $k \simeq 0.5\, h\,$\,Mpc$^{-1}$. We discuss the input of non-linear information as well as the possibility of overfitting the data. A comparison between the two and three loop results is performed and we show that there is improvement in the addition of the three loop effects for scales up to $k \sim 0.4\, h\,$\,Mpc$^{-1}$, although the increase in accuracy is somewhat marginal. At the same time we show there is circumstantial evidence of the asymptotic nature of the perturbative expansion in the EFTofLSS. We elaborate on this point in sec. 5, where we discuss in more detail the properties of the perturbative series up to three loop orders. In contrast to the `on-shell' prescription of \cite{Foreman:2015lca}, throughout this paper we use a generalized renormalization scheme which incorporates the {\it running} (or scale dependence) of the individual counter-terms, thus allowing us to extend the reach of the perturbative expansion. In appendix B we analyze the power spectrum using the on-shell scheme. We reproduce their results up to two loop orders and show that our results to three loop orders are essentially unmodified using the on-shell prescription. 
\section{Matching procedure \label{sec:fit}} 

In order to match the analytic computations within the EFT approach to numerical data\footnote{We use the (new) `Horizon' Simulation data given in \cite{horizon}.} we introduce the $\chi^2$ of the fit, given by
\be
\label{eq:def_chi_data}
\chi^2(c_i(\Lambda),k_{\rm fit}) \equiv \sum^{N}_{n=1} \left(\frac{P_{\rm data}(k_n) - P^{\rm EFT}_{\ell{\text -}\rm loop}
\left(k_n,c_i(\Lambda)\right)}{P_{\rm data}(k_n)} \right)^2 \, ,
\ee
which we minimze to a given fitting value, $k_{\rm fit}\equiv k_N$, with $N \gg 1$. $P^{\rm EFT}_{\ell{\text -}\rm loop}$ is the power spectrum computed in the EFTofLSS up to $\ell$-loop orders with a cutoff scale $\Lambda$, whose dependence is absorbed into a series of $i$-independent coefficients, $c_{i}(\Lambda)$, each one with up to $\ell$ contributions, depending on the loop order at which each counter-term first enters.\vskip 4pt

Before we proceed, let us stress an important point. Our procedure to fit the EFTofLSS to numerical data will be somewhat different than in previous studies at two loops, e.g. \cite{Foreman:2015lca}. Unlike what it has been done in the literature, we will not fix the higher loop order counter-terms, $c_{i (2)},\cdots, c_{i (\ell)}$, as a function of $c_{i (1)}$ through a renormalization condition. Instead, we allow for independent contributions from each one of them. The reason is twofold. First of all, because we use numerical results for power spectrum, it is difficult to cleanly disentangle the (cutoff dependent) counter-term which cure the mistake of SPT from the (finite) renormalized part which accounts for the true non-linear information. Therefore the size of the effect from the counter-terms does not respect the naive (loop) power-counting, while the renormalized parameters do. This is related to the second reason, which is the fact that we use a cutoff to define the EFT and therefore the size of the effect of the counter-terms at high loop orders may be as large as lower loop contributions, simply from the existence of `power-law' cutoff dependence.\footnote{This could be resolved by working with an approximate analytic expression for the power spectrum, as in \cite{Carrasco:2013mua}, such that we can explicitly remove the polynomial $\Lambda$ dependence in the counter-terms.} As a consequence, in what follows we do not distinguish between the counter-terms and the renormalized contributions and treat the extra parameters in the EFT as matching coefficients (although we will loosely refer to them as counter-terms in general). Since in our approach all counter-terms will be fixed by matching, each one of the $c_{i(\ell)}$'s loose a physical significance, and only the sum
\be
c_i = c_{i(1)}+ c_{i(2)} + \cdots + c_{i(\ell)}\,,
\ee
will be physically relevant.\footnote{The reader should keep in mind that different Wilson coefficients at leading order (i.e. {\it tree-level}), and beyond, contribute at various loop orders in the SPT counting.}  In particular, we will see that each one of them is typically very sensitive to small changes to the (simulated) power spectrum, in particular for wave numbers close to the upper limit of the fitting range, while the sum remains much more stable.\vskip 4pt

Fixing the lowest order counter-term by matching, while removing higher order contributions by the renormalization prescription, is akin to an on-shell scheme in quantum field theory. However, in principle any other prescription is equally viable, as long as there is convergence. (In general, we expect the difference between various schemes to be a higher order effect in the systematic expansion of the EFT, much like different renormalization schemes.)  On the other hand, since one of the focus of our work is to understand precisely the properties of the perturbative expansion within the EFT framework to high loop orders, addressing whether we have convergence or not should not hinge upon a particular choice of renormalization. In order to bypass this issue, we will not commit to a specific choice and instead allow for running of the individual $\ell$-loop order counter-terms when fitting to the numerical data.\footnote{For example, for the sound speed, the two loop parameter $c_{s(2)}$ may be fixed in terms of the one loop counter-term, $c_{s(1)}$, by the requirement that the power spectrum remains unchanged at two loop orders in the limit of small wave number \cite{Carrasco:2013mua}. In our case we will instead keep both $c_{s(1)}$ and $c_{s(2)}$ as independent parameters. While this increases our freedom, we will also show that the sum, which is the relevant quantity, remains remarkably stable to variations of $k_{\rm fit}$.} In this sense our choice is conservative, yet we also suffer from potential `over-fitting', something which was already emphasized in \cite{Foreman:2015lca}. We will attempt to address this issue throughout the paper. In any case, following our motivation to assess the best results the EFTofLSS can possibly achieve, we will not refrain from allowing the most freedom the effective theory can offer. For completeness, we analyze in appendix~\ref{appB} the three loop result using the renormalization prescription of \cite{Foreman:2015lca}.\vskip 4pt

Another important point for our matching procedure is to determine the uncertainty bands for a given counter-term, and linear combinations thereof. In order to do so, we will minimize
\be
\label{eq:def_delta_chi}
\Delta \chi^2 \equiv c_{i(\ell)} \frac{\partial^2\, \chi^2}{\partial c_{i(\ell)} \partial c_{i(\ell')}} c_{i(\ell')} \, ,
\ee
while varying the linear combination of counter-terms we are interested in. We will display error bands which correspond to variations in $\chi^2$ of order $\Delta\chi^2 \simeq 10^{-3}$. While, realistically when fitting data we are subject to larger errors, we are guided by the self-consistency of EFT approach to three loop orders for a given choice of matching coefficients.\footnote{For other (perhaps more realistic) choices, e.g. $\Delta\chi_0^2 = 0.1$, the error bands would be a factor of $10$ thicker.} 

\vskip 4pt
Not all of the $c_i$ coefficients are equally determined by this procedure. The reason is twofold. First of all, at low values of $k$ their contribution is highly suppressed. Secondly, when they start to become relevant at higher values of $k$, the effect of higher order coefficients which we do not include (see below), together with the failure of the loop expansion, significantly impairs their determination. This leaves some of our Wilson coefficients poorly constrained. We will return to this point in sec.~\ref{sec:disc}.


\section{Power spectrum to three loops \label{sec:setup}} 
In this section we discuss the construction of the EFTofLSS in Euler space, and in particular the number of relevant 
counter-terms which will be needed to renormalize the SPT computations up to three loop order. We will also implement the infrared (IR) resummation inherited from Lagrangian space \cite{left,Senatore:2014via,Cataneo:2016suz,Leofin}, which  improves the predictability of the theory, more radically at three loops.\vskip 4pt

\begin{figure}[h]
\centering
  \includegraphics[width=0.3\textwidth]{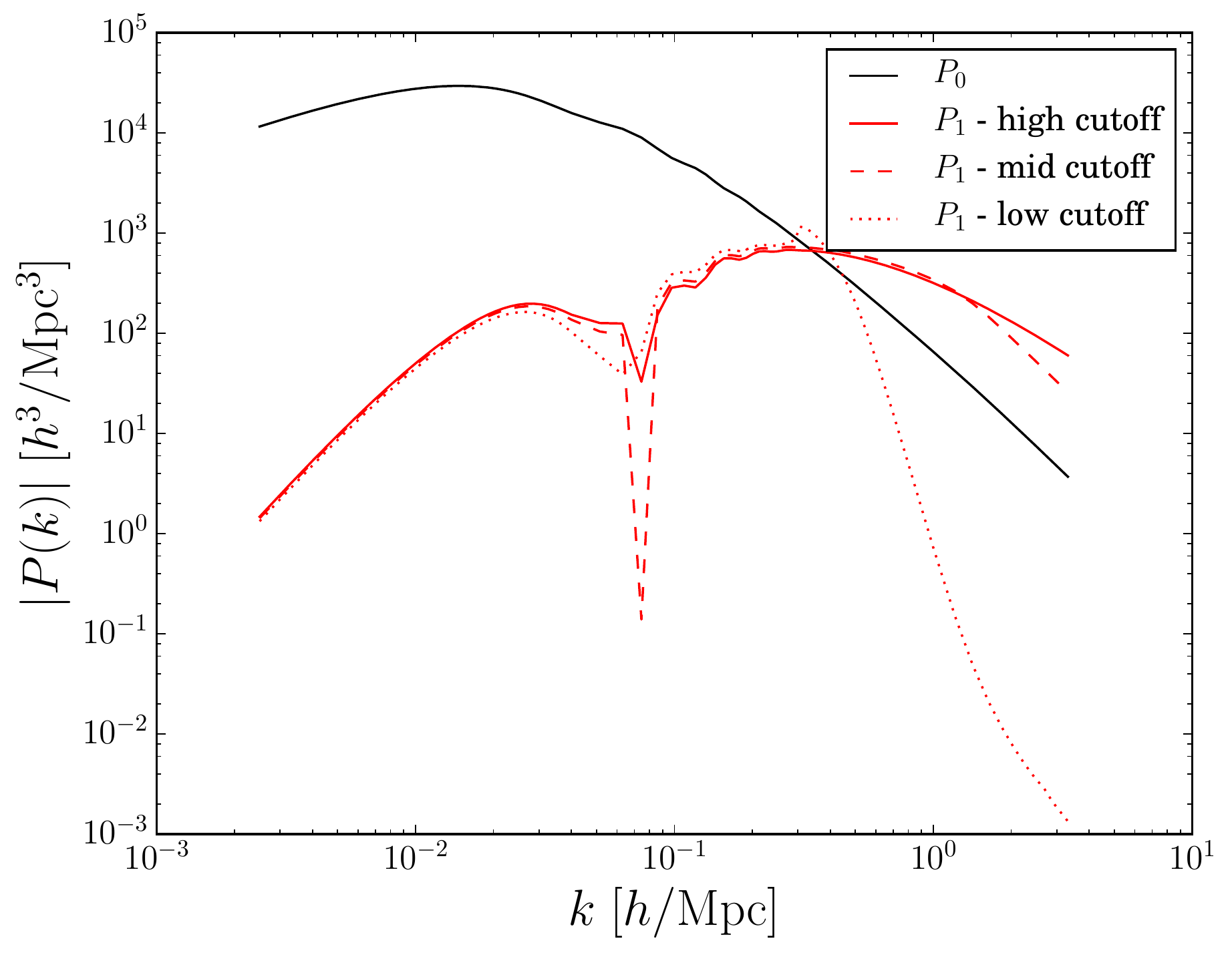}
  \includegraphics[width=0.3\textwidth]{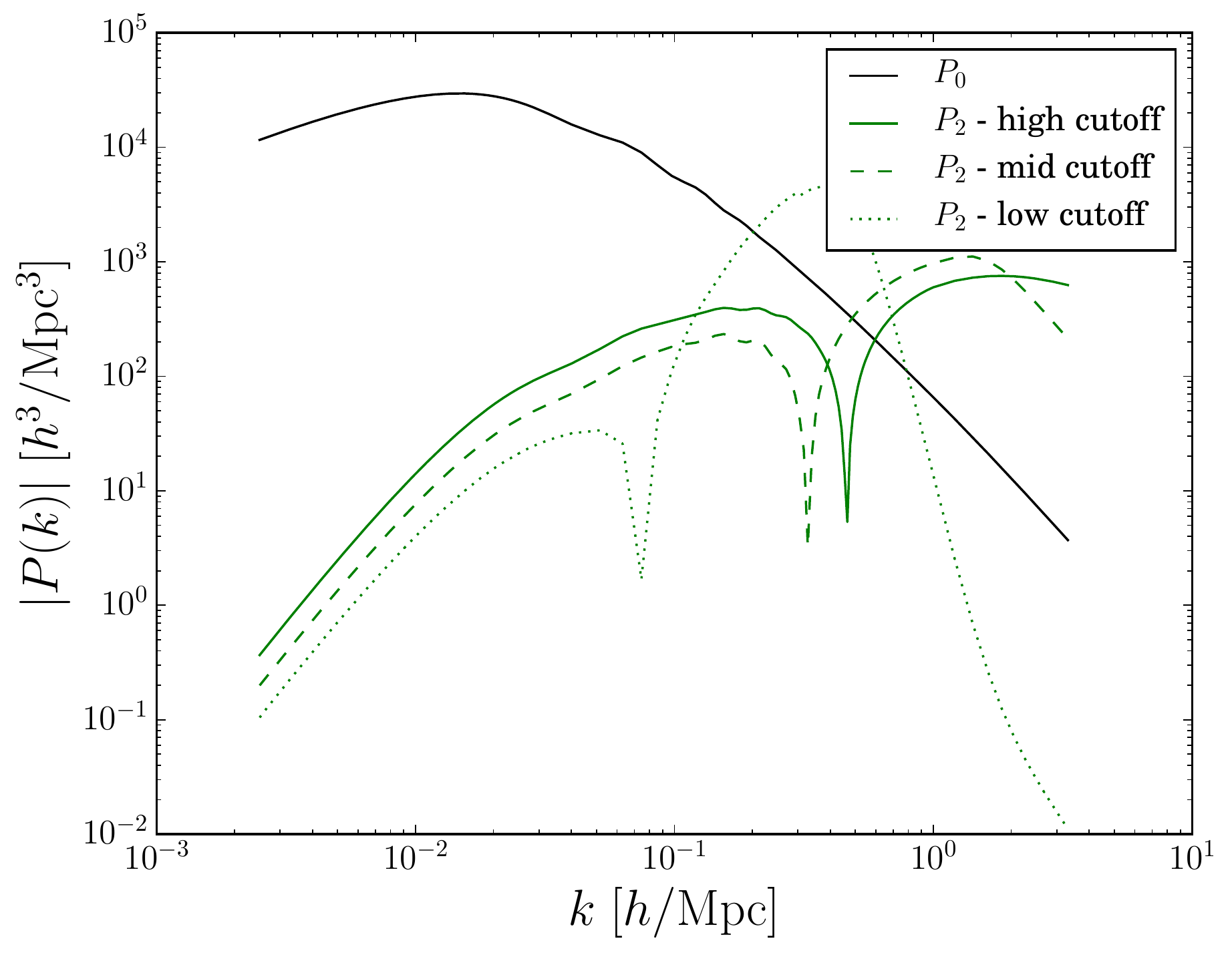}
  \includegraphics[width=0.3\textwidth]{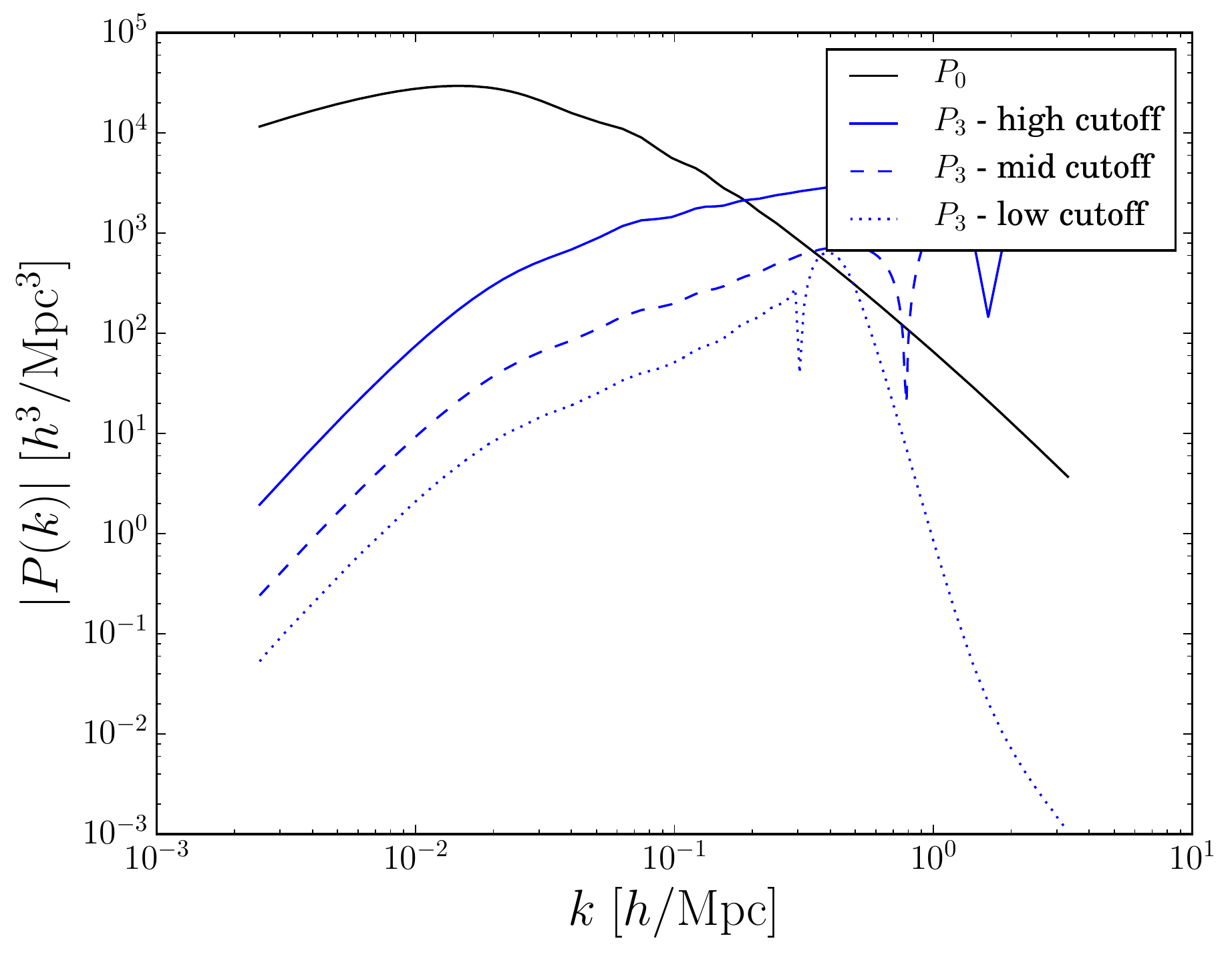}
  \includegraphics[width=0.3\textwidth]{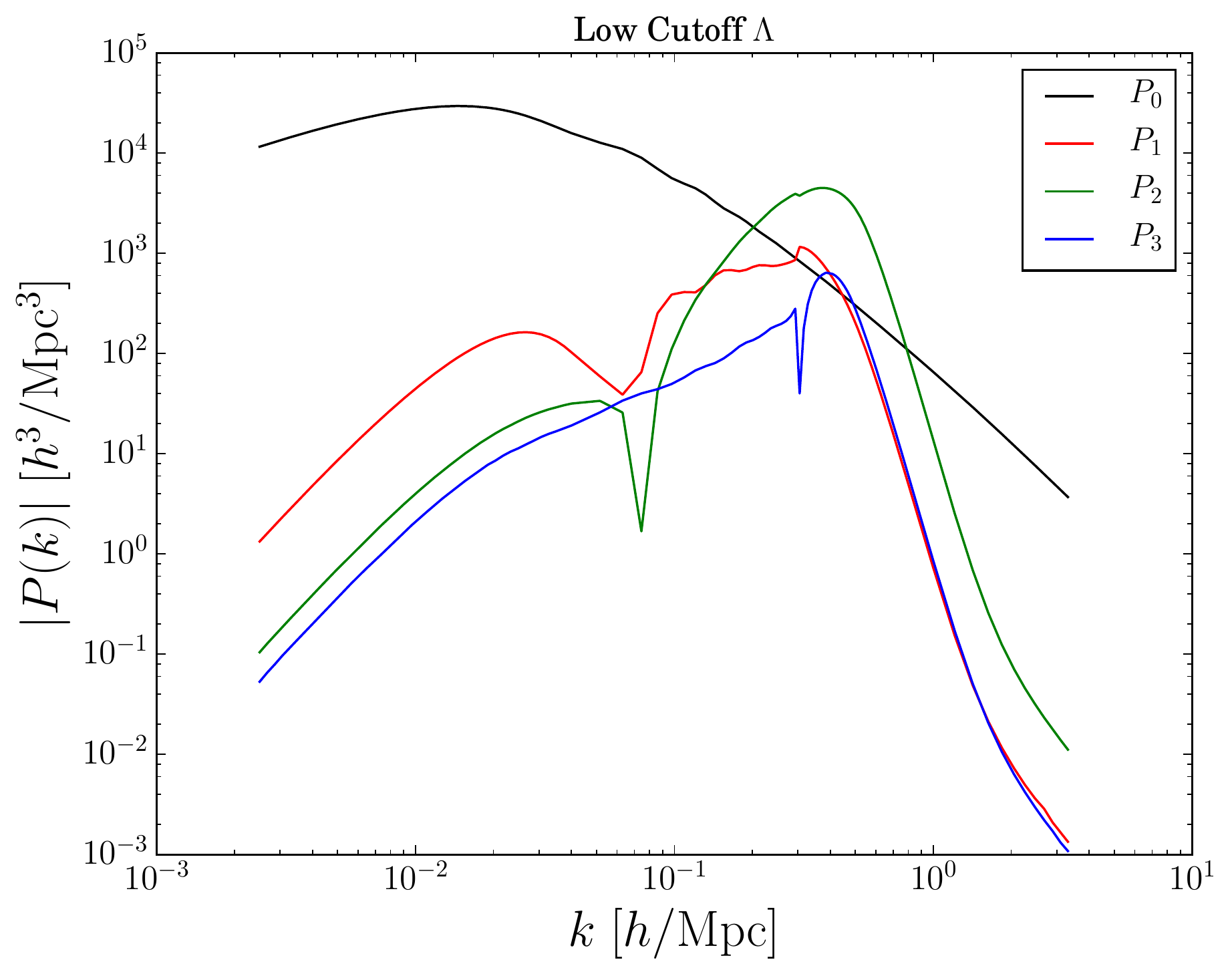}
  \includegraphics[width=0.3\textwidth]{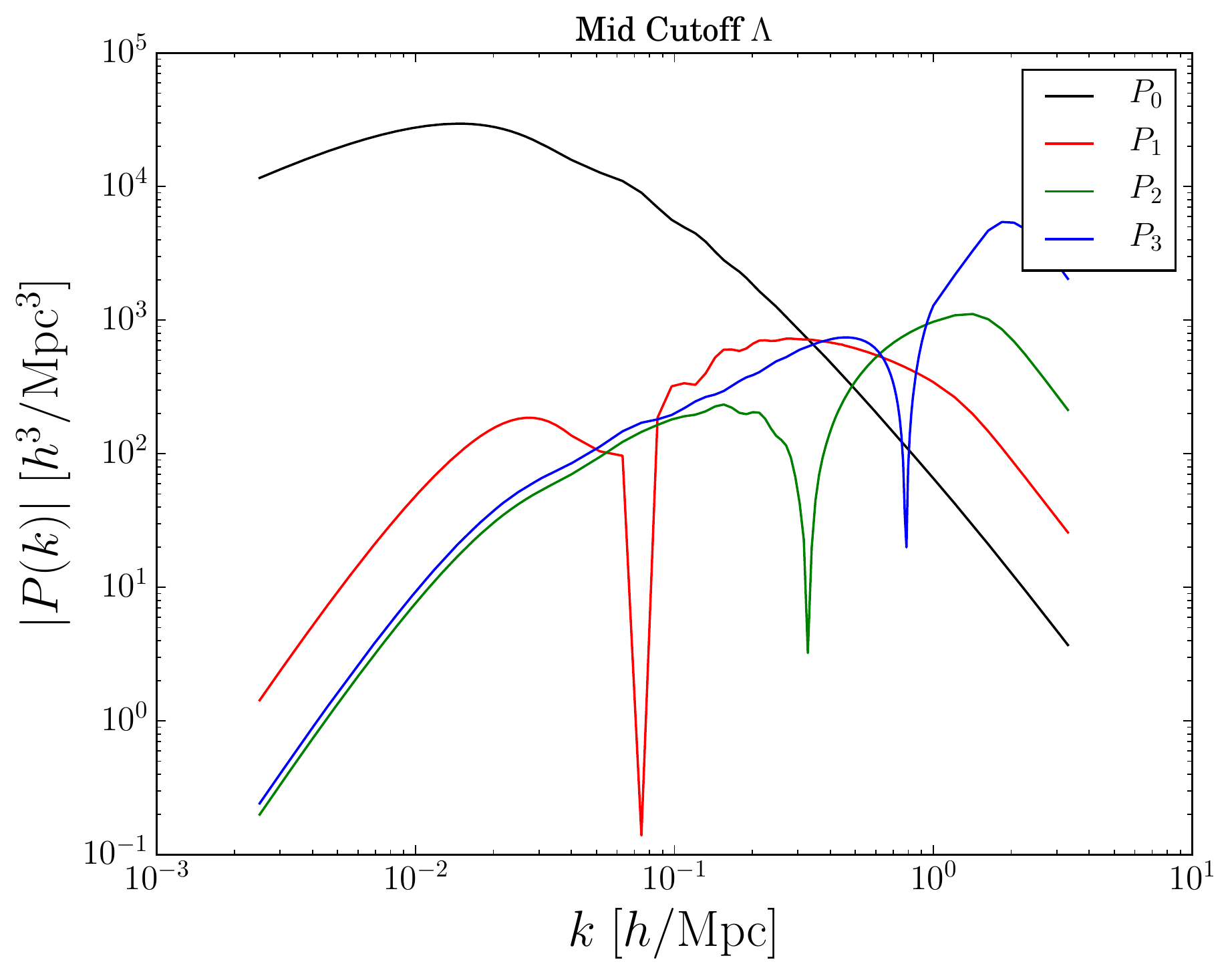}
  \includegraphics[width=0.3\textwidth]{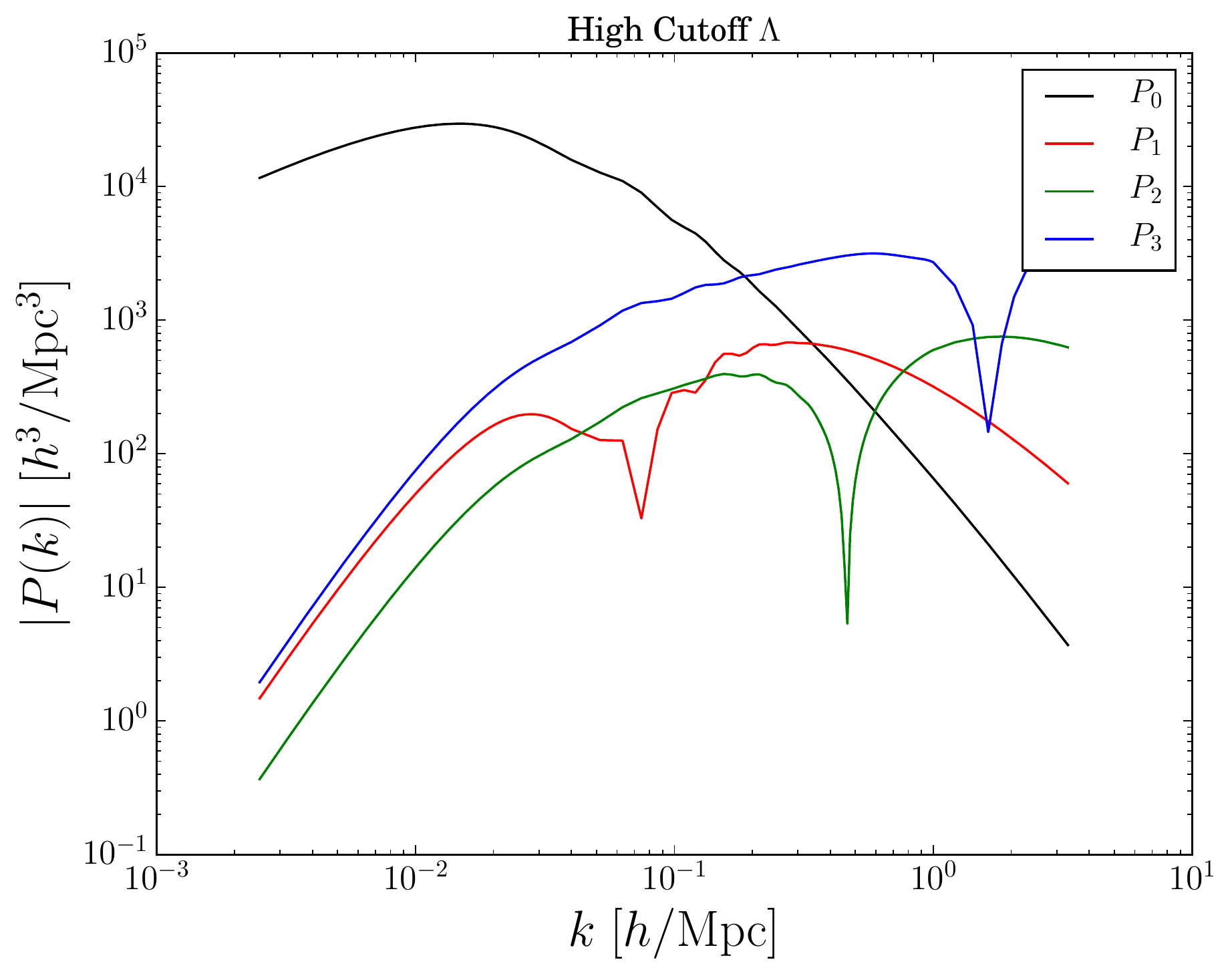}
\caption{\label{fig:Ps}
\small The SPT power spectrum at a given loop order ($P_{\ell}$) is plotted in the first row, calculated with high ($\Lambda = 60 \,h\, \rm{ Mpc}^{-1}$), moderate ($\Lambda = 0.7 \, h\,\rm{ Mpc}^{-1}$) and low ($\Lambda = 0.3 \,h\,  \rm{ Mpc}^{-1}$) cutoffs. The second row compares the size of each loop term using different cutoffs. The kink in the plot is due to a change in sign in the $P_{\ell}$'s, while we plot the absolute value. The reader will notice that $P_3$ dominates over the other contributions at moderate values of $k$, and only becomes (somewhat) smaller in the low cutoff case. This feature, as we shall see, anticipates much of our conclusions in our paper.}
\end{figure}

\subsection{Cutoff our ignorance}

The basic idea of the EFT approach is to introduce a cutoff scale, $\Lambda$, below which the perturbative expansion is under control, and above which some UV information is needed. The cutoff dependence is absorbed into Wilson coefficients, $c_i(\Lambda)$, up to a given loop order. The theory becomes predictable once a finite set of coefficients, which also incorporate the true non-linear information, are read off either from data or a realization of the full theory. For the SPT computations at $\ell$-loop order we have
\be
\label{eq:pspt}
P^{\rm SPT}_{\ell{\text -}\rm loop}(k,\Lambda)  = P_0 + P_1(k,\Lambda) + \cdots + P_{\ell}(k,\Lambda)\,,
\ee
where $P_{\ell}$ denotes the SPT prediction of the power spectrum up to $\ell$ loops, with the integrals appearing in the perturbative expansion cut off at $k< \Lambda$.\footnote{As it has been repeatedly emphasized, the finiteness of SPT integrals for a given cosmology does not imply the errors are necessarily small, let alone under theoretical control \cite{review}.} We will use both, a high ($\Lambda = 60 \,h\, \rm{ Mpc}^{-1}$) and a moderate ($\Lambda = 0.7 \, h\, \rm{ Mpc}^{-1}$) cutoff, the latter used as a proxy for the non-linear scale. Figure~\ref{fig:Ps} shows the SPT results at one $(P_1)$, two ($P_2$) and three ($P_3$) loop orders. The reader will immediately notice that $P_3$ is somewhat larger than $P_2$ even for a moderate cutoff, and it becomes smaller only when $\Lambda \lesssim 0.3\, h\, \rm{ Mpc}^{-1}$, which we also display for comparison in Fig.~\ref{fig:Ps}.  As we shall see, this fact turns out to have important implications for the convergence of the EFTofLSS, even after all the counter-terms are included.

\subsection{Counter-terms and UV test}\label{sec:UV}
Once the counter-terms, $c_i(\Lambda)$, are added the power spectrum in the EFTofLSS takes the form, schematically:
\be
P^{\rm EFT}_{\ell{\text -}\rm loop}(k) = P^{\rm SPT}_{\ell{\text -}\rm loop}(k,\Lambda) + {\rm counter}\text{-}{\rm terms}(\Lambda)\,, 
\ee
where $P^{\rm SPT}_{\ell{\text -}\rm loop}(k,\Lambda)$ is given in \eqref{eq:pspt}. There are several counter-terms which contribute at three loop orders. However, one can show many of which are less important to reproduce the data. For concreteness, the following set can be singled out:\footnote{There are various notations in the EFT literature for the Wilson coefficients. We will keep using $c_s$ for the sound speed. For terms involving $n$-derivatives we introduce a $c_n$ parameter. For terms quadratic in the perturbations we will  also use a `quad' label, as in \cite{Foreman:2015lca}.}
\bea
\label{eq:eft_basis}
&\Big\{ 
4 \pi c_s^2 k^2 P_0(k), \quad
8 \pi^2 \bar c_4 k^4 P_0(k), \quad
4 \pi c_{2, \rm quad} k^2\, P_{\rm quad}(k),\quad  & \\
&  4 \pi^2 c_{\rm stoch} k^4, \quad
16 \pi^3 \bar c_6 k^6 P_0(k),  \quad
8 \pi^2 c_{4,\rm quad} k^4 P_{\rm quad}(k)  
\Big\} & , \nonumber
\eea
where $P_0(k)$ is the linear power spectrum, and
\bea
\bar c_4 = -\frac12 c_s^4 + c_4 \, , \nonumber \\
\bar c_6 = - c_s^2\, c_4  + c_6 \, ,
\eea
(see appendix~\ref{appA} for more details). As we discuss momentarily, these coefficients are sufficient to deal with the UV-sensitivity of the SPT computations. All the above expressions must be expanded up to the relevant loop order. For instance, the sound speed $c_s$ is inherently of one loop order, but it also enters at higher orders and therefore 
\be 
\label{csum} c^2_s = c^2_{s (1)} + \cdots + c^2_{s (\ell)}\,, 
\ee
to $\ell$-loop order. On the other hand, the coefficients $c_4$, $c_{2,\rm quad}$ and $c_{\rm stoch}$ are naturally of two loop order, hence $(\ell-1)$ coefficients would be needed, and so on. Notice that, at two loops and beyond, we encounter counter-terms which descend from loop corrections to higher $n$-point functions, such as $c_{2,\rm quad}$ and $c_{4,\rm quad}$. The leading contribution comes from a contraction with the three-point function, see appendix~\ref{appA}. Higher order corrections may be obtained from the expansion of the stress tensor at higher orders in the density field, see e.g. \cite{Foreman:2015lca}.\footnote{In principle, given our set in \eqref{eq:eft_basis}, $P_{\rm quad}$ should be computed to next-to-leading order, giving rise to another counter-term, i.e. $c_{2,{\rm quad} (2)}$. However, we will show that the leading order $P_{\rm quad}$ is sufficient to renormalize the theory. The same applies to other counter-terms to three loop orders. In fact, one can show that $c_{4,\rm quad}$ has little impact at three loops, and therefore it will be omitted when we fit to the numerical data. This will also reduce the number of free parameters in the EFT approach.} \vskip 4pt


In the EFTofLSS we then have one new term at one loop, $c^2_{s (1)}$, four extra coefficients at two loops, $\{ c^2_{s (2)}, c_{4(1)}, c_{2,{\rm quad}}, c_{\rm stoch} \}$, and three more to three loops, $\{c^2_{s (3)}, c_{4 (2)}, c_6\}$.  A first check of this choice consists on explicitly checking whether the UV sensitivity of the SPT results can be absorbed into the counter-terms.  For this `UV test' we use the high cutoff SPT result and renormalize it using the EFTofLSS, and then compare against the moderate cutoff answer, as a proxy of the non-linear scale, see Fig.~\ref{fig:Ps}. As a measure for the fit we use the residuals
\be
\label{eq:def_chi}
\chi^2_{\rm UV\text{-}test} = \sum_n  \left(\frac{P^{\rm SPT}_{\ell \text{-}\rm loop}(k_n,\Lambda=0.7 \, h\,{\rm Mpc}^{-1}) - P^{\rm EFT}_{\ell \text{-}\rm loop}(k_n,\Lambda=60 \, h\,{\rm Mpc}^{-1})}{P^{\rm SPT}_{\ell \text{-}\rm loop}(k_n,\Lambda=0.7 \, h\,{\rm Mpc}^{-1})}\right)^2 \, , 
\ee
to a given loop order. This is introduced in analogy to the measure in \eqref{eq:def_chi_data} we use to test the quality of the EFT fit to data. 
\begin{figure}[ht]
\centering
  \includegraphics[width=0.3\textwidth]{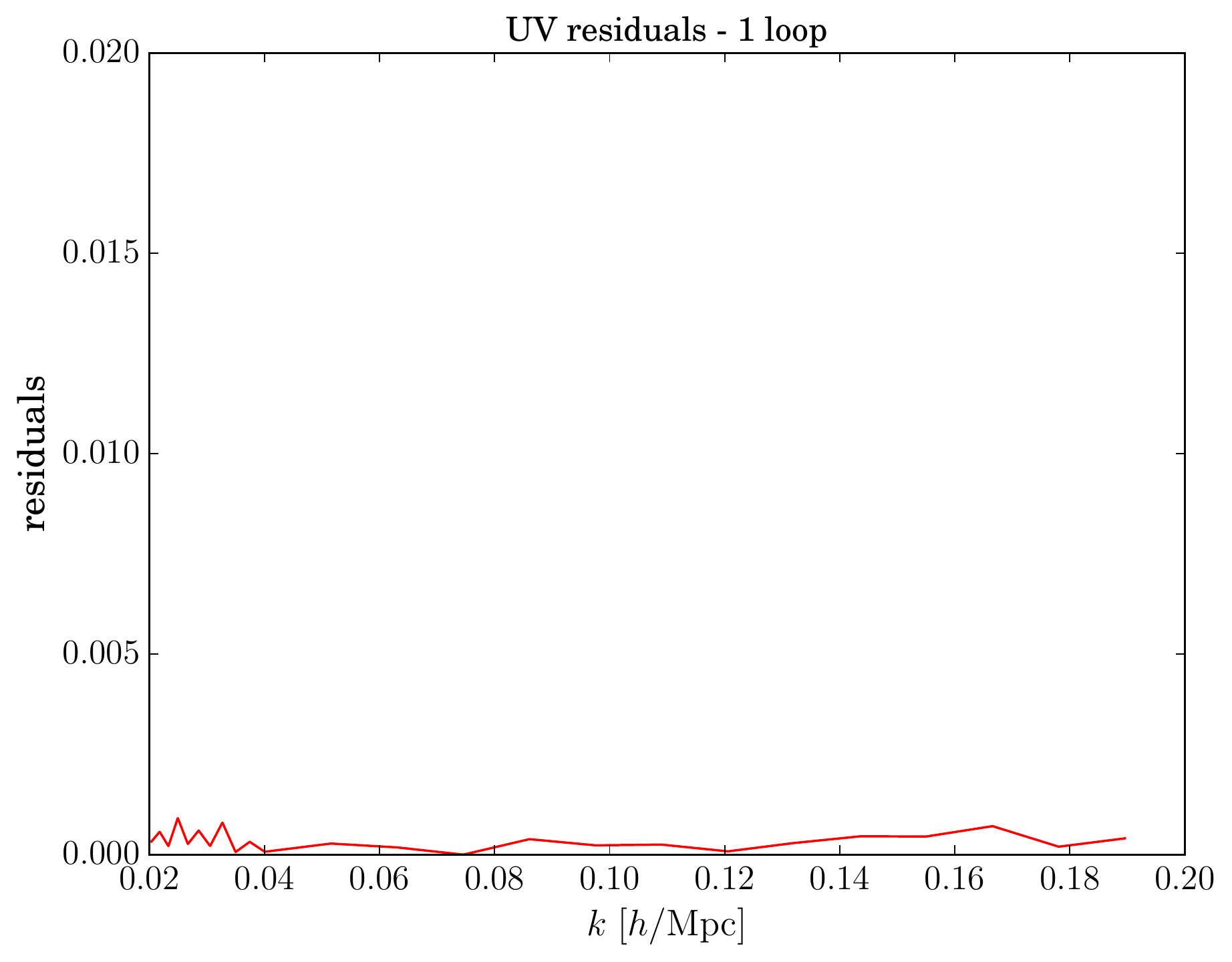}
  \includegraphics[width=0.3\textwidth]{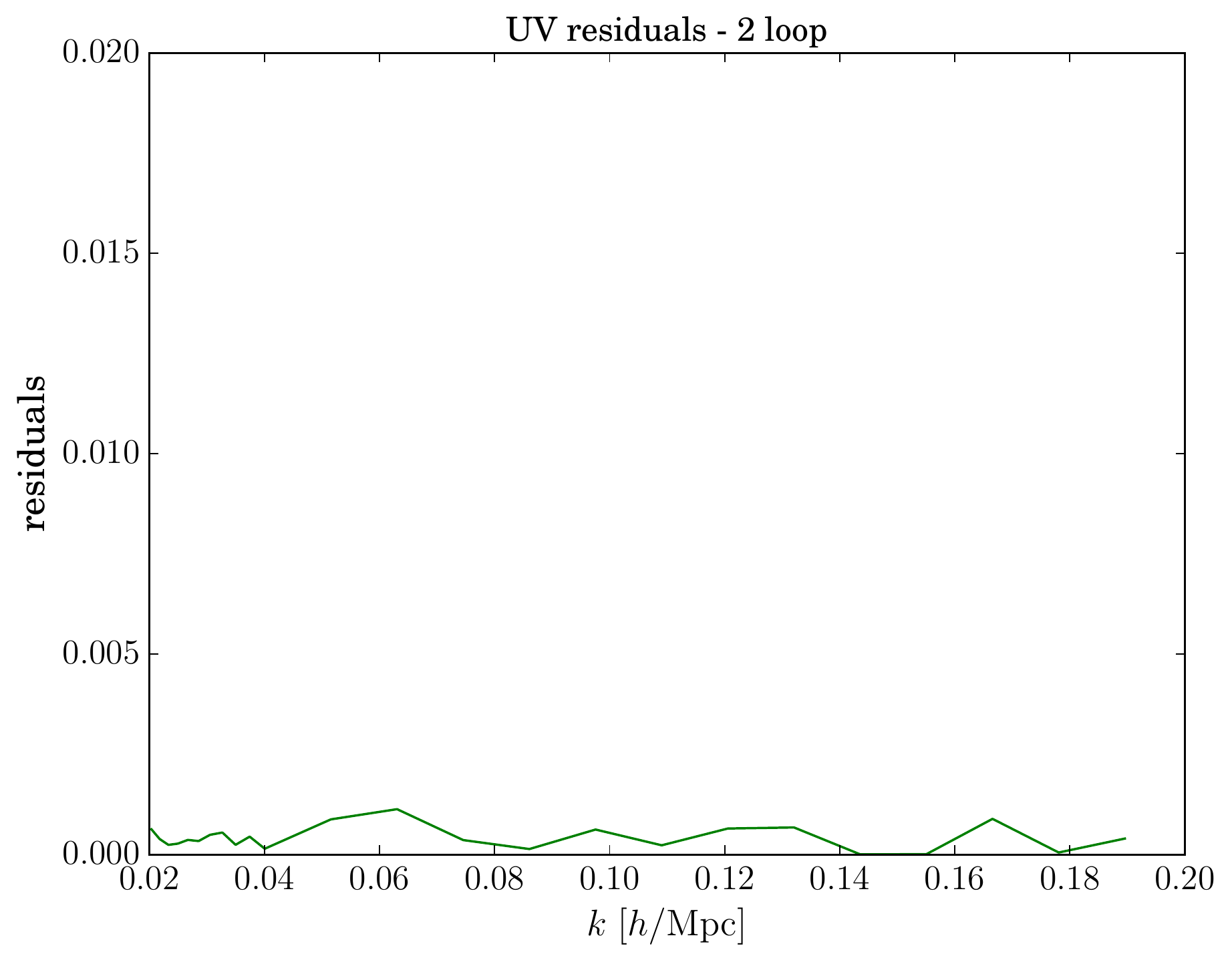}
  \includegraphics[width=0.3\textwidth]{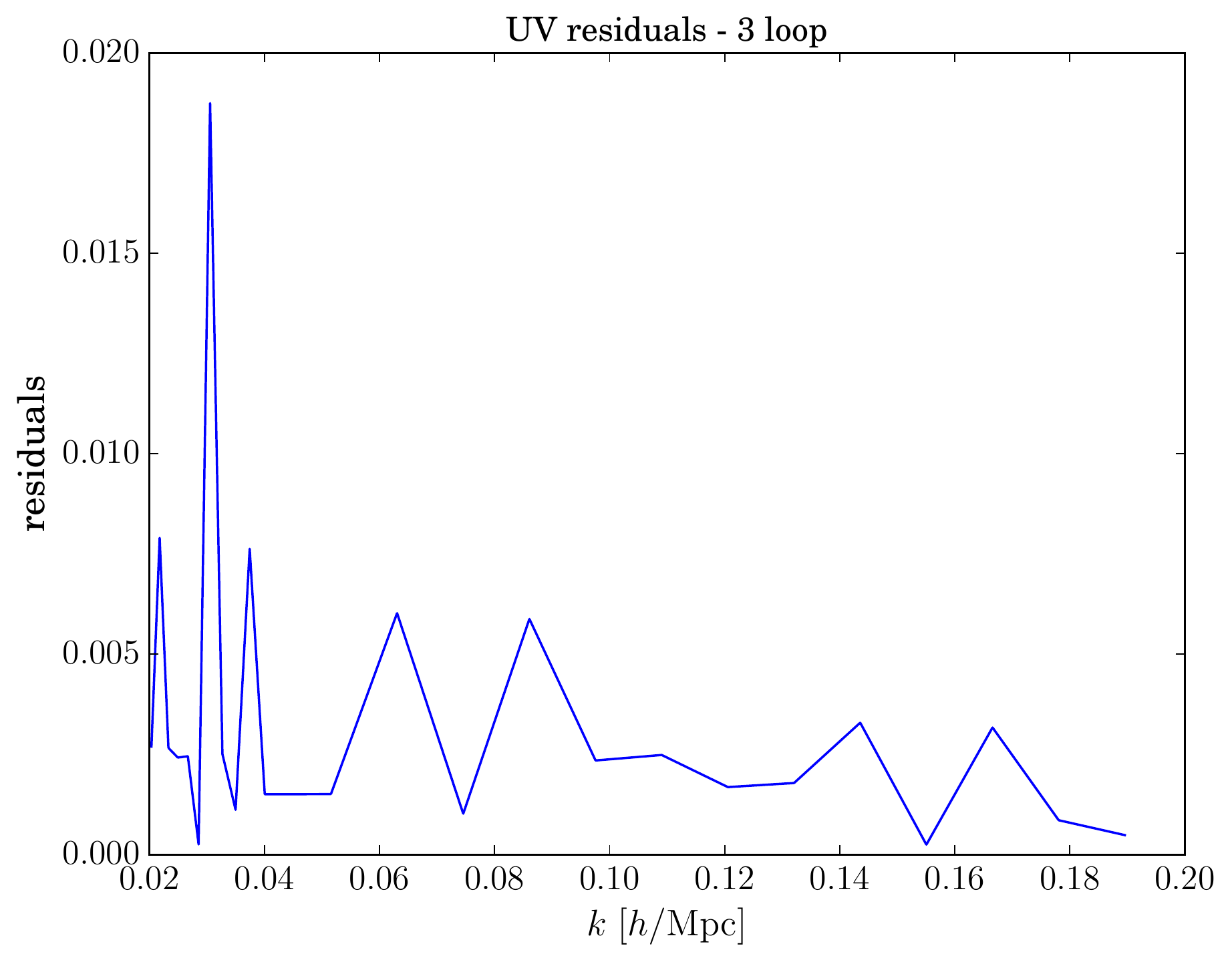}
  \includegraphics[width=0.3\textwidth]{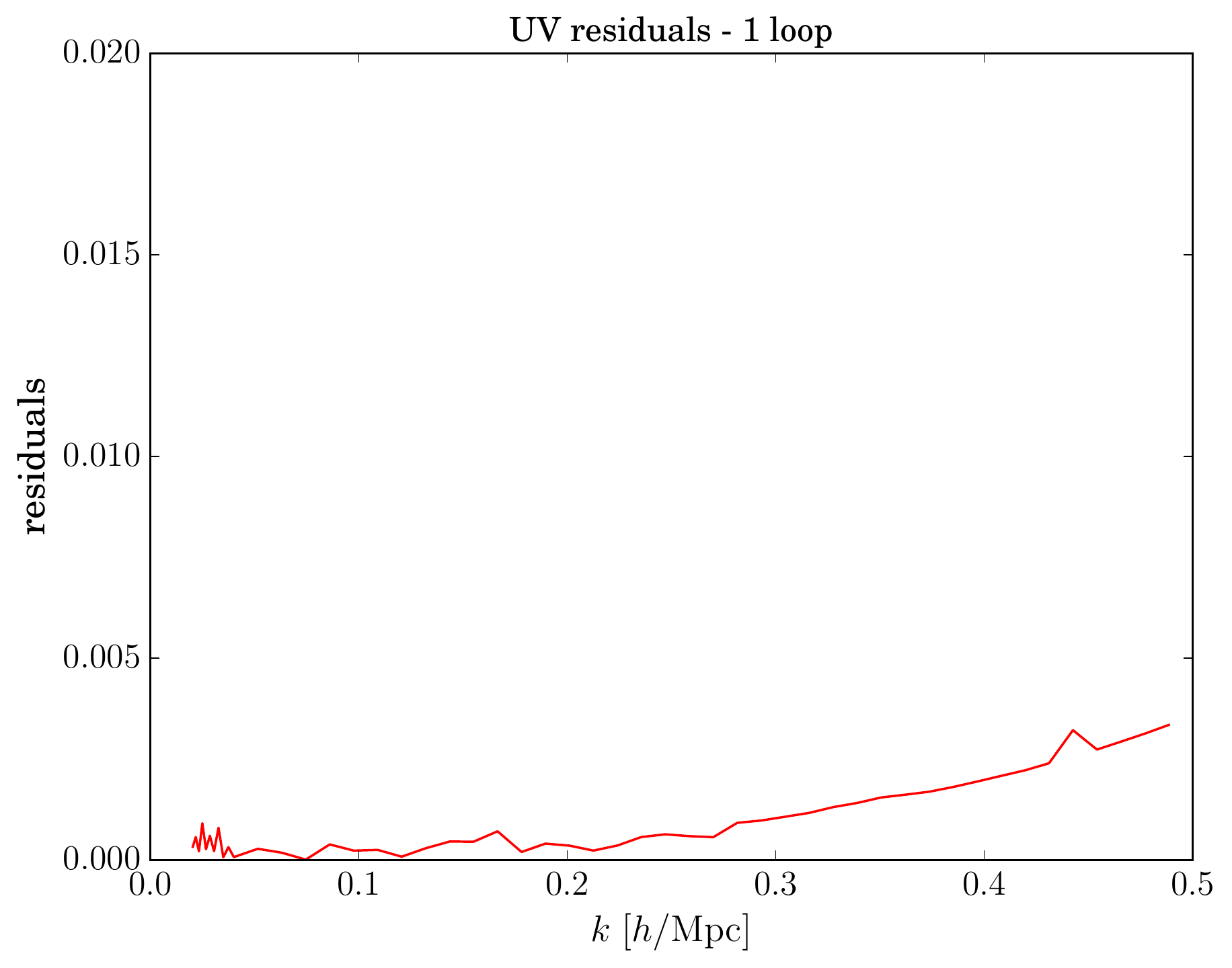}
  \includegraphics[width=0.3\textwidth]{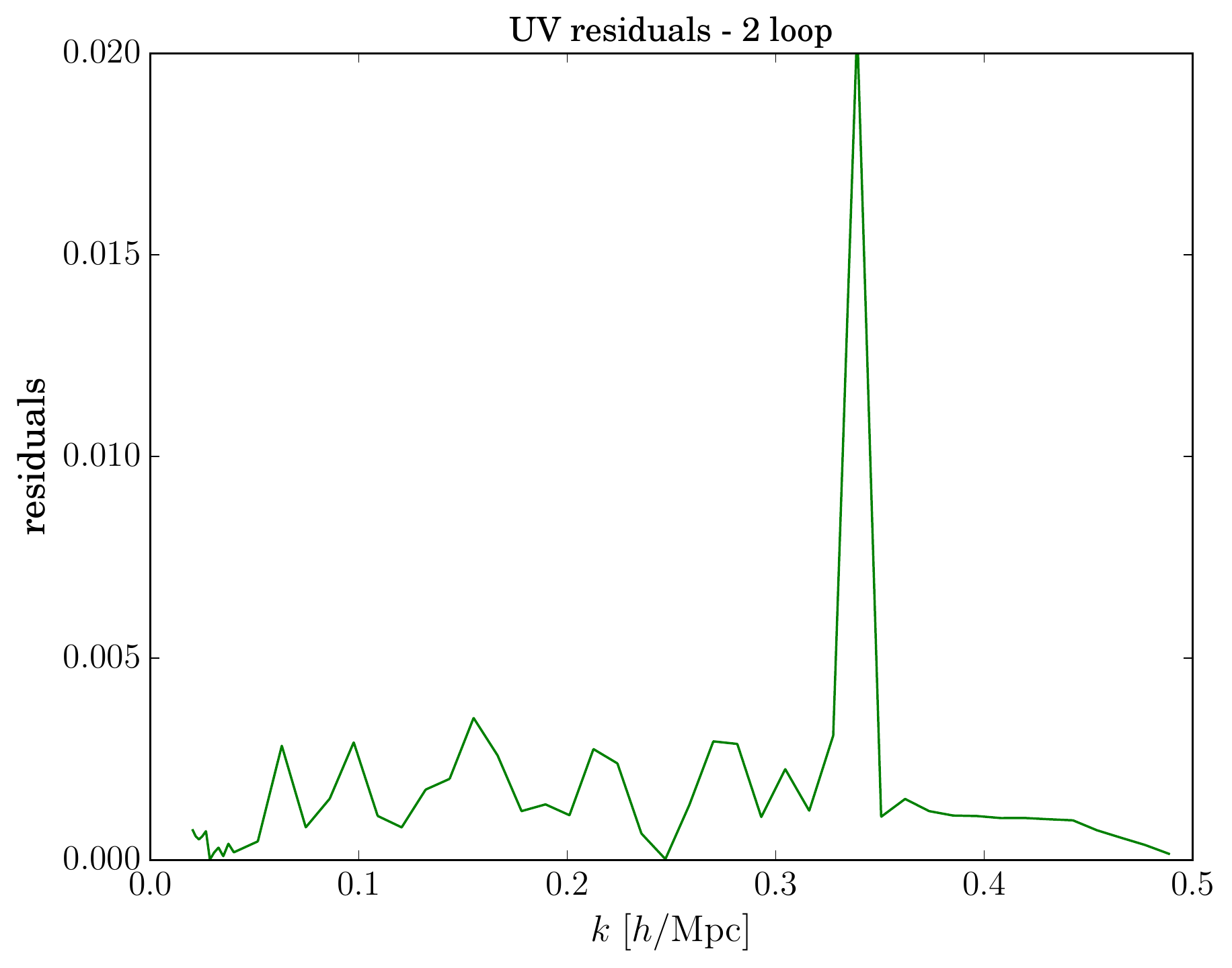}
  \includegraphics[width=0.3\textwidth]{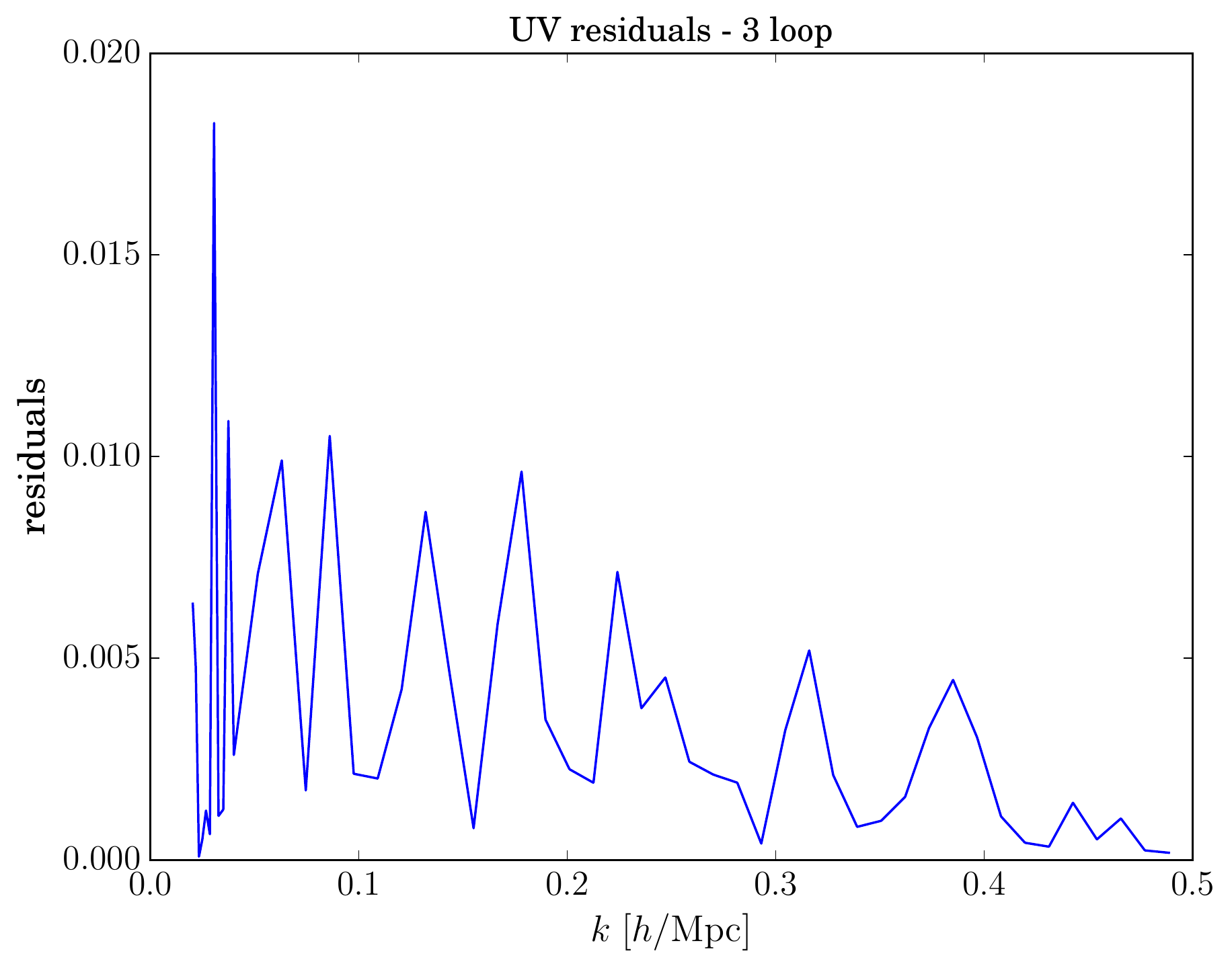}
\caption{\label{fig:UVresiduals}%
\small The residuals from fitting the EFTofLSS with the appropriate counter-terms to the SPT spectra with a moderate cutoff. The top row uses a fit in the momentum range $k \in [0.02,0.2]$ Mpc$^{-1}$, while for the bottom row
we use $k \in [0.02,0.5]$ Mpc$^{-1}$. The three columns show the one, two and three loop results, respectively.
}
\end{figure}
\begin{figure}[ht]
\centering
  \includegraphics[width=0.45\textwidth]{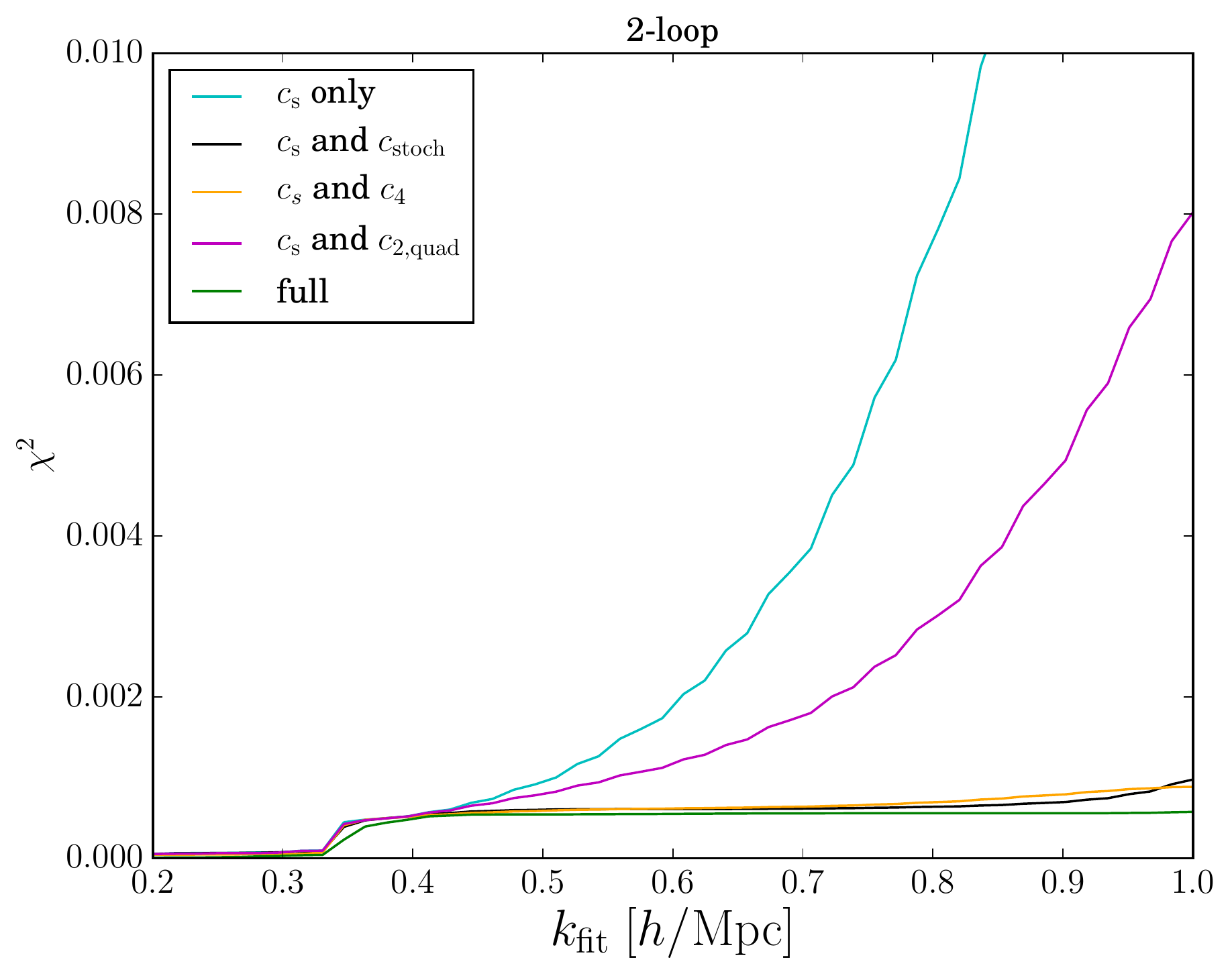}
  \includegraphics[width=0.45\textwidth]{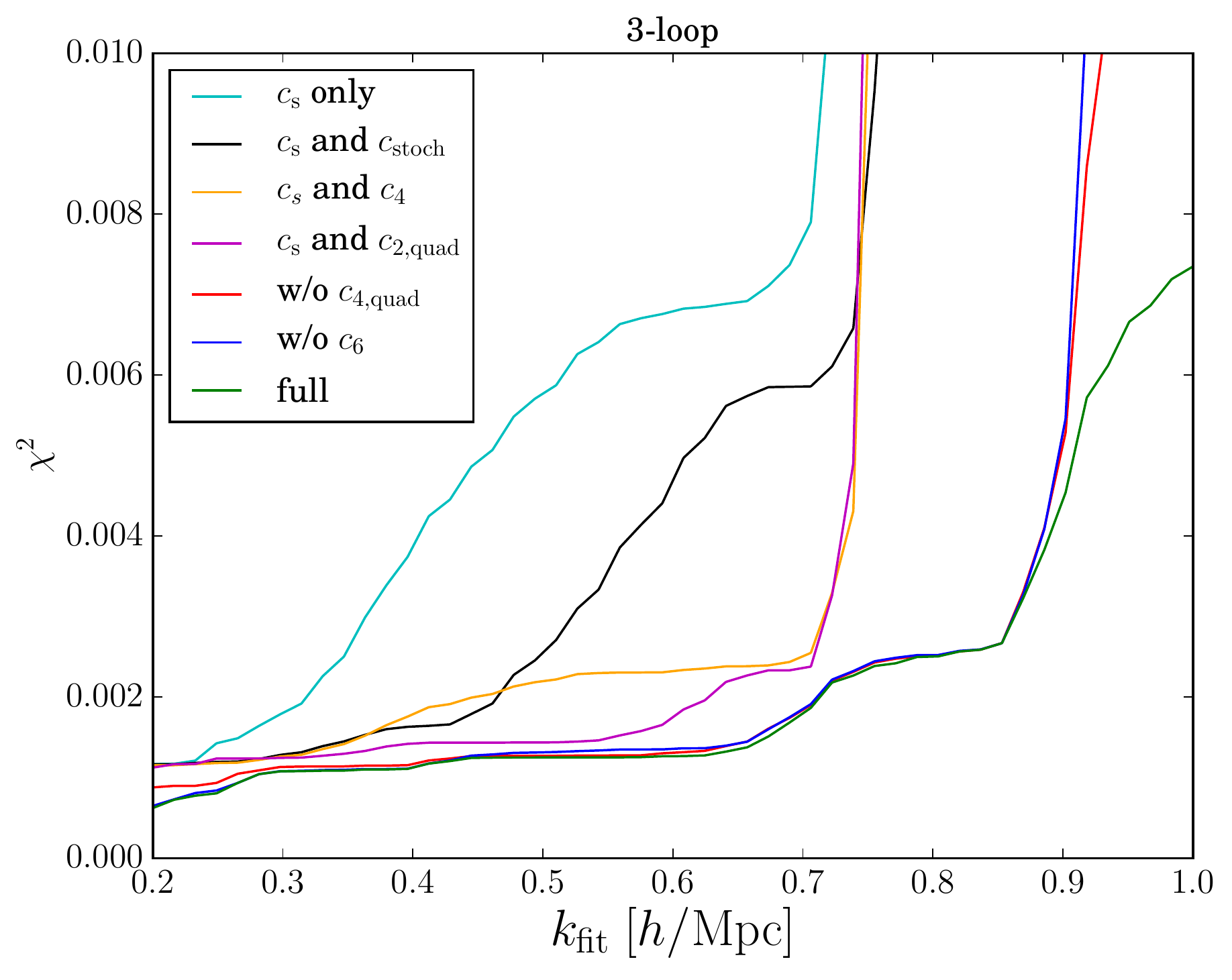}
\caption{\label{fig:UVmodels}%
\small The $\chi^2$ for the UV test (EFT with high cutoff vs. SPT with moderate cutoff) for different sets of Wilson coefficients. The left and right plots are for two and three loops respectively.
}
\end{figure}

\vskip 4pt Figure~\ref{fig:UVresiduals} shows the results for the one, two and three loops for two different fitting regions, $k \in [0.02,0.2]$ Mpc$^{-1}$ (top)  and $k \in [0.02,0.5]$ Mpc$^{-1}$ (bottom). In all cases, the residuals are below $1\%$. (This is particularly surprising in the bottom row, since the fitting range comes rather close to the cutoff scale, $\Lambda = 0.7 \, \rm{ Mpc}^{-1}$.) Notice the spike in $\chi^2$ in the 2-loop results. This is due to a zero in the denominator of the measure introduced in \eqref{eq:def_chi}, which can be easily removed from the fit to get a better agreement. Figure~\ref{fig:UVmodels} displays the results of the UV test, for models where different counter-terms are removed from the fit.\footnote{Notice that the step in the accumulated $\chi^2$ results from the spike in Fig.~\ref{fig:UVresiduals}, due to the normalization, as mentioned before.} At two loops the UV dependence can be absorbed into $c_s, c_4$ and $c_{\rm stoch}$. For this UV test, counter-term associated with $c_{2,\rm quad}$ has no noticeable effect to two loops. As we will see later, this changes when we perform the matching to the numerical simulations. This means that a renormalized finite contribution will be required to fit the data. At three loop level, the UV dependence can be removed by adding the sound speed as well as $c_4$ and $c_6$. 
The remaining counter-terms have little impact on the UV dependence in the momentum range we are interested in. Yet, similarly to the two loop case, a renormalized contribution may be needed.\vskip 4pt 

As we mentioned, because we cannot separate the counter-term from the renormalized (cutoff independent) contributions, we will match for all of the $c_i$'s without distinction. The UV test, however, tells us that we have included all the needed  counter-terms in \eqref{eq:eft_basis} to account for the UV dependence in the SPT computations. We will show shortly that we have also included those which capture (most of) the true non-linear information in the numerical data. In principle, one could include additional counter-terms (from extra contributions to the stress-energy tensor in Euler equation). This can potentially improve the fit to the data. However, while the perturbative expansion is under control,  we do not expect their impact to be large. The reason is the following. Any new term in the EFT expansion with a parameter without a significant cutoff dependence (such as renormalized contributions carrying true short-distance information as opposite to pure counter-terms) will be down by powers of the expansion parameter (in our case the density perturbation) relative to those with counter-terms (or lower orders). If adding these new parameters we find large contributions without a counter-term part (needed to fix the SPT UV behavior), this would be an indication of the failure of the perturbative series. Therefore, we will fit the numerical data assuming that is not the case, and resort only to the counter-terms needed to remove the cutoff dependence of the perturbative expansion, and their correspondent finite (renormalized) parts.

\section{Comparison with simulations\label{sec:results}} 

We present now the results of the fit to numerical data at redshift $z=0$, and the subsequent determination of the counter-terms of the EFTofLSS using the SPT results with the large cutoff.
\begin{figure}[t!]
\centering
  \includegraphics[width=0.6\textwidth]{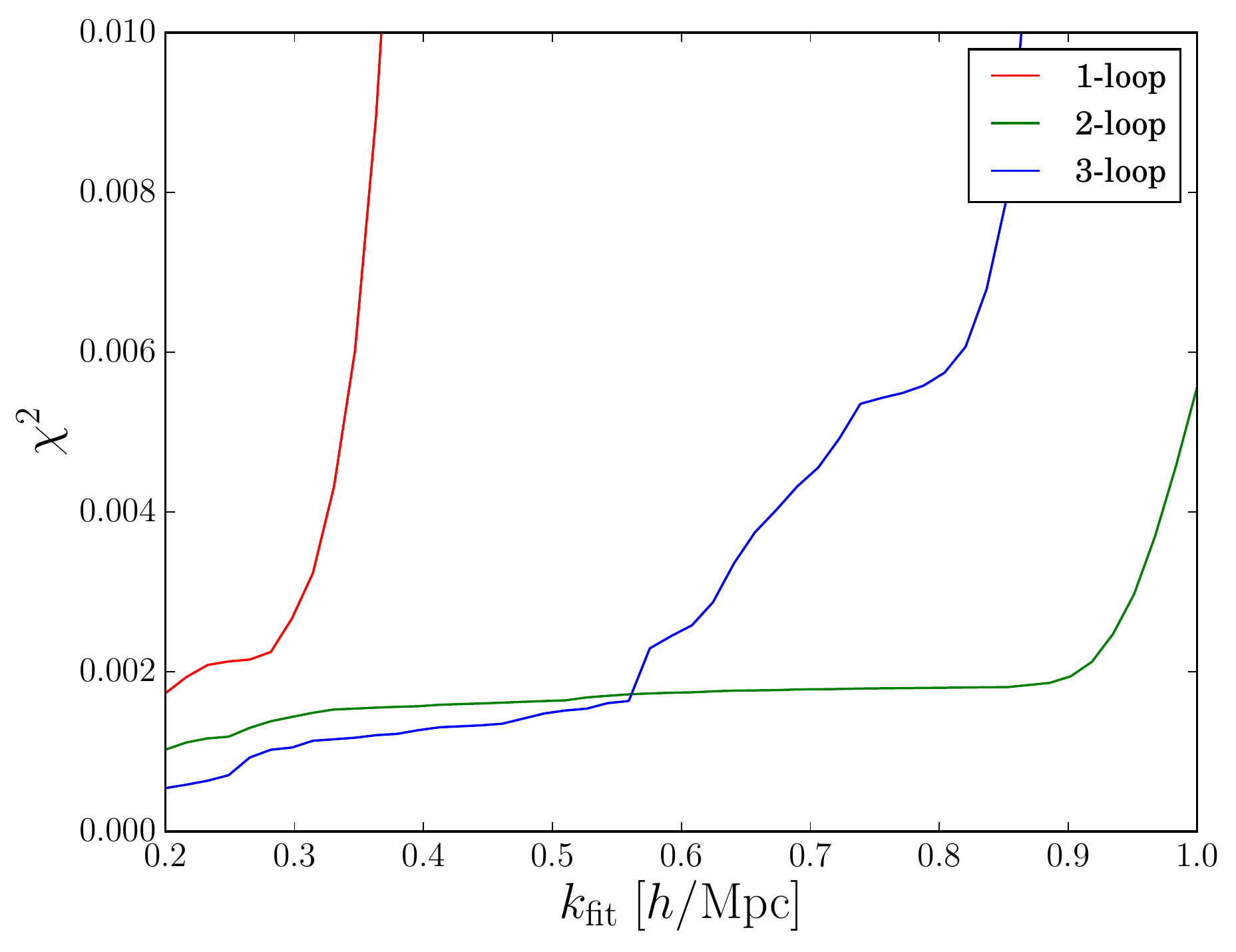}
\caption{\label{fig:chi2summary}%
\small The cumulative $\chi^2$ as a function of the upper bound of the fitting range for the $\ell$-loop EFT results, with $\ell = 1,2,3$.}
\end{figure}
In Figure~\ref{fig:chi2summary} we display the resulting $\chi^2$ including all our available EFT parameters at one, two and three loop orders, respectively. The range of $\chi^2$ chosen is such that the prominent increase shown in the plot is indicative of residuals moving above the $1\%$ level, which is the accuracy we are aiming at in this paper. We immediately see that the EFTofLSS to two loops provides an excellent fit to the data up to wavenumbers of order $k\simeq 0.9\, h\,$Mpc$^{-1}$. (There is, however, some reliance on an unphysical value (negative) for the stochastic term for $k \gtrsim 0.7\, h$ Mpc$^{-1}$.) From the plot, we also conclude that the EFTofLSS to three loop orders shows some improvement with respect to the two loop results  in the regime $0.2 \lesssim k \lesssim 0.4\,h\,$Mpc$^{-1}$. Nevertheless, the increase in accuracy is somewhat marginal. At the same time, around $k \simeq 0.55\, h\,$Mpc$^{-1}$, the power spectrum to three loops starts to be in tension with the numerical data. We will return the reasons for the discrepancy in sec.~\ref{sec:disc}.\vskip 4pt

\subsection{Non-linear information}

\begin{figure}[ht]
\centering
  \includegraphics[width=0.45\textwidth]{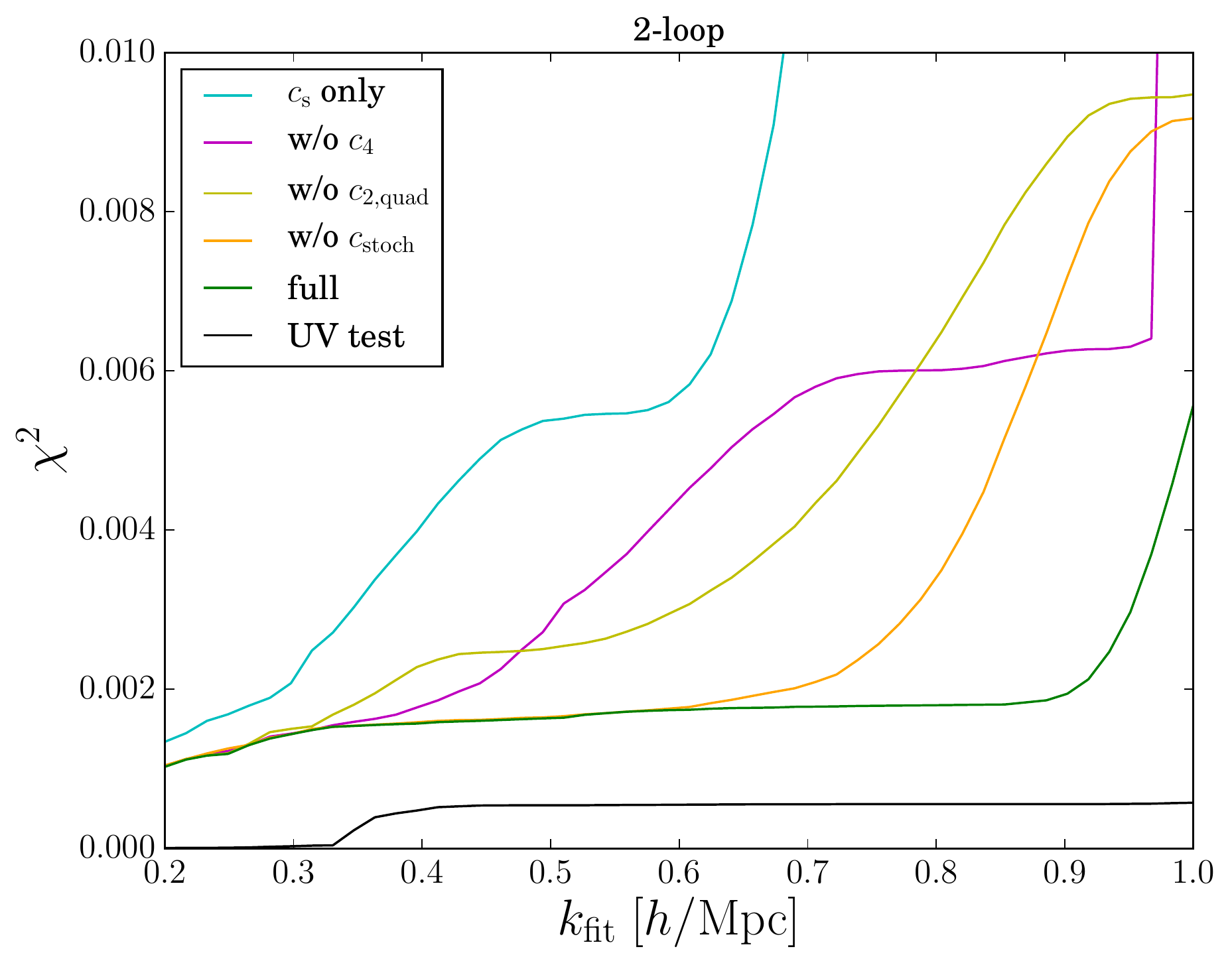}
  \includegraphics[width=0.45\textwidth]{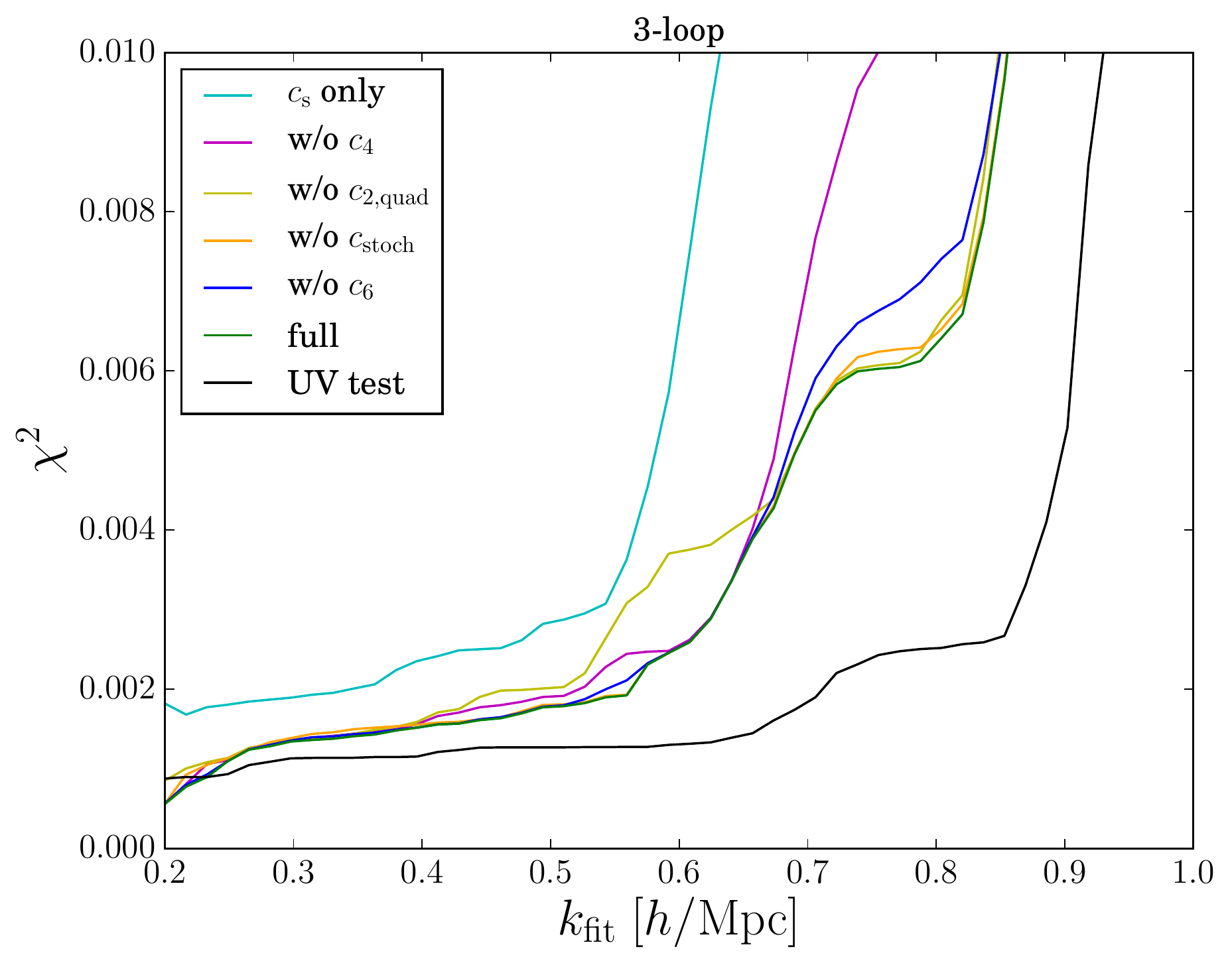}
\caption{\label{fig:chi2models}%
\small Same as Figure~\ref{fig:chi2summary}, but for different combinations of counter-terms. For comparison, we also include the result for the UV test (see text).}
\end{figure}

In Fig.~\ref{fig:chi2models} we show several EFT models at two and three loop orders with only a subset of Wilson coefficients. For comparison, we also provide the $\chi^2$ associated with the UV test, which gave us the necessary counter-terms to remove the short-distance sensitivity of SPT. The difference between the UV test and the fit to the simulations accounts also for the non-linear information in the numerical data. As we see, the relative difference is small, which means the extra matching coefficients to two loop orders not only correct the SPT results, they also incorporate all of the relevant non-linear information to a very good level of accuracy.\vskip 4pt

For the two loop results, in the left panel of Fig.~\ref{fig:chi2models}, we immediately realize that the sound speed is no longer sufficient, and the fit to the power spectrum beyond one loop order can be significantly improved, in principle up to $k \simeq 0.9 h\,$Mpc$^{-1}$, by including more EFT coefficients. We notice, however, that at this high $k$ the value of the stochastic term starts to influence the results (see e.g. \cite{error}). We comment on this below. To three loops, shown in the right panel of Fig.~\ref{fig:chi2models}, amusingly adding just the sound speed provides a relatively good fit to the data up to $k \simeq 0.55 h\,$Mpc$^{-1}$. As we noticed earlier, when we performed the UV test, $c_s$~alone is not sufficient to remove the (incorrect) UV portion of the SPT calculation. Therefore, the reason for the apparent success of the fit is twofold. Firstly, there is non-linear information in the numerical data which resembles the $k^2$ behavior of the $c_s$ counter-term and, secondly, there are partial cancelations between the part of the counter-terms which correct the SPT result with high cutoff, and the true non-linear information which is also encoded in the matching coefficients. Notice that, after all, the fit does improve the more parameters are added. Nonetheless, as anticipated in Fig.~\ref{fig:chi2summary}, and starting around the same value, $k \simeq 0.55 h$ Mpc$^{-1}$, the result to three loops starts to deviate from the data, even when all the counter-terms from the UV test are included. As we argue here, the reason for the mismatch is not because of the UV sensitivity, or lack of counter-terms, but rather due to SPT contributions with loop momenta of order of the external momentum of the power spectrum.

\subsection{(Over)fitting the data}

\begin{figure}[ht]
\centering
\includegraphics[width=0.39\textwidth]{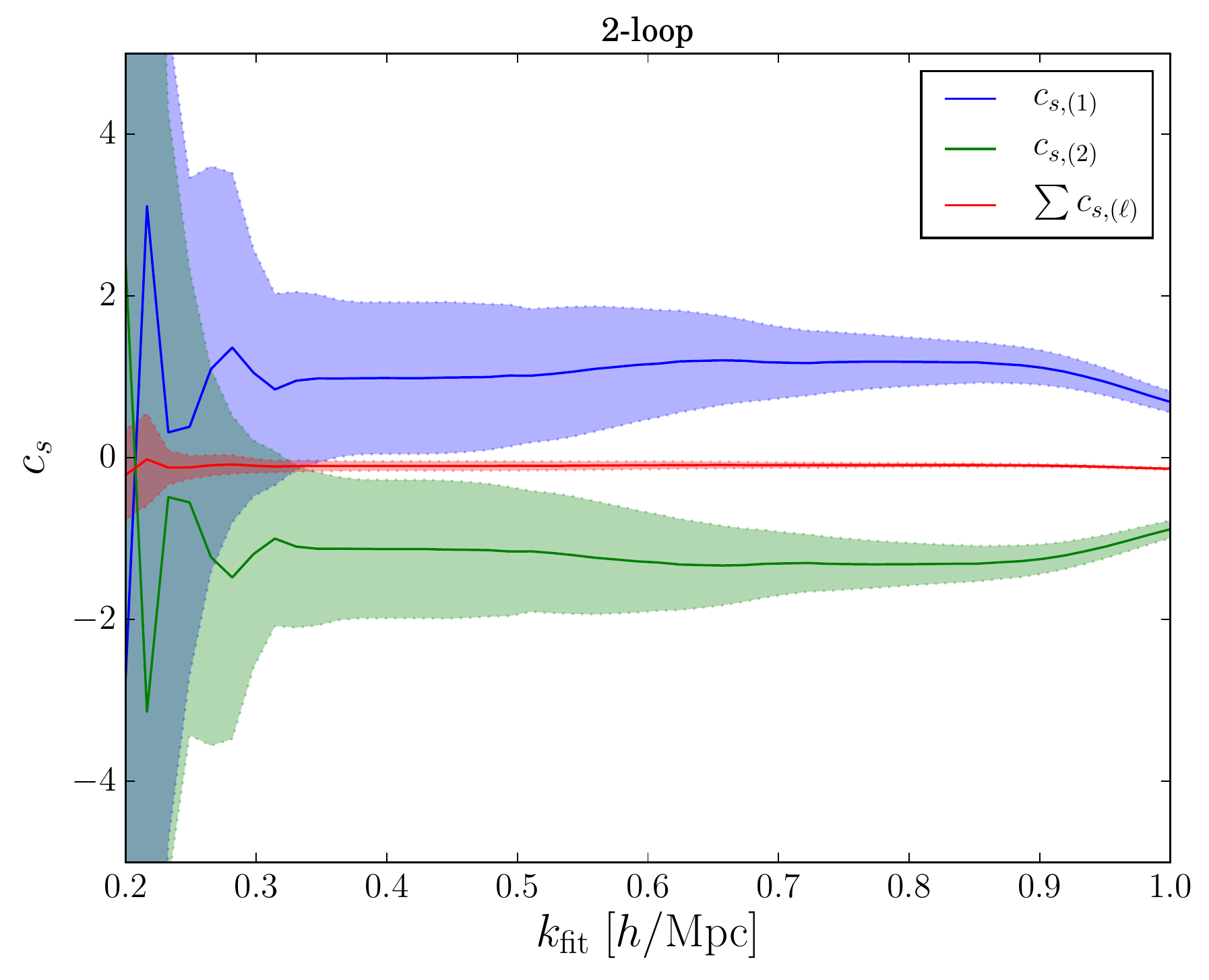}
\includegraphics[width=0.39\textwidth]{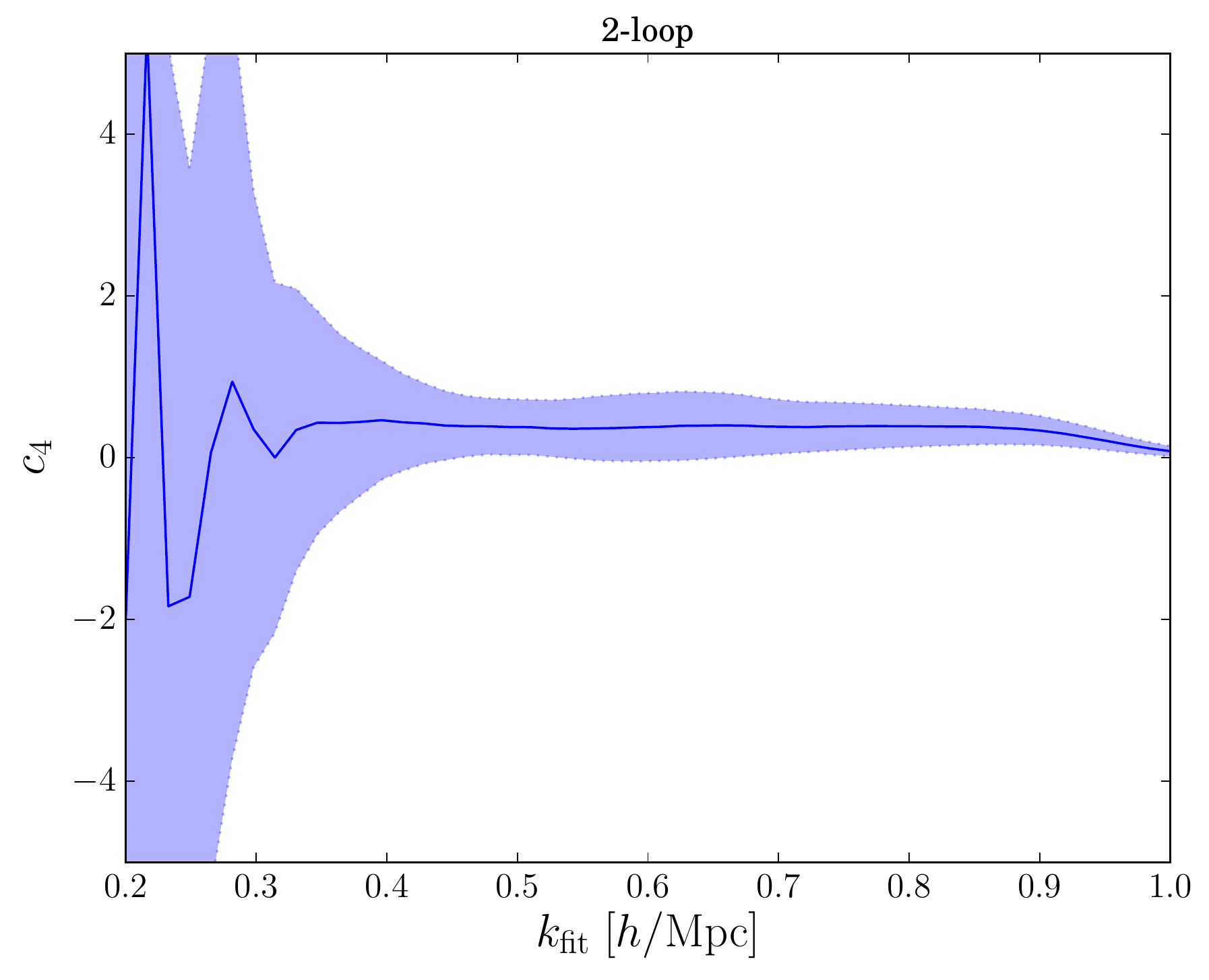}
\includegraphics[width=0.39\textwidth]{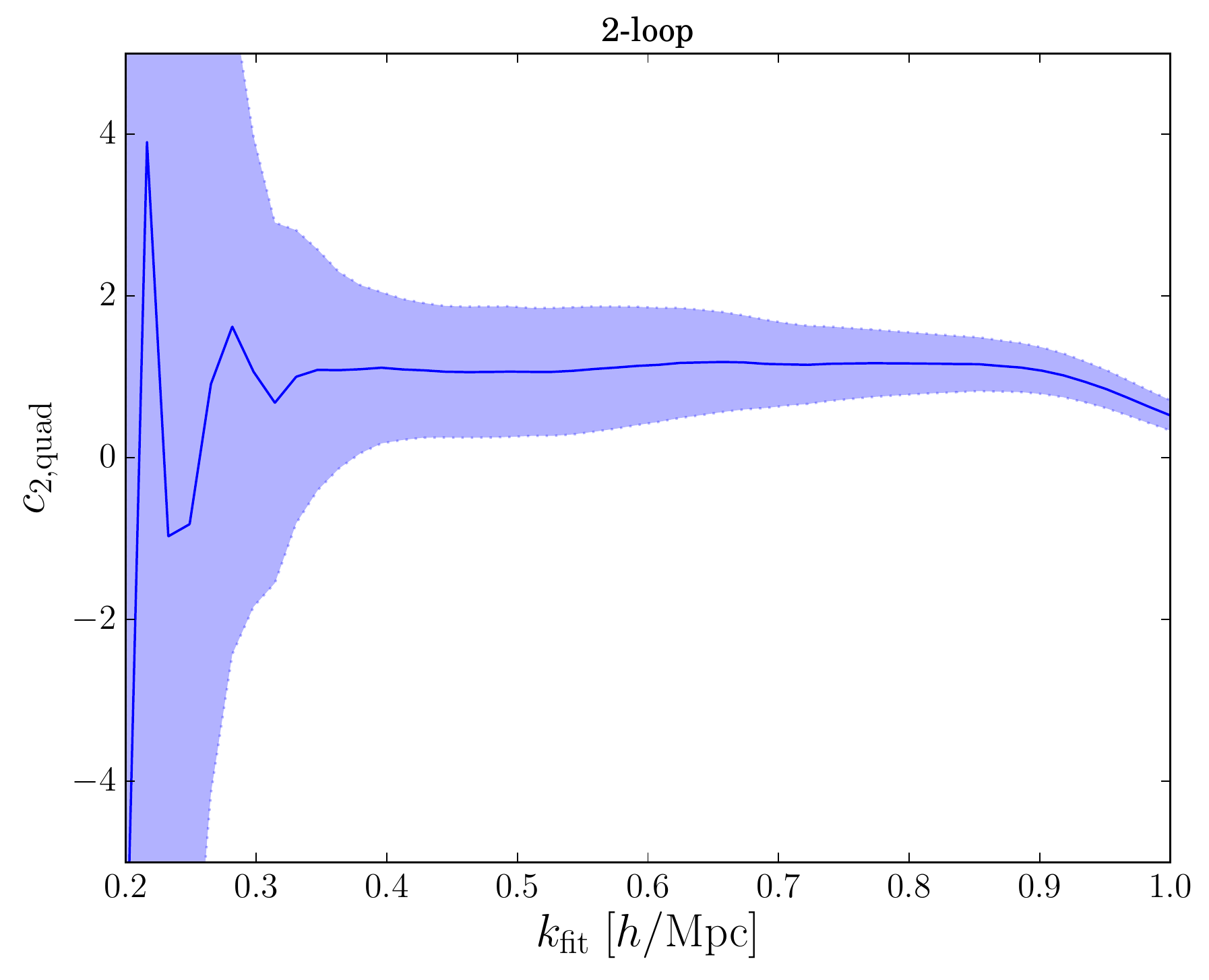}
\includegraphics[width=0.39\textwidth]{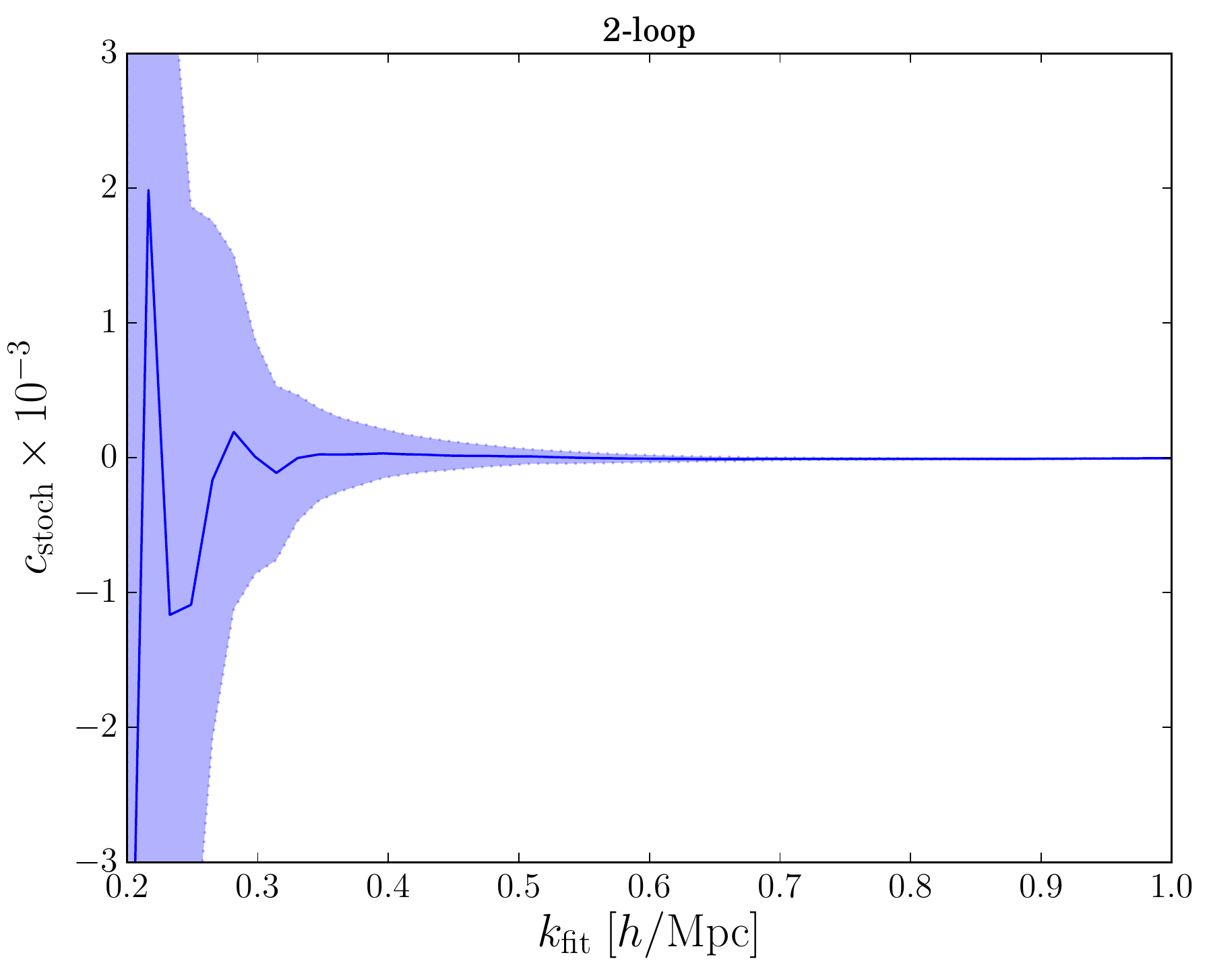}
\caption{\label{fig:cs2L}%
\small The fit for the relevant EFT coefficients up to two loops in the range $k_{\rm fit} \in [0.2,1]\, h/$Mpc. The error bands are determined as explained in the text.  }

\end{figure}

\begin{figure}[ht]
\centering
\includegraphics[width=0.39\textwidth]{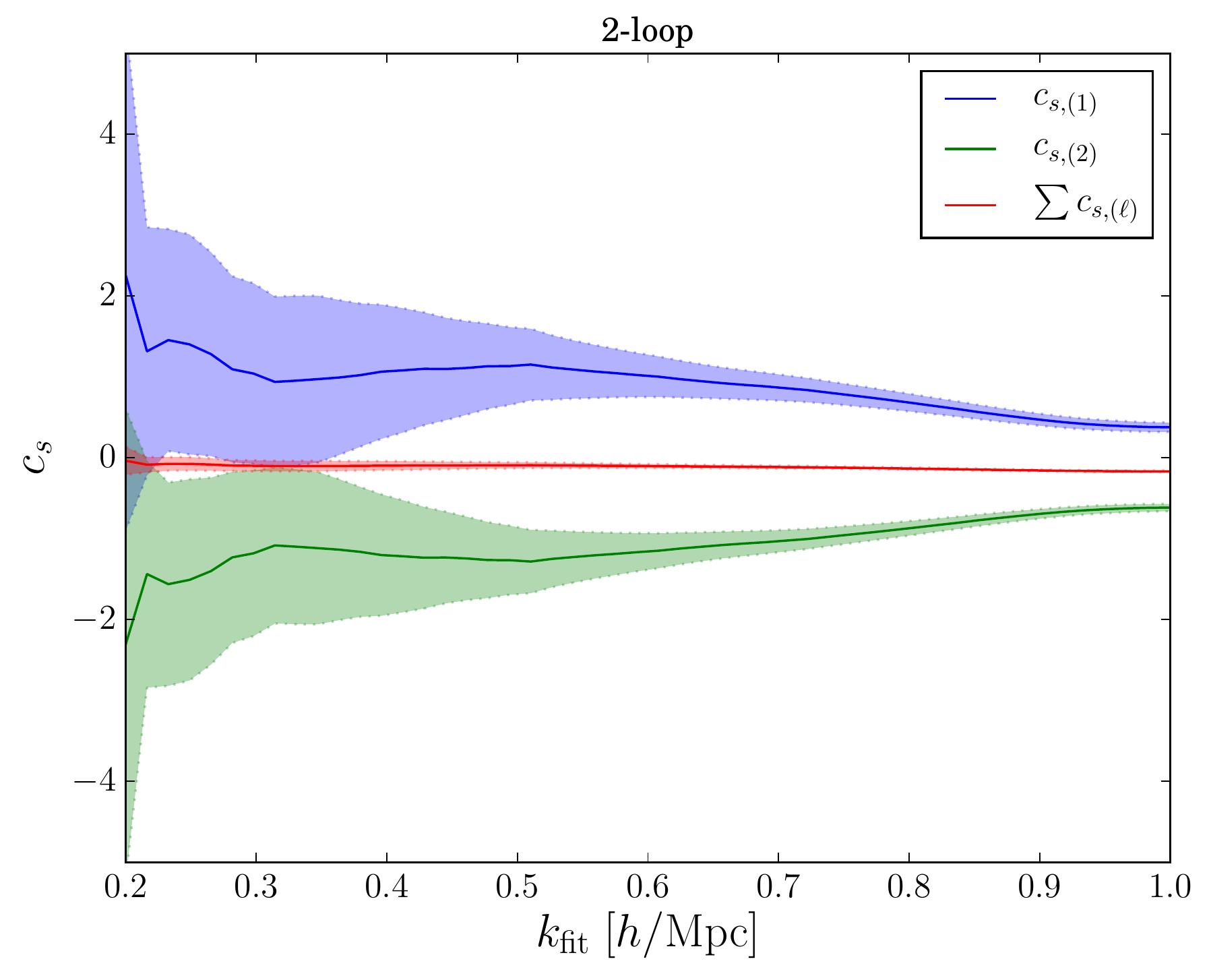}
\includegraphics[width=0.39\textwidth]{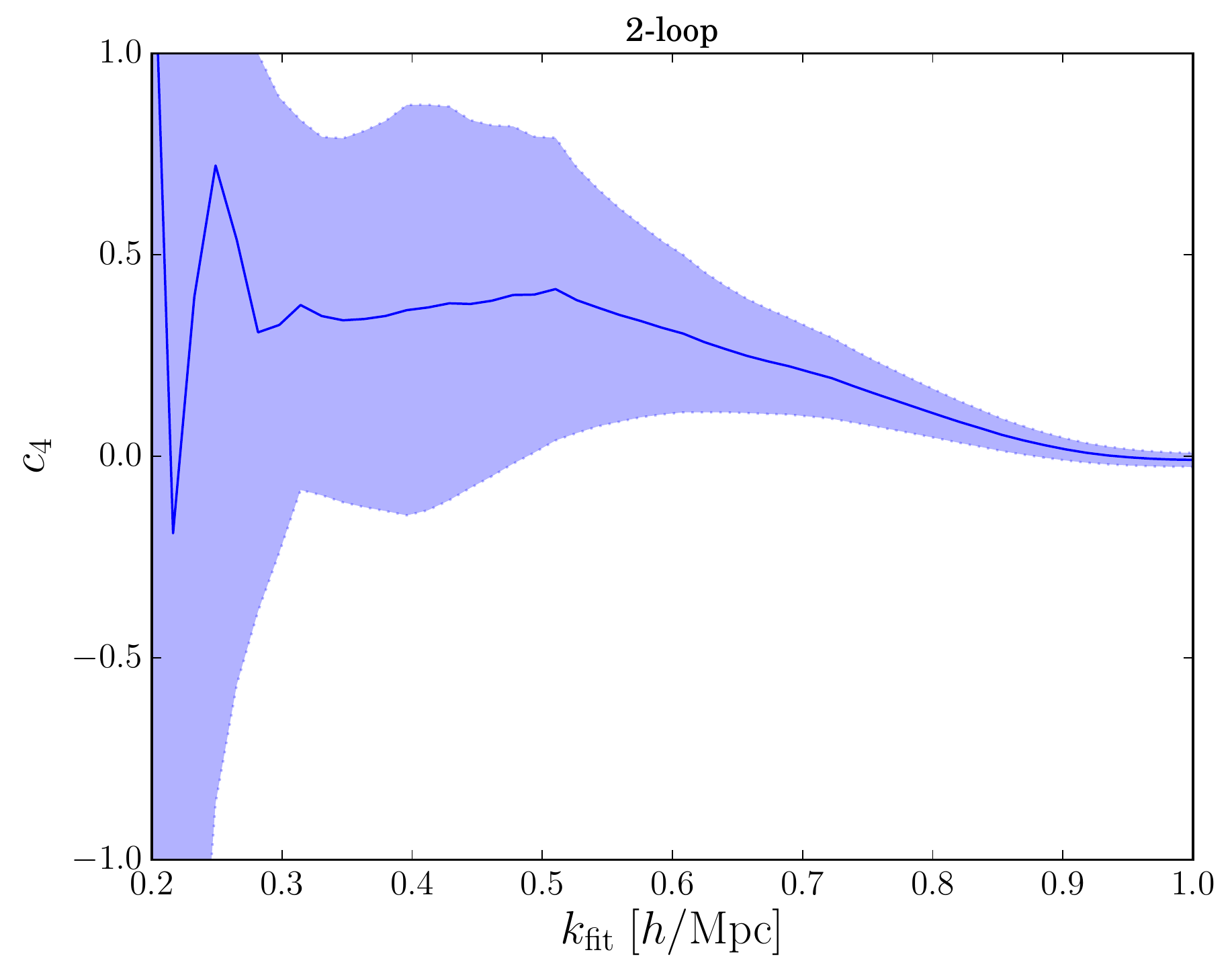}
\includegraphics[width=0.39\textwidth]{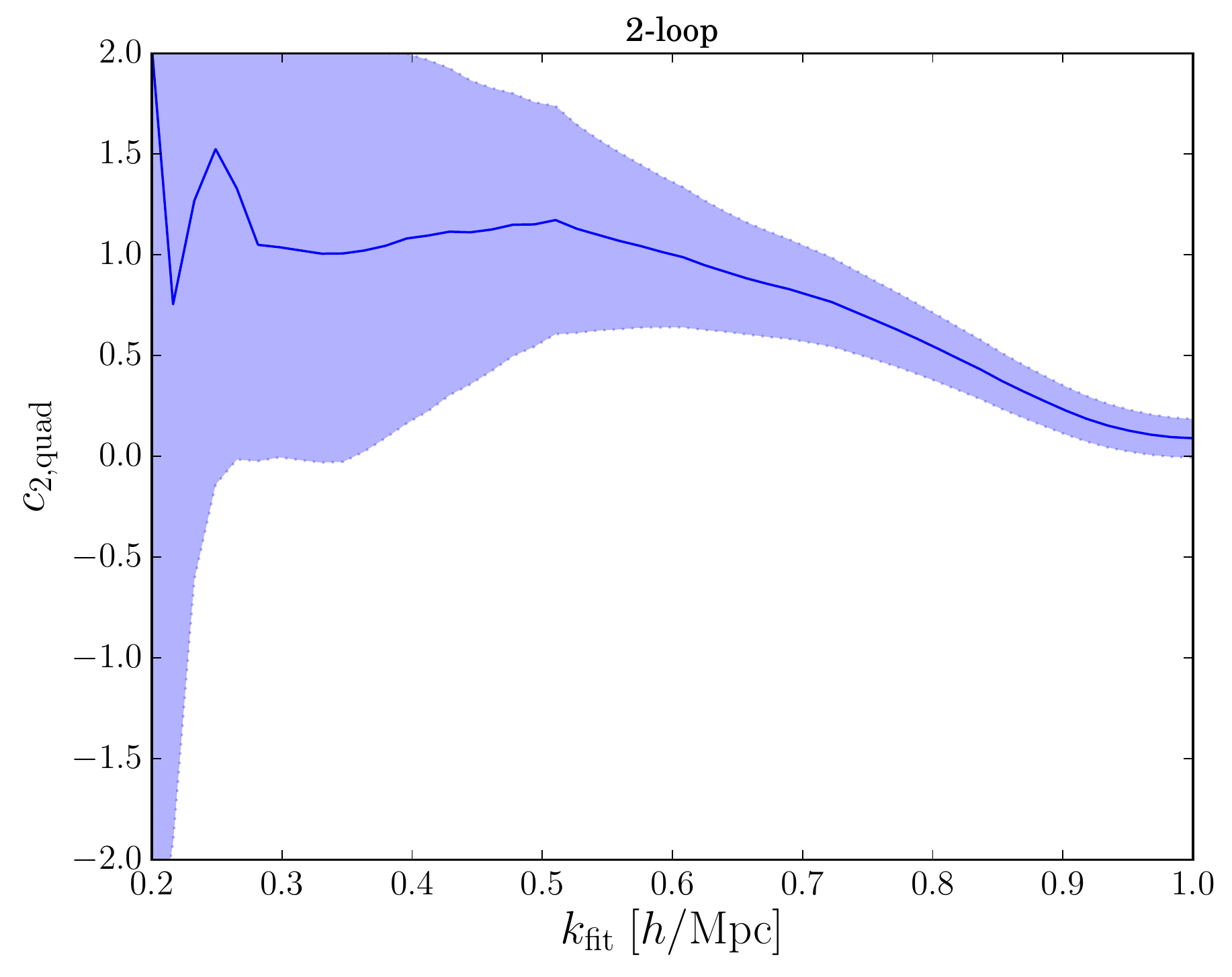}
\caption{\label{fig:cs2L_wostoch}%
\small Same as in Fig.~\ref{fig:cs2L}, but without the stochastic term.}

\end{figure}

\begin{figure}[ht]
\centering
\includegraphics[width=0.39\textwidth]{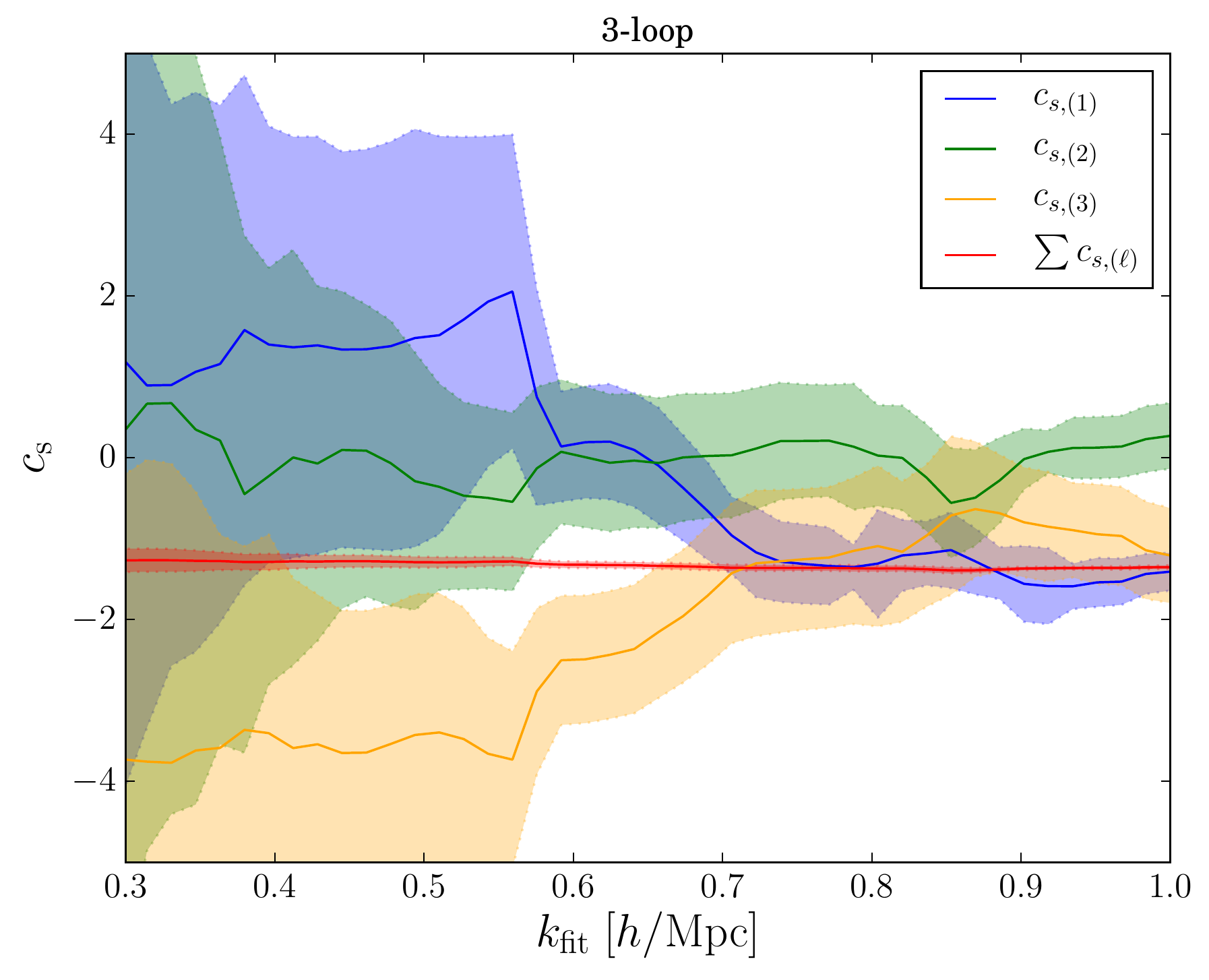}
\includegraphics[width=0.39\textwidth]{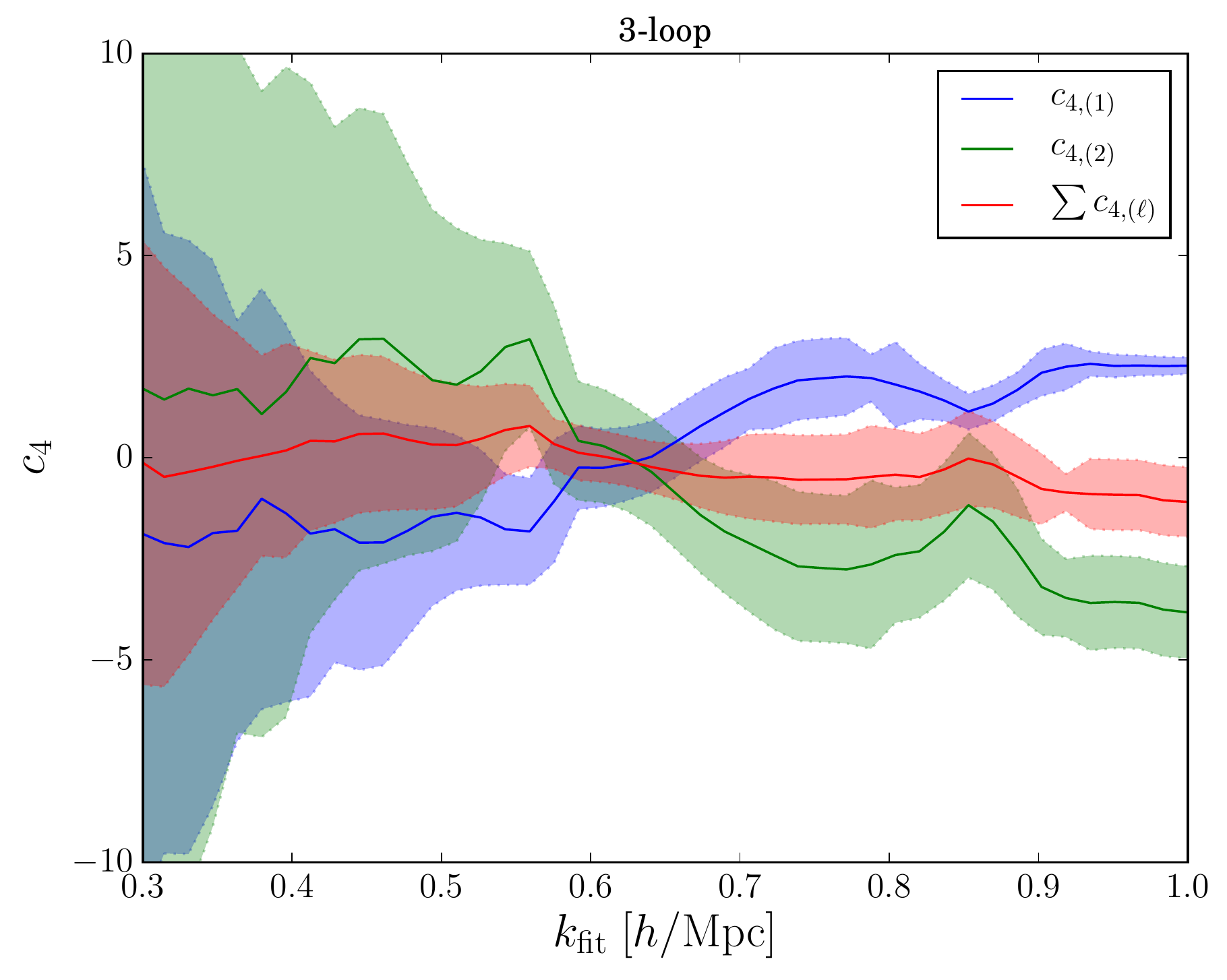}
\includegraphics[width=0.39\textwidth]{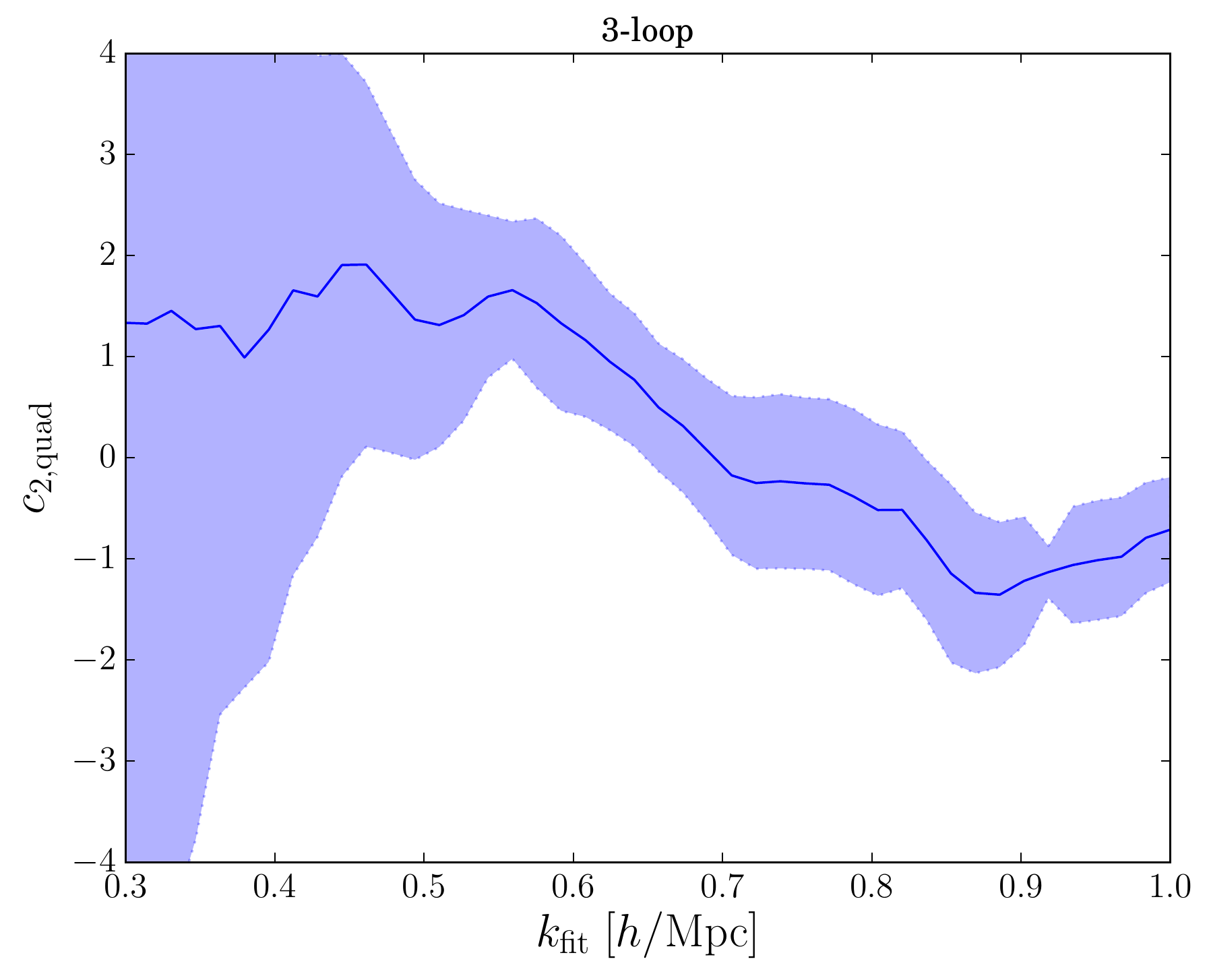}
\includegraphics[width=0.39\textwidth]{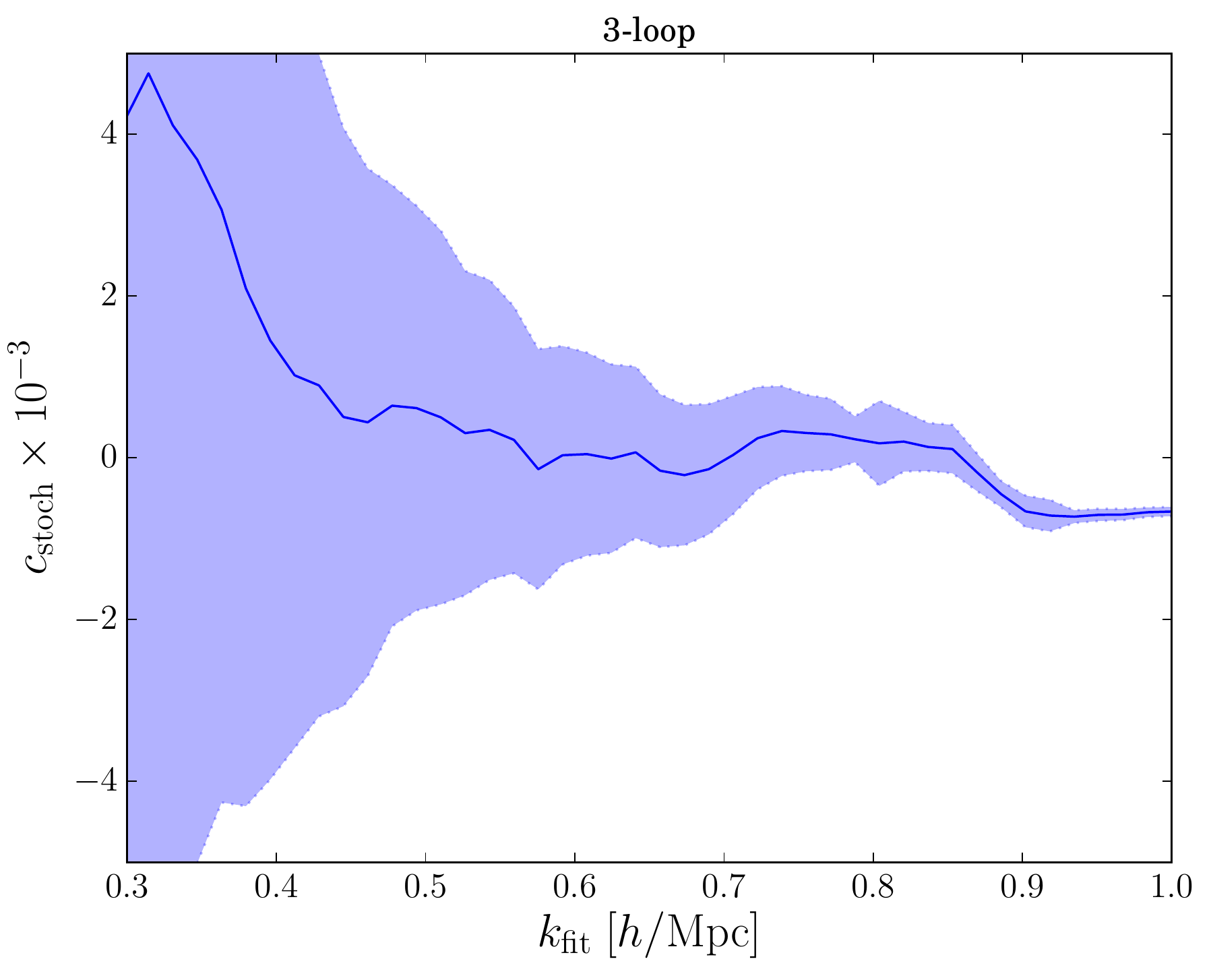}
\includegraphics[width=0.39\textwidth]{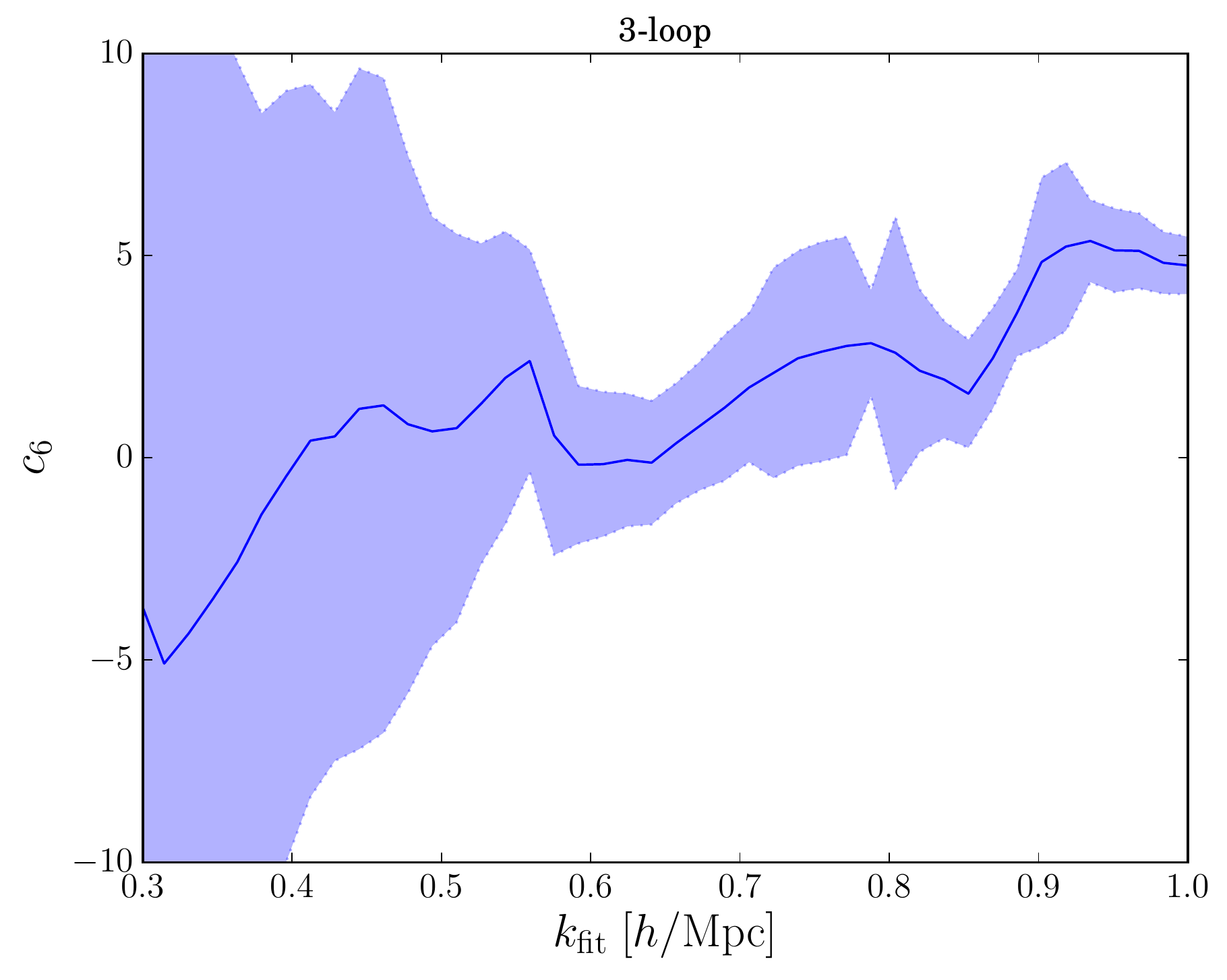}
\caption{\label{fig:cs3L}%
\small Same as Fit.~\ref{fig:cs2L} for the three-loop counter terms. Due to large errors and degeneracies we display their values in the range $k_{\rm fit} \in [0.3,1]\, h/$Mpc.}
\end{figure}

As we mentioned earlier, our approach to fit the numerical data gives us more leeway for the value of the matching coefficients at a given loop order. Therefore, in principle, we are subject to some degree of overfitting. In order to explore whether that is the case, we studied the behavior of each of the $\ell$-loop order contributions for the EFT coefficients as a function of the fitting range. In contrast, we also provide the value for the total sum, see \eqref{csum}. The results at two loops are shown in Fig.~\ref{fig:cs2L}, and in Fig.~\ref{fig:cs2L_wostoch} without the stochastic term, and in Fig.~\ref{fig:cs3L} at three loops. The errors are computed as explained in sec.~\ref{sec:fit}.\vskip 4pt Unfortunately, since we are unable to disentangle the counter-term from the true non-linear information, the contributions from the Wilson coefficients become important at values of $k$ lower than expected by power-counting, for the sole purpose of removing the SPT (miss-)behavior. Hence, our ability to determine each coefficient from the power spectrum alone deteriorates.\footnote{In cases where the linear theory can be solved exactly, without resorting to numerical input, the perturbative expansion is written in terms of analytic functions. The counter-term and renormalized coefficients can be then properly disentangled. Once the counter-terms cancel the unwanted contribution from the loop integrals, the perturbative expansion obeys a well-defined power-counting in coupling constant(s) of the theory, as long as they remain small.} This is exacerbated by the many new degeneracies between counter-terms at high loop~orders. In~spite of these caveats, up to two loops the situation is relatively under control. This is partially due to the remarkable determination of the (total) sound speed at lower values~of~$k$. As we see in the plots, the other coefficients start to become relevant at $k \simeq 0.3-0.4 \, h$ Mpc$^{-1}$, and their values remain relatively stable until $k \simeq 0.9 h\,$Mpc$^{-1}$, where there is a noticeable shift. This is consistent with the behavior of the $\chi^2$ seen in Fig.~\ref{fig:chi2summary}. We also notice the reduction of the error bars, which is due to the scaling with $k$ of each of the (derivatively coupled) counter-term. In the spirit of  \cite{Foreman:2015lca}, we interpret the sudden change in the counter-terms as an indication of overfitting.\footnote{Notice, in contrast, the kink in the matching coefficients occurs at a value of $k$ higher than the one reported in \cite{Foreman:2015lca}. We attribute the different limiting values to the different choice of renormalization scheme. We have also studied the renormalization scheme introduced in \cite{Foreman:2015lca}, and reproduce their results at two loops. We also analyzed the power spectrum to three loop orders using their approach to fix the counter-terms, see appendix~\ref{appB}.}
We should emphasize, however, that there is a non-trivial sensitivity to the value of the stochastic term for 
$k \gtrsim 0.6\, h$ Mpc$^{-1}$. Moreover, the best-fit value for $c_{\rm stoch}$ becomes negative at high values of $k$, even after taking into account the part due to the UV counter-term (see Fig.~\ref{fig:UVmodels}). This goes against the expectation of a positive (renormalized) stochastic coefficient \cite{Foreman:2015lca}. Nonetheless, we have checked that omitting the stochastic term produces similar results for the other coefficients and overall fit up to $k \simeq 0.7\, h$ Mpc$^{-1}$, where potential overfitting starts to show up, see Fig.~\ref{fig:cs2L_wostoch}. This is consistent with Fig.~\ref{fig:chi2models}, which suggests that while the UV test requires a stochastic term to correct SPT in the hunt for high accuracy, the EFT matching to the power spectrum alone does not entirely capture the correct renormalized contribution. In view of the lack of additional information, our results provide a very good fit to the data up to $k \simeq 0.7\, h$ Mpc$^{-1}$ to two loops.\vskip 4pt

For the three loop results, on the other hand, the failure of a systematic loop expansion in SPT, already at relatively low values of $k$, complicates the independent determination of the counter-terms and extraction of the true non-linear information. While  the (total) sound speed remains very well determined, the other coefficients are less constrained by the data from the power spectrum, leading to large uncertainties at low values of $k$ (not shown).  It is only a somewhat small window, around $k \in [0.3,1]\, h$ Mpc$^{-1}$, that the Wilson coefficients (other than the sound speed) can be determined more accurately, which is displayed in Fig.~\ref{fig:cs3L}. In particular, we notice that the terms which first appear at two loops, $c_4$ and $c_{2,\rm quad}$, are much better extracted than $c_6$.\footnote{We have included the loop correction, $c_{4(2)}$, but we found that adding $c_{4,{\rm quad}}$ did not change our results significantly.} This is not surprising, given the high power of $k$ involved for the terms associated with the latter coefficients, which only become relevant for $k \gtrsim 0.5\, h $Mpc$^{-1}$. This is consistent with what we observed in Fig~\ref{fig:UVmodels}, where the additional term is required for the UV test but only at higher values of $k$. Similarly to the results at two loops, but somewhat earlier, there's a clear shift in the value of the Wilson coefficients, more prominently for $c_4$ and $c_{2,\rm quad}$, around $k \simeq 0.55 h\,$Mpc$^{-1}$. There is also another  large variation near $k \simeq 0.85 h\,$Mpc$^{-1}$, more prominently for $c_6$. These results are consistent with our findings for the overall $\chi^2$ in Fig.~\ref{fig:chi2summary}. We have also checked that omitting the stochastic term the plots for the Wilson coefficients are essentially unaltered up to $k \simeq 0.55 h\,$Mpc$^{-1}$. We will return to this point in sec.~\ref{sec:disc}.

\subsection{IR-resummation}

An important aspect in implementing the EFT approach in Euler space is the so-called IR-resummation, introduced in \cite{Senatore:2014via,Cataneo:2016suz, Senatore:2017pbn} following the construction of the EFT in Lagrangian space in \cite{left} (where the resummation is manifest).
\begin{figure}[ht]
\centering
  \includegraphics[width=0.6\textwidth]{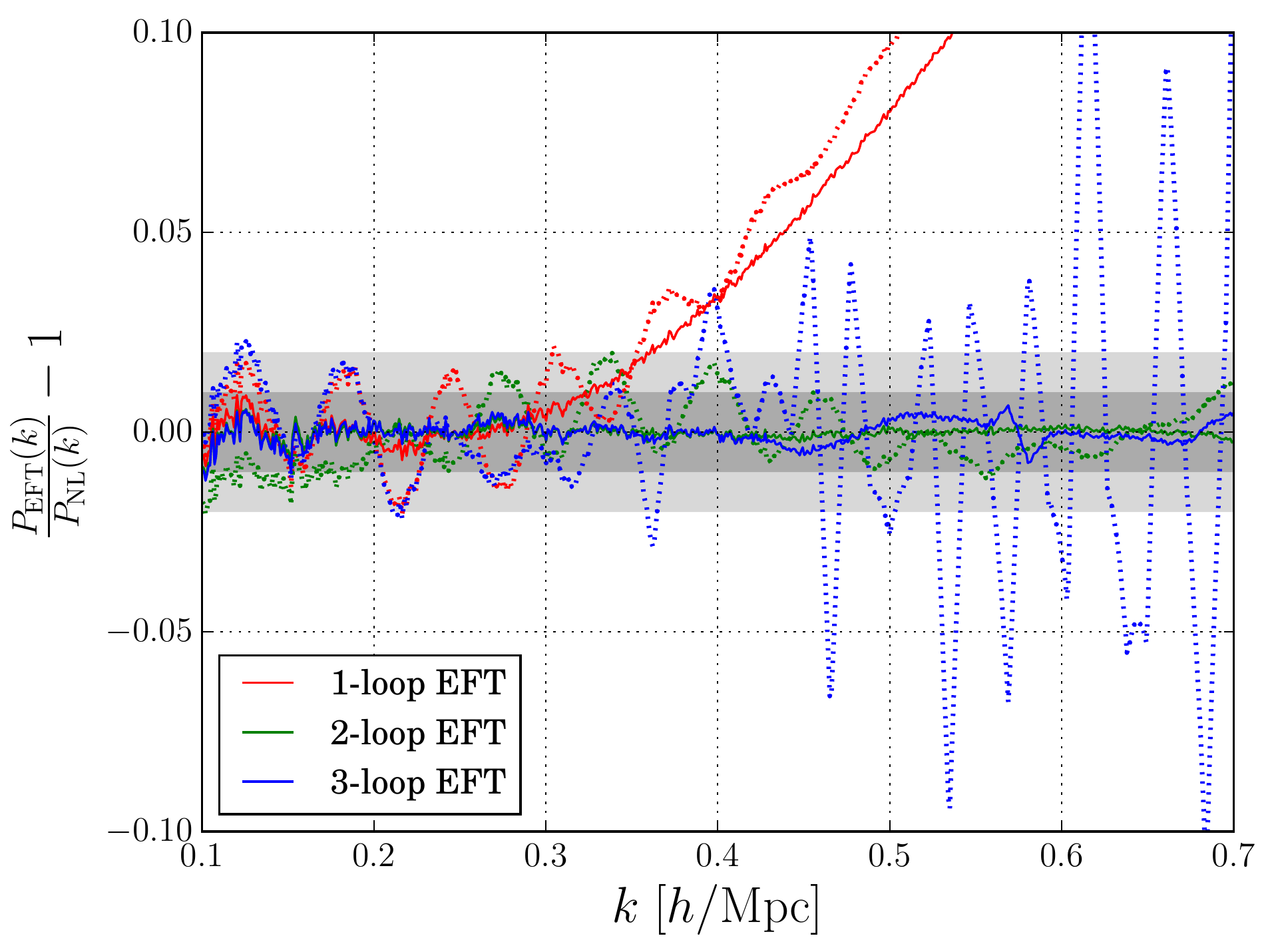}
\caption{\label{fig:IR_resum}%
\small The result of the best fit with (continuous line) and without (dashed line) IR-resummation at different loop orders.
}
\end{figure}
In Figure~\ref{fig:IR_resum}, we show the impact of the IR-resummation procedure in the EFT approach. The continuous lines indicate the result after fitting to the simulation {\it with} the IR-resummation, while the dashed lines display the best fit {\it without} performing the resummation. The values for the counter-terms do not vary significantly, with or without resummation. However, the oscillatory behavior is nicely removed by the procedure. The reader will note that the amplitude of the oscillations increases, the higher the loop order in the expansion. At one loop, they are of the order of $1\%$, while they reach $2\%$ at two loops and as high as $10\%$ for the three loop results, mainly at high values of $k$. Notice that some of these features are due to integration noise in the three-loop results. Even though the accuracy of the three-loop results is by itself on the level $10^{-3}$, a large portion of the three-loop result is absorbed into the counter terms what increases the noise to a few percent (c.f.~Fig.~\ref{fig:Ps}).   
In short, the IR-resummation and IR-safe integrals were vital at three loops to improve the quality of the matching. This provides yet another clue that the long-distance behavior, as opposite to the non-linear dynamics, is failing in SPT (without resummation) at high loop orders.

\section{Discussion \label{sec:disc}} 

\begin{figure}[t!]
\centering
  \includegraphics[width=0.6\textwidth]{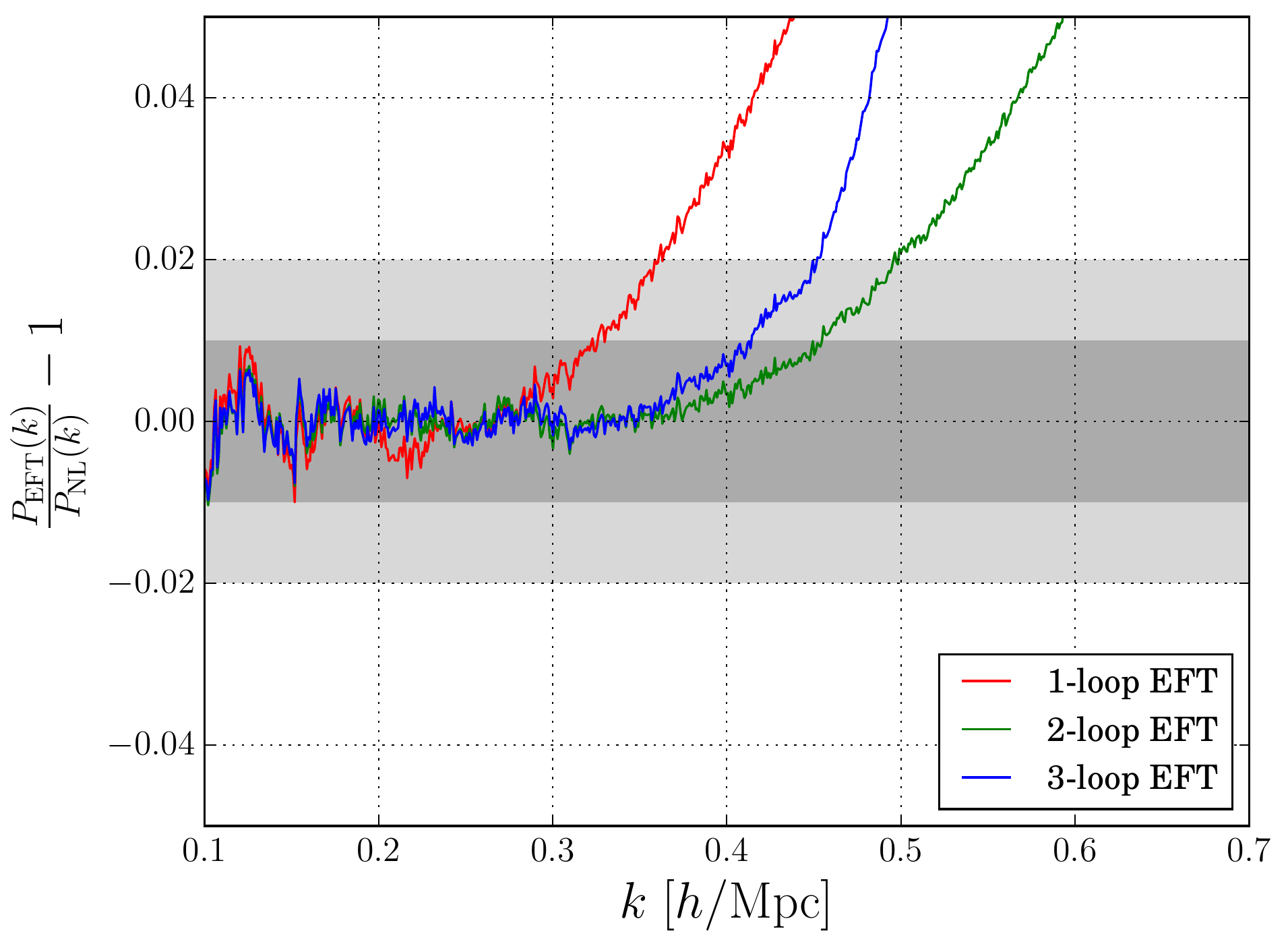}
\caption{\label{fig:residuals}%
\small The best fit to the data with $k_{\rm fit} = 0.35 \, h$ Mpc$^{-1}$ using our renormalization scheme up to two (green) and three (blue) loop orders, respectively. The one loop result (in red), in contrast, is fitted in the range $k \in [0.1, 0.3]\, h$ Mpc$^{-1}$. The results are essentially unaltered by the presence of a stochastic term, which is omitted here. Notice that while the three loop results provide a somewhat better fit to the data in the region $k< k_{\rm fit}$, the two loop power spectrum (with our renormalization scheme) is better behaved beyond that point (see text).}
\end{figure}

We have computed the power spectrum in the EFTofLSS to three loop orders. We show the residuals in Fig.~\ref{fig:residuals} for the best fits to the data using our renormalization procedure, both for the two and three loops, with $k_{\rm fit} = 0.35\, h$~Mpc$^{-1}$ at redshift $z=0$. The values for the counter-terms are given by:
\begin{eqnarray}
c_{s(1)}^2 &=&1.60 \left(\frac{1}{h \textrm{Mpc}^{-1}}\right)^2, \quad c_{s(2)}^2 = -0.27   \left(\frac{1}{h \textrm{Mpc}^{-1}}\right)^2,  \quad c_{s(3)}^2 = -3.54   \left(\frac{1}{h \textrm{Mpc}^{-1}}\right)^2,\nonumber\\
\quad c_{4(1)} &=& -1.82\, \left(\frac{1}{h \textrm{Mpc}^{-1}}\right)^4, \quad c_{4(2)} = 2.67  \left(\frac{1}{h \textrm{Mpc}^{-1}}\right)^4\,,\nonumber \\ \quad c_{2,\rm quad} &=& 1.66  \left(\frac{1}{h \textrm{Mpc}^{-1}}\right)^2, \quad c_{6} =  1.94\left(\frac{1}{h \textrm{Mpc}^{-1}}\right)^6\,,
\end{eqnarray}
for the three loop results.\footnote{For the best fit to two loops the counter-terms are given by: \begin{eqnarray}
c_{s(1)}^2 &=&0.97 \left(\frac{1}{h \textrm{Mpc}^{-1}}\right)^2, \quad c_{s(2)}^2 = -1.11   \left(\frac{1}{h \textrm{Mpc}^{-1}}\right)^2\,, \nonumber\\ 
\quad c_{4} &=& 0.34\, \left(\frac{1}{h \textrm{Mpc}^{-1}}\right)^4, \quad c_{2,\rm quad} = 1.0  \left(\frac{1}{h \textrm{Mpc}^{-1}}\right)^2\,.\nonumber
\end{eqnarray} } 
We notice the level of accuracy to three loop orders increases with respect to two loops at redshift $z=0$, although the improvement is somewhat marginal. We also notice that adding the three loop order makes the best fit to rapidly deviate from the numerical data, unlike the two loop result which is much better behaved, see also Fig.~\ref{fig:chi2summary}. We interpret this as an evidence of the asymptotic nature of the loop expansion in the EFTofLSS, as we discuss here in more detail.\vskip 4pt

\subsection{Perturbative expansion}

\begin{figure}[t!]
\centering
\includegraphics[width=0.6\textwidth]{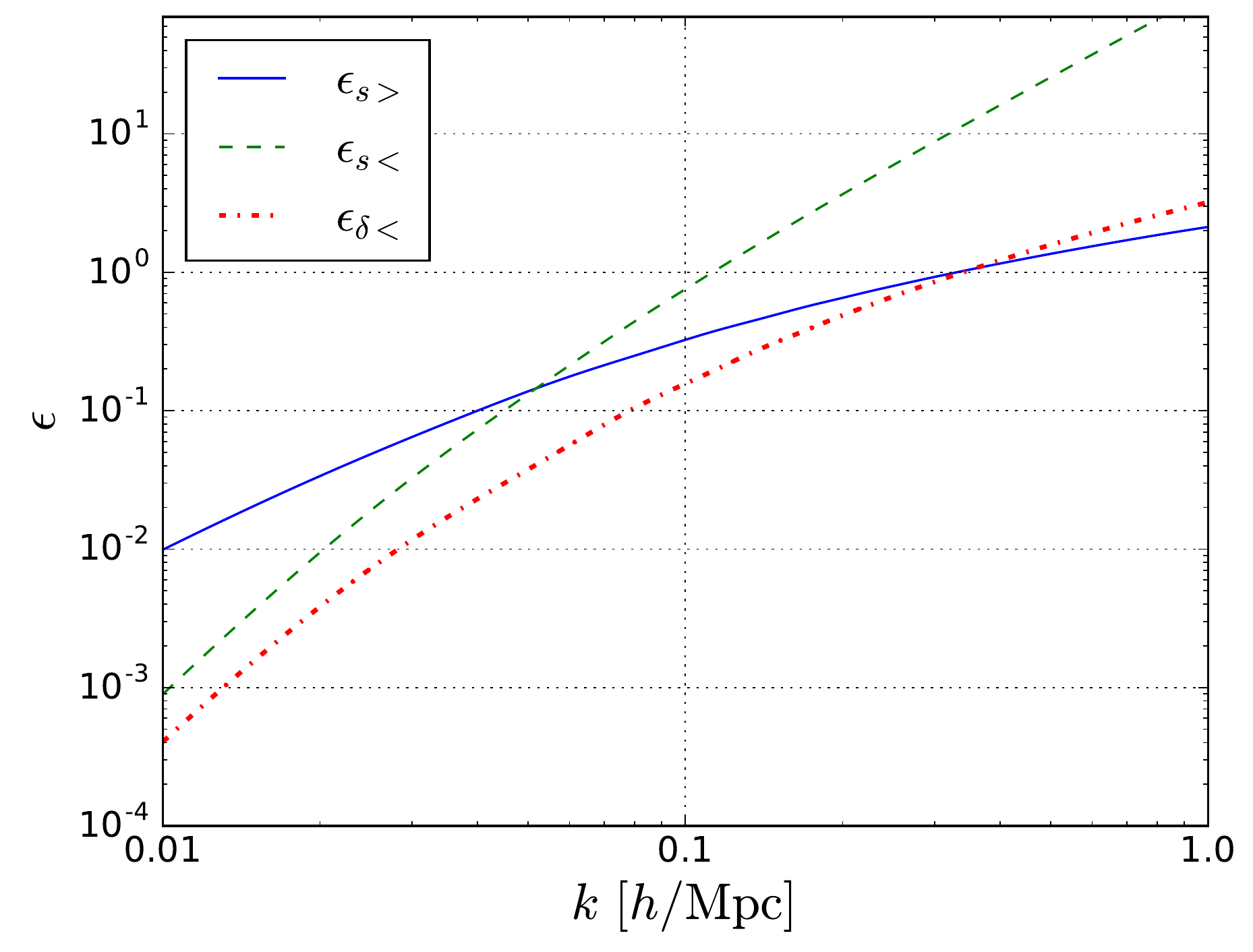}
\caption{\label{fig:eps}%
\small Parameters measuring the amplitude of non-linear corrections on a mode of wavenumber
$k$, computed for our universe at $z=0$. They quantify the motions created by modes longer, $\epsilon_{s >}$, and shorter, $\epsilon_{s <}$,  than $k$ and the tides, $\epsilon_{\delta <}$, from larger scales. See \cite{left,review} for more details.}
\end{figure}

The parameters that control the EFTofLSS are given by \cite{left}
\bea
\epsilon_{s>}(k) &=& k^2\int_{k}^{\infty} \frac{d^3p}{(2\pi)^3} \frac{P_0(p)}{p^2}\,, \nonumber \\
\epsilon_{s<}(k) &=& k^2\int_{0}^k \frac{d^3p}{(2\pi)^3} \frac{P_0(p)}{p^2} \,,\\
\epsilon_{\delta<}(k) &=& \int_{0}^k \frac{d^3p}{(2\pi)^3}P_0(p)\,,
\eea
where $P_0$ is the leading order (tree-level) power spectrum. For instance, let us take the power spectrum at one loop, $P_1$, which includes the sum of two diagrams, $P_{13}$ and $P_{22}$. In the limit the loop momenta, $p$, is higher than the external momenta, $k$, we find 
\begin{equation}
P_1(k) = P_{13}(k) + P_{22}(k) \xrightarrow{k \ll p}
 P_0(k) \, \epsilon_{s>}(k),
\end{equation}
while in the opposite limit,
\begin{equation}
P_{1}(k)   \xrightarrow{k \gg p} P_0(k)\epsilon_{\delta<}(k)\,.
\end{equation}
In terms of the EFTofLSS, the behavior due to short-distance modes, encapsulated in $\epsilon_{s>}$, is absorbed into a series of {\it local} (derivatively-coupled) terms and Wilson coefficients. The latter include each a {\it finite} piece responsible for the true non-linear information, but also a counter-term to correct the SPT contribution beyond the non-linear scale. Notice that $\epsilon_{s<}$ does not appear in the final expression. This is due to a partial cancellation between the two contributions at equal time. In our universe, however, the parameter $\epsilon_{s<}$ is indeed the largest (see Fig.~\ref{fig:eps}). This means that a proper treatment of the perturbative expansion is mandatory, in order to correctly account for all physical effects. In Lagrangian space, the perturbative expansion does not rely on $\epsilon_{s<}$ being small and it is automatically resummed. The IR-resummation of \cite{Senatore:2014via} translates this feature into Euler space, removing the oscillatory behavior due to the BAO scale (see Fig.~\ref{fig:IR_resum}).\vskip 4pt
Finally, $\epsilon_{\delta <}$ is the long-distance expansion parameter of the effective theory. For example, at $\ell$-loop order the renormalized result scales as:\footnote{The ellipses include logarithmic corrections which may also play an important role to determine the precise region of analytic control.}
\be
\label{eq:al}
P_{\ell}(k)   \xrightarrow{k \gg p} a_{\ell}\, P_0(k)\, (\epsilon_{\delta<}(k))^{\ell}+\cdots\,,
\ee
and therefore $\epsilon_{\delta<}(k)$ must remain somewhat small in order to have a well-defined perturbative series. We see in Fig.~\ref{fig:eps} that around $k \simeq\, 0.4\, \text{-}\, 0.6 \,h\, \rm{ Mpc}^{-1}$ we have $\epsilon_{\delta<} \simeq 1$. However, at these values of $k$ the non-linear behavior plays an important role to determine the precise location of the scale at which perturbation theory breaks down, depending on the convergence properties of the series expansion. This is, after all, one of the main motivations behind the EFTofLSS. Yet, we observe that the perturbative series in the EFTofLSS to three loop orders already starts to deteriorate significantly for $k \gtrsim 0.55\, h$ Mpc$^{-1}$ at $z=0$, which also translates in the much more rapid deviation from the data seen in Fig.~\ref{fig:residuals} at lower values of $k$, with $k_{\rm fit} \simeq 0.35\, h$ Mpc$^{-1}$. The sharp increase in the $\chi^2$ is also in the region where $\epsilon_{\delta <}$ is order one (see Fig~\ref{fig:eps}). However, at the same time, the computation up to two loops can be extended, somewhat surprisingly, to higher values of $k$ with our (more generic) renormalization scheme. Since, as we see in Fig.~\ref{fig:eps}, the dependence on $k$ of $\epsilon_{\delta <}$ is rather mild for $k \gtrsim 0.1\, h$Mpc$^{-1}$, the precise region of analytic control could in principle be at higher values of $k$, depending on the exact location of the (non-linear) scale at which the perturbative expansion ought to break down. In general, the behavior of the series expansion as a whole, even after renormalization, depends on the form of the $a_{\ell}$'s in \eqref{eq:al}, e.g.~\cite{Blas:2013aba,Sahni:1995rr,Pajer:2017ulp, Pietroni:2018ebj}. In an EFT approach these coefficients commonly growth at each loop order, leading to an asymptotic behavior for the series expansion. We confirm this expectation in the present~paper.\vskip 4pt
\begin{figure}[t!]
\centering
\includegraphics[width=0.6\textwidth]{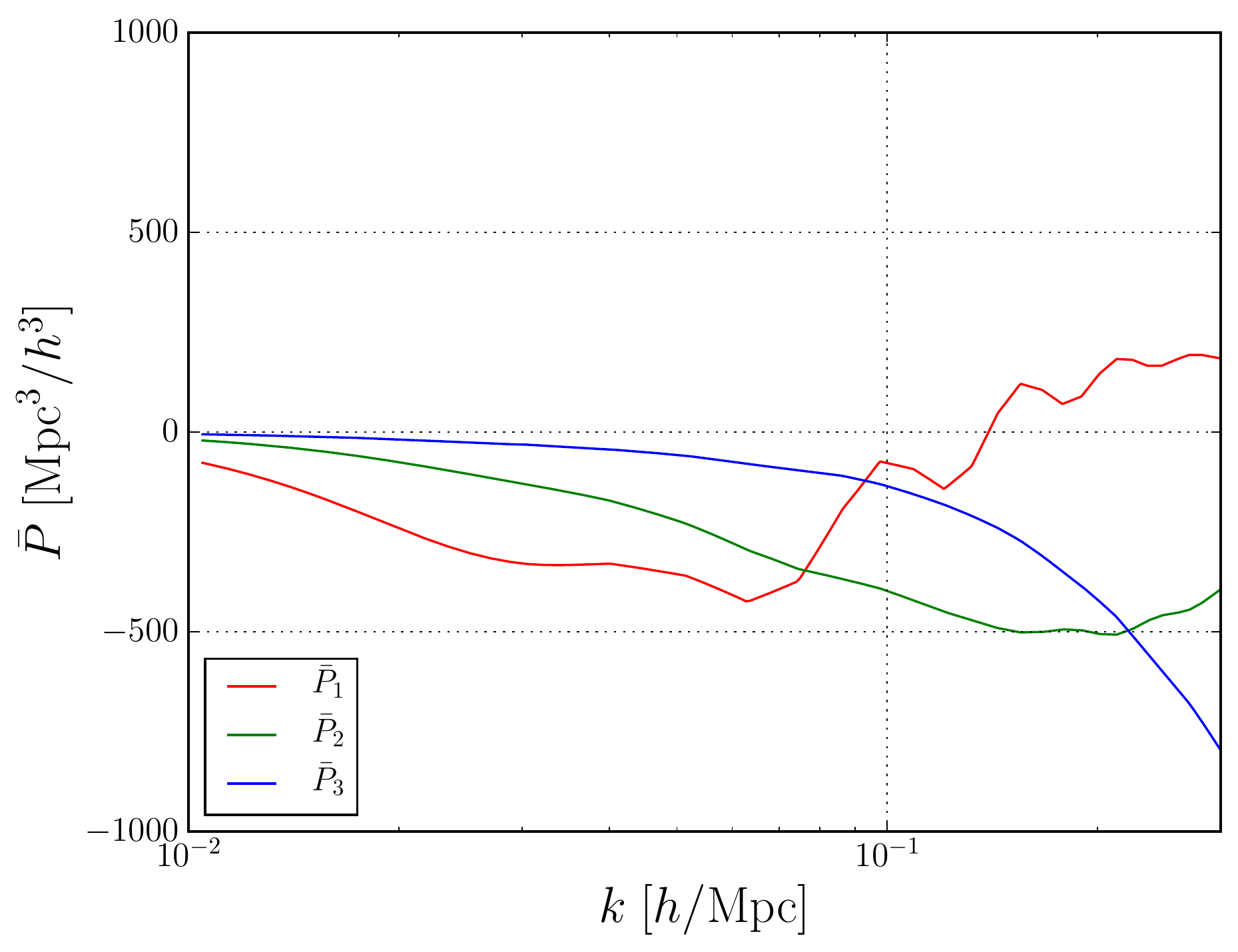}
\caption{\label{cstest}%
\small We plot the power spectrum in SPT at $\ell$-loop order computed with high cutoff after removing the associated (leading) $c^2_{s(\ell)}(\Lambda) k^2 P_0$ counter-term found by fitting in the low $k$ region, which we denote as $\bar P_{\ell}$. As expected, the naive power counting is restored, yet there is a significant increase in $\bar P_3$ at  $k \gtrsim 0.1\, h\,$Mpc$^{-1}$. As we see in Fig.~\ref{fig:Ps} with a low cutoff, this behavior is due to large IR contributions rather than the need of additional UV counter-terms.}
\end{figure}

The reason for the (putative) asymptotic behavior behind the renormalized perturbative series is {\it not} due to an incorrect treatment of short-distance modes --- which are naturally incorporated in the EFTofLSS --- but instead due to the nature of the perturbative expansion itself.\footnote{As an example, let us consider a scaling universe: $P^{\rm s}_0(k) = A k^n \simeq \left(k/k_{\rm NL}\right)^{n+3}$, such that $\epsilon_{\delta<}(k)\simeq \left(k/k_{\rm NL}\right)^{3+n}$. The power spectrum at $\ell$-loops, after renormalization, scales like
\be
P^{\rm s}_{\ell}(k) \simeq a^{\rm s}_\ell \left(k/k_{\rm NL}\right)^{(n+3)\ell} P^{\rm s}_0(k)\,.
\ee
A proxy for the value of the $a_\ell$'s for a scaling universe can be obtained from the one dimensional case where $a^{\rm s}_\ell \simeq \left(a_n\ell\right)^{- n\ell}$, with $a_n$ a numerical factor depending on $n$ \cite{Foreman} (see Eq. 3.24 in \cite{Pajer:2017ulp}). A good approximation of the power spectrum near the non-linear scale is given by $n\simeq -2$ \cite{Carrasco:2013mua}, and therefore $a^{\rm s}_\ell  \simeq \tilde a_{-2}^\ell \sqrt{\ell} (2\ell)!$, after using Stirling's approximation and absorbing numerical factors into $\tilde a_n$. This scaling with $\ell$ is paramount of asymptotic series, with an {\it optimal} number of loops given by $\ell_{\rm opt} (k) \simeq 1/|\tilde a_{-2}\epsilon_{\delta<}(k)|$. Since at the scales of interest we have $\epsilon_{\delta<} \simeq 1$, it is not surprising then to find $\ell_{\rm opt} \simeq 3$ for our universe.} Physically, the renormalized power spectrum is dominated by loop momenta in the range between the peak of the linear spectrum and a mildly non-linear scale, where lower loop orders provide the best approximation to the data. We notice that, after removing the leading $c_s$ counter-term to each one of the $P_{\ell}$'s (see Fig.~\ref{cstest}), the naive power counting is restored at low $k$, yet there is a rapid increase in the contribution at three loops as we move to higher wavenumbers. Some of this behavior is cured by extra counter-terms, however, an important comes from modes near the non-linear scale.\footnote{This is consistent with the expectation in the soft limit discussed in \cite{Ben-Dayan:2014hsa,Garny:2015oya}.}  This is also evidenced in Fig.~\ref{fig:Ps}, where the relative size between $P_{\ell}$ and $P_{\ell+1}$ is reduced at higher loop orders even after implementing a low cutoff ($\Lambda \simeq 0.3\, h$ Mpc$^{-1}$).\vskip 4pt

We should emphasize, nonetheless, that for wavenumbers below $k \simeq 0.4\, h$ Mpc$^{-1}$ the EFT results to three loop orders (slightly) outperform the two loop results, see Figs.~\ref{fig:chi2summary}.  At the same time, it is not too surprising that an EFT approach with more Wilson coefficients can improve the situation in this regime.\footnote{\label{footv} Other counter-terms, which we have not included in the matching, can in principle impact the resulting $\chi^2$. For instance, we have checked that adding the leading $c_{4,\rm quad}$, or  velocity-dependent counter-terms (see appendix~\ref{appA}), both improve the fit with respect to the two loop results, up to $k \simeq 0.5\, h$ Mpc$^{-1}$, but only by a small amount.} Yet, this conclusion turns out to be independent of the renormalization scheme, and therefore quite robust. In fact, as we show in appendix~\ref{appB}, similar wavenumbers can also be reached to three loop orders with the on-shell prescription of \cite{Foreman:2015lca}, and with no additional counter-term other than~$c^2_{s(3)}$. This is an indication that long-distance effects are becoming important before higher order Wilson coefficients have a chance to kick in, which we would have expected to play an important role at high values of $k$ according to the UV test (see Fig.~\ref{fig:UVmodels}).\footnote{We already commented on the fact that $c_s$ alone provides a relatively good fit to data to three loops, see Fig.~\ref{fig:chi2models}. This feature appears to translate to the on-shell scheme, where adding $P_3$ and only $c^2_{s(3)}(c^2_{s(1)},k_{\rm ren})$ outperforms the two loop result with the same number of independent parameters. However, once again, we observe similar behavior as in our analysis with a more generic scheme, and around the same scale (see appendix~\ref{appB}).} Hence, while we find that we can extend the reach of perturbation theory previously found in \cite{Foreman:2015lca}, our analysis using a more generic renormalization procedure demonstrates that (not only the accuracy of the two loop results was somewhat underestimated in \cite{Foreman:2015lca}) the value $k \simeq 0.4\, h$ Mpc$^{-1}$ is plausibly the best we can achieve in terms of accuracy of the EFTofLSS~at~$z=0$.\vskip 4pt In conclusion, our results suggest that we do not expect higher orders to change our main results, but rather to further deteriorate the matching to the data, such that the perturbative EFT derivation will start to depart from the true answer at even lower values of~$k$. This is made worse by our inability to disentangle the finite part (carrying the $\epsilon_{s>}$-dependent renormalized part) of the Wilson coefficient, together with the failure of SPT manifesting itself at relatively lower values~of~$k$ (than expected by power-counting). As a result, the explicit counter-term part of the Wilson coefficient that removes the unwanted contributions from SPT will be needed earlier than expected (except perhaps for those involving higher powers of $k$). This introduces large degeneracies and errors which will hinder the quality of the fit. Hence, as a consequence of all of the above, we believe that the EFTofLSS to three loop orders already provides the best approximation to the dark matter power spectrum in the weakly non-linear~regime.\vskip 4pt It is conceivable that a (Borel or Pade) resummation in $\epsilon_{\delta <}$ (similar to the attempt in \cite{Blas:2013aba} for SPT) could in principle improve the level of precision in the region of analytic control of the EFTofLSS. We leave this possibility open for exploration.\footnote{While convergence beyond the non-linear scale is doomed to fail, see e.g. \cite{review}, a Pade-type resummation --- which in practice amounts to an ansatz of the form $P(x)/Q(x)$ (with two polynomials) to replace the (asymptotic) series $\sum_n a_n x^n$ --- could in principle provide a better approximation in the mildly non-linear regime. Nevertheless, most likely a Pade approximation will not be able to reproduce the true answer at fully non-linear~scales.} The reader must keep in mind, however, that our computation only includes the deterministic part of the evolution equations, and a careful treatment of the stochastic term is required to properly address the ultimate reach of perturbation theory, see e.g. \cite{Baldauf:2015aha,Baldauf:2015zga,Baldauf:2015tla}. We also postpone for future work the study of the redshift dependence of the EFT results, which were largely ignored here. This will have important implications for future surveys, which can in principle reach up to high values (e.g. $z \gtrsim 2$ \cite{Laureijs:2011gra}). In that case, it is (very) possible higher loop orders can still play an important role.

\section*{Acknowledgments}
We would like to thank Matias Zaldarriaga for helpful comments and discussions. The authors acknowledge support from the Deutsche Forschungsgemeinschaft (DFG, German Research Foundation) under Germany's Excellence Strategy (EXC 2121) `Quantum Universe' (390833306). R.A.P. acknowledges financial support from the ERC Consolidator Grant ``Precision Gravity: From the LHC to LISA"  provided by the European Research Council (ERC) under the European Union's H2020 research and innovation programme (grant agreement No. 817791).  R.A.P. would like to thank the Simons Foundation and FAPESP for support through Young Investigator Awards during the early stages of this work. R.A.P. also thanks the participants of the `Simons Foundation Symposium: Amplitudes meet Cosmology'\footnote{\url{https://www.simonsfoundation.org/event/amplitudes-meet-cosmology-2019/}} for helpful discussions. H.R. acknowledges support from the FAPESP grants 2016/08700-0 and 2017/09951-9. H.R. also acknowledges the Universidade de S\~ao Paulo, for all the support during this work.

\appendix

\section{Wilson coefficients to three loops}\label{appA}

\subsection{Density perturbations}
We have considered counter-terms which have the following structure:
\begin{equation}
k^2\delta, \quad k^2\delta^2, \quad k^4 \delta, \quad k^4 \delta^2
\end{equation}
with $\delta$ the density perturbation. The computation of the power spectrum in the EFTofLSS to three loop orders then takes the form:
 \begin{eqnarray}
P^{\rm EFT}_{3\text{-}\rm loop} &=& P^{\rm SPT}_{3\text{-}\rm loop} - 
  2(2\pi) ({c_{s(1)}^2}+{c_{s(2)}^2}+{c_{s(3)}^2)} k^2P_0
  -2(2\pi) ({c_{s(1)}^2}+{c_{s(2)}^2}) k^2P_1
  -2(2\pi) {c_{s(1)}^2} k^2P_2  
  \nonumber\\
&+&  (2\pi)^2 \Big(({c_{s(1)}^2})^2 + 2{c_{s(2)}^2c_{s(1)}^2} \Big)k^4P_0
  +(2\pi)^2 ({c_{s(1)}^2)^2} k^4P_1 
  \nonumber\\
  &-& 2 (2\pi)({c_{2, {\rm quad}(1)}} + \red{c_{2,{\rm quad} (2)}}) k^2 P_{\rm quad}   -2 (2\pi) c_{2, {\rm quad}(1)}k^2P_{\rm quad(2)}   -2 (2\pi)^{2}({c_{4(1)}} +{c_{4(2)}})k^4 P_0
  \nonumber\\
  &-& 
  2 (2\pi)^{2}{c_{4(1)}}k^4 P_1 +
   2(2\pi)^{3}{c_{s(1)}^2c_{4(1)}} k^6P_0 +
  2(2\pi)^2{c_{s(1)}^2c_{2, {\rm quad}(1)}} k^4P_{\rm quad}+(2\pi)^2{c_{\rm stoch}}k^4 \nonumber \\
  &-& 
   2(2\pi)^2 \red{c_{4,\rm quad}}k^4P_{\rm quad} -  2(2\pi)^3{c_{6}}k^6P_{0}.\label{eq3loopeft}
 \end{eqnarray}
We also used the definitions:
\begin{gather}
  P_{\rm quad} = \int_q  P_0(|\vec{k}-\vec{q}|)P_0(q)F_2(\vec{k}-\vec{p},\vec{p}), \\
  P_{{\rm quad}(2)} = 6k^2 \int_{p_1,p_2} P_0(p_1)P_0(p_2)P_0(|\vec{k}-\vec{p}_1|) F_3(\vec{p}_2,-\vec{p}_2,\vec{k}-\vec{p}_1)F_2(\vec{k}-\vec{p}_1,\vec{p}_1)  + \nonumber \\
  8k^2\int_{p_1,p_2} P_0(p_1)P_0(|\vec{p}_1+\vec{p}_2|)P_0(|\vec{k}-\vec{p}_1-\vec{p}_2|)\times \nonumber \\
   F_2(-\vec{p}_1,\vec{p}_2+\vec{p}_1)F_2(\vec{p}_1,\vec{k}-\vec{p}_1-\vec{p}_2)F_2(\vec{p}_2+\vec{p}_1,\vec{k}-\vec{p}_1-\vec{p}_2).
 \end{gather}

The coefficients in red, which in principle enter at three loops, have been ignored when fitting to the numerical data. We have found, and shown explicitly for the case of $c_{4,{\rm quad}}$ in sec.~\ref{sec:UV}, that they are not needed for the UV test. Moreover, we have confirmed they have a small impact on the fit for the power spectrum. Therefore, in order to reduce the number of additional parameters, we have omitted them from the analysis in the main text. In principle, more observables would be needed to assess their true contribution to encode non-linear data. 

\subsection{Velocity Terms}

In principle the new terms in the evolution equations can include the (renormalized) velocity field, $\theta \equiv \nabla\cdot \vec v$. Additional terms may arise also from contractions involving (renormalized) composite operators such as $\delta^2$, see e.g. \cite{Assassi:2014fva}. Since we have $\theta^{(1)} \propto \delta^{(1)}$ at linear order, the leading order effects in the velocity are degenerate with the density perturbations and therefore can be absorbed into the existent counter-terms. At second order, one can also show that 
\begin{equation}
\nabla^2 \Phi^{(2)}_g - \nabla^2 \Phi^{(2)}_v = -\frac{2}{7}\mathcal{G}_2(\Phi^{(1)}_g),
\end{equation}
with  $\Phi_{g,v}$ the gravitational and velocity potentials, and  $\mathcal{G}_2(\Phi_g)$ is a second-order `Galileon' operator, see \cite{Assassi:2014fva}. The latter, which not renormalized at leading order in derivatives, is proportional to $P_{\rm quad}$, which we have already included in our analysis. At cubic order in $\Phi_v$ we start to see new counter-terms which would enter through a two loop diagram, and therefore may be needed to renormalize the power spectrum at three loops. (Recall the tree-level $c_s$ counter-term is needed at one loop order and so on and so forth.)  We have checked that, while the new counter-term(s) are not degenerate with those of the density, they have a somewhat minor impact on the fit (see footnote~\ref{footv}).

\section{{\it On-shell} Renormalization scheme}\label{appB}
In our approach we have used a more generic renormalization scheme, allowing for more freedom in the determination of the EFT parameters. Partly, our chose was driven by the desire to explore the main features of the perturbative series, as generically as possible. However, in the renormalization scheme used in \cite{Carrasco:2013mua,Senatore:2014via,Foreman:2015lca} the additional parameters are fit using an `on-shell' renormalization scheme. For instance, the sound speed at two loops, $c_{s(2)}$, is obtained by imposing the condition: 
\be
\label{kren1}
 P^{\rm EFT}_{2\text{-}\rm loop}(k_{\rm ren}) = P^{\rm EFT}_{1\text{-}\rm loop}(k_{\rm ren}) \,, 
\ee
 at a renormalization scale (originally at $k_{\rm ren} \simeq 0.2\, h\, \textrm{Mpc}^{-1}$). 
As a consequence, one has
\be
c^2_{s(2)}(c^2_{s(1)},k_{\rm ren}) = \frac{P_{2}(k_{\rm ren})}{2(2\pi)k_{\rm ren}^2P_{0}(k_{\rm ren})}-c_{s(1)}^2\frac{ 
 P_{1}(k_{\rm ren})}{P_{0}(k_{\rm ren})} 
+ \pi (c_{s(1)}^2)^2k_{\rm ren}^2,
\ee
effectively having only one free parameter for the sound speed. Similarly, one can also constrain $c_{s(3)}$ by this procedure. In this appendix we analysis the three loop results using the on-shell renormalization scheme. We will see that our conclusions are essentially unmodified by the choice of scheme, even though the increase in accuracy between two and three loop results appears more prominently in the on-shell scheme.

\subsection{Two loops} 
\begin{figure}[ht]
\centering
  \includegraphics[width=0.6\textwidth]{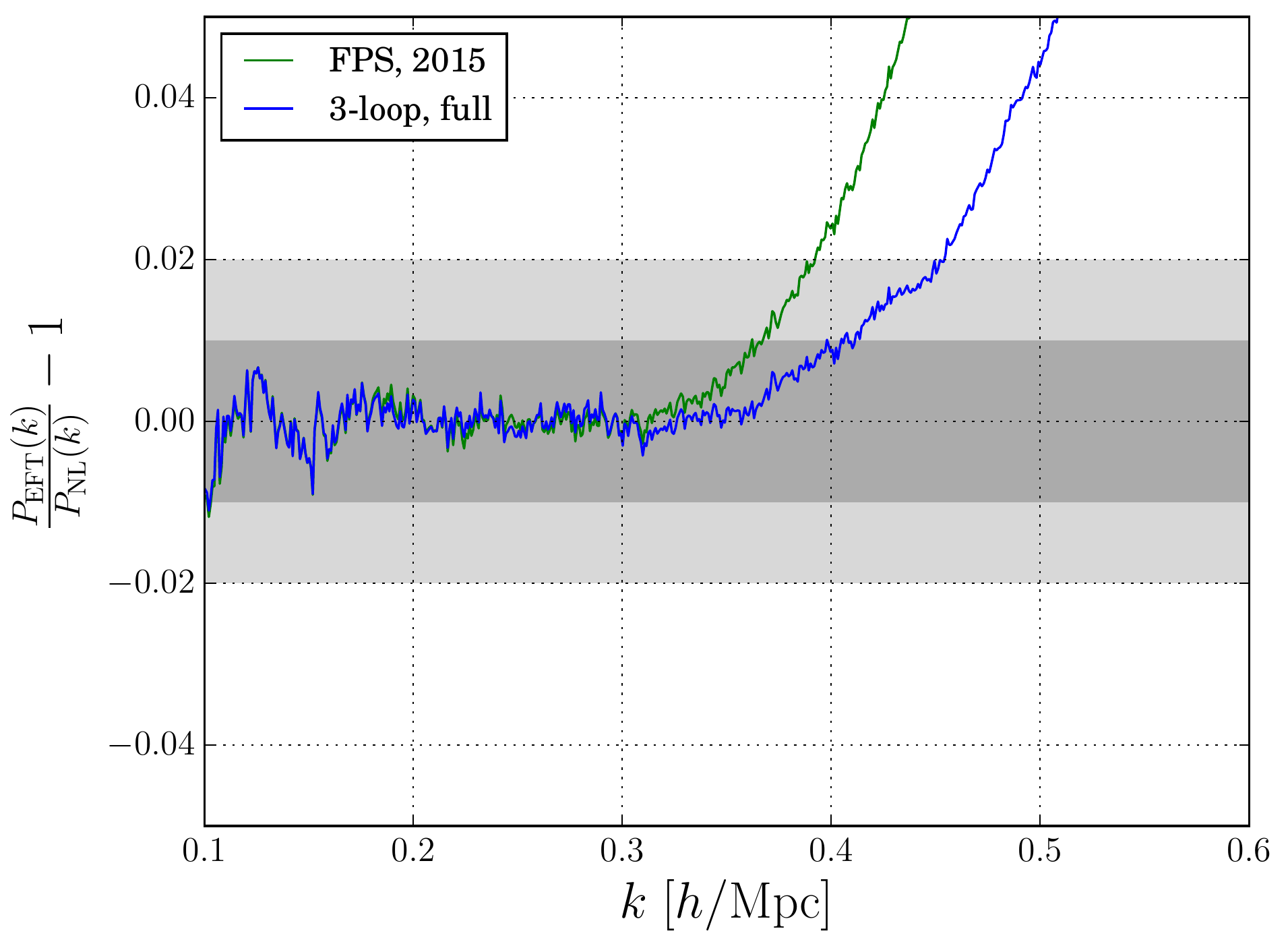}
\caption{\label{fig:2loop_other}%
\small Residuals for the best fit for the power spectrum using the on-shell prescription of  \cite{Foreman:2015lca}. For the two loop result (green) the reach is approximately to $k \simeq 0.35\, h$Mpc$^{-1}$ with small variations of the counter-terms, as discussed in \cite{Foreman:2015lca}, while to three loop order (blue) we find the reach can be extended to $k \simeq 0.4\, h$Mpc$^{-1}$ at percent level, with $k_{\rm fit} \simeq 0.35\,h$Mpc$^{-1}$.}
\end{figure}

\begin{figure}[ht]
\centering
\includegraphics[width=0.6\textwidth]{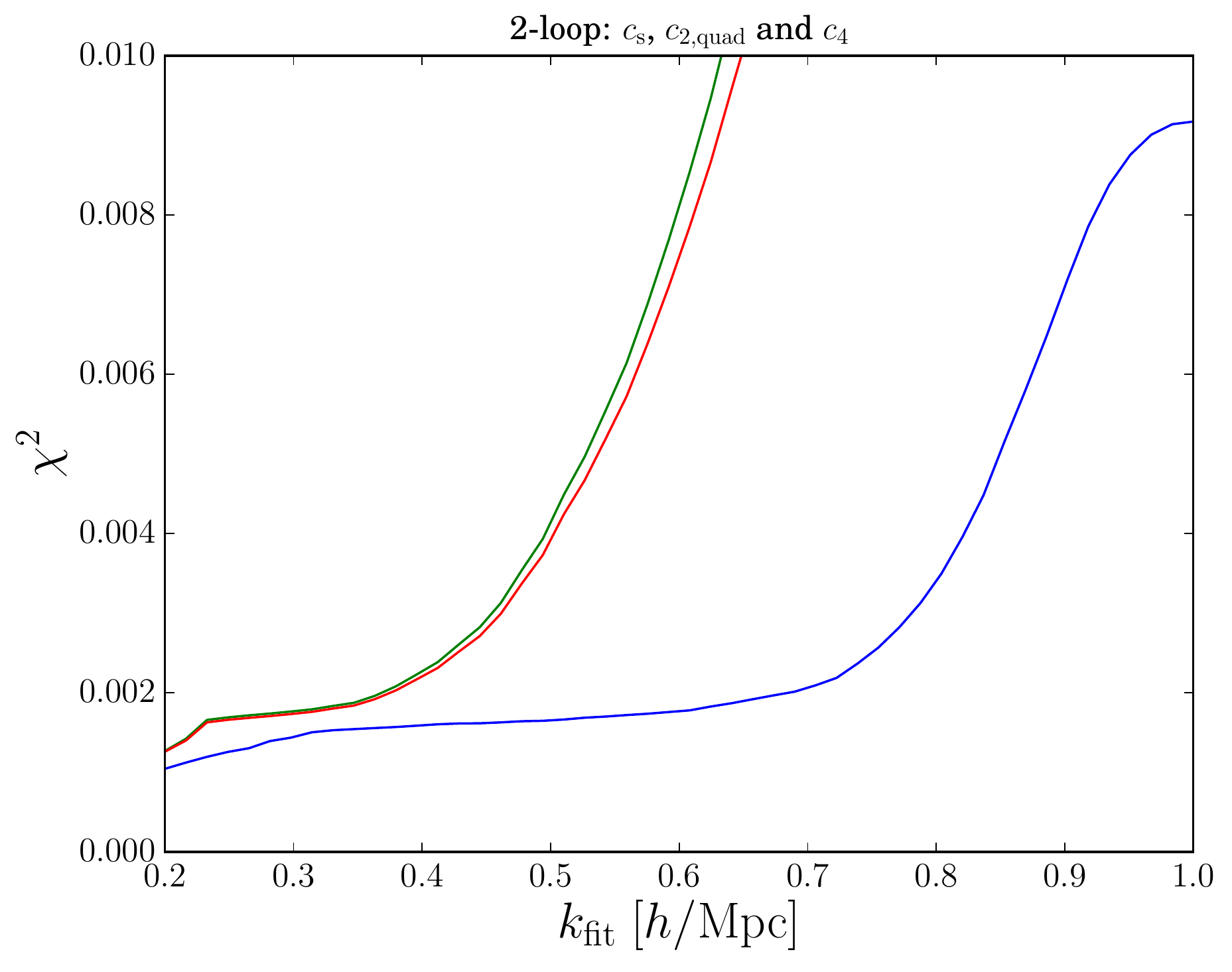}
\caption{\label{fig:residues_kren_2loop}%
\small The value of $\chi^2$ for the on-shell prescription, for two values of the renormalization scale: 
$k_{\rm ren}=\{0.005,0.01\}\, h\, \textrm{Mpc}^{-1}$ (in green and red), against the scheme used in this paper (in blue).}
\end{figure}

In order to warm up, let us first reproduce the results previously derived at two loops. Here we will follow the choices made in  \cite{Foreman:2015lca}, where the IR-resummation was also implemented. The renormalization scale was chosen at: 
\begin{gather}\label{eq:renorm_scale}
 P^{\rm EFT}_{2\text{-}{\rm loop}}(k_{\rm ren}) = P^{\rm EFT}_{1\text{-}{\rm loop}}(k_{\rm ren}) \quad \textrm{at} \quad k_{\rm ren} = 0.005h\textrm{Mpc}^{-1}\,.
\end{gather}
The range for the best fit was obtained by looking for islands of stability of the Wilson coefficients as a function of $k_{\rm fit}$. They found that the percent-level accuracy can be achieved up to $k \simeq  0.34\, h\, \textrm{Mpc}^{-1}$. We reproduce their results in Figure~\ref{fig:2loop_other}, with the following (central) values for the counter-terms: 
\begin{equation}
c_{s(1)}^2 = 0.54\left(\frac{1}{2h\textrm{Mpc}^{-1}}\right)^2, \quad c_{4(1)} = 6.06\left(\frac{1}{2h\textrm{Mpc}^{-1}}\right)^4, \quad c_{2, \rm quad} = 0.82\left(\frac{1}{2h\textrm{Mpc}^{-1}}\right)^2\,,\nonumber
\end{equation}
where we used the normalization used in \cite{Foreman:2015lca} (with the change $c_1 \to c_{2,\rm quad}$). These values are comparable with those obtained in \cite{Foreman:2015lca} (see table 2 in sec. 5.2).\footnote{The sign difference in $c_{2,\rm quad}$ and $c_4$ is due to the convention we used in this paper (see \eqref{eq3loopeft}).} We attribute the small changes to the use of different cosmologies.\vskip 4pt
In Fig.~\ref{fig:residues_kren_2loop} we show the resulting $\chi^2$ for two choices of $k_{\rm ren}$, and in comparison with our renormalization scheme. Neither includes the stochastic term. Notice the spike at $k \simeq  0.3-0.4\, h\, \textrm{Mpc}^{-1}$ anticipated in \cite{Foreman:2015lca} with the on-shell scheme. In contrast, the renormalization procedure in this paper can be extended to higher wavenumbers, see Fig.~\ref{fig:residuals}.

\subsection{Three loops}
In Fig.~\ref{fig:2loop_other} we include also the result for the power spectrum to three loop order using the on-shell prescription. We used one renormalization scale at $k_{\rm ren}= 0.005\, h\, \textrm{Mpc}^{-1}$, as in \eqref{kren1}, and a second one, $\tilde k_{\rm ren}$, for the three loop result:
\be
\label{eq:renorm_scale_2}
 P^{\rm EFT}_{3\,\text{-}\rm loop}(\tilde k_{\rm ren}) = P^{\rm EFT}_{2\,\text{-}\rm loop}(\tilde k_{\rm ren}) \quad \textrm{at} \quad \tilde k_{ren}= 0.01 \, h\, \textrm{Mpc}^{-1}\,.
\ee
For simplicity we omit $c_{4,(2)}$ and $c_6$ as well as the stochastic term, such that only $c_{s(3)}$ is added to the fit and determined as explained above.\footnote{We find that including these additional counter-terms does not significantly improve the fit in the realm of convergence of the EFT expansion.} We find for the counter-terms the following values for the best fit:
\begin{eqnarray}
c_{s(1)}^2 =0.22 \left(\frac{1}{h \textrm{Mpc}^{-1}}\right)^2,\,
\quad c_{4(1)} = -0.14\, \left(\frac{1}{h \textrm{Mpc}^{-1}}\right)^4, \, \quad c_{2,\rm quad} = 0.53  \left(\frac{1}{h \textrm{Mpc}^{-1}}\right)^2\,.
\end{eqnarray}

\begin{figure}[ht]
\centering
\includegraphics[width=0.6\textwidth]{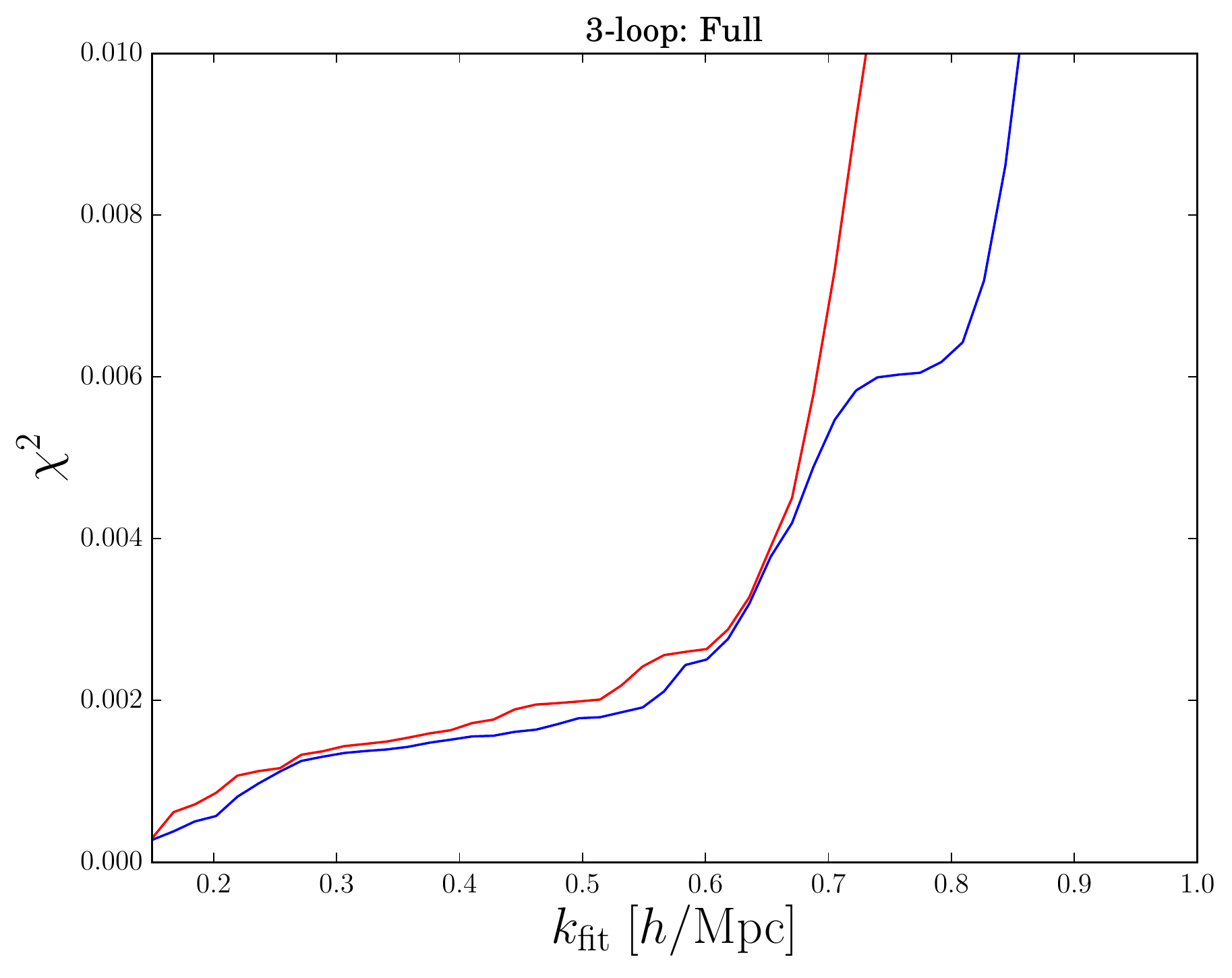}
\caption{\label{fig:residues_kren}%
\small The value of $\chi^2$ for the on-shell prescription (in red), in comparison with the scheme used in this paper (in blue). Unlike what we see in Fig.~\ref{fig:residues_kren_2loop}, we find that both produce essentially the same results to three loop orders. We have omitted the effects of the stochastic term in this plot.
 }
\end{figure}

For comparison, in Fig.~\ref{fig:residues_kren} we present the $\chi^2$ in the EFT to three loops using the two different prescriptions (without the stochastic term). Notice that the resulting $\chi^2$'s are essentially the same, starting to rapidly change around a similar value, $k \simeq 0.5\,h\,$Mpc$^{-1}$. This supports the asymptotic nature of the perturbative series, regardless of the renormalization scheme. We also notice, comparing with Fig.~\ref{fig:residues_kren_2loop}, that the result to three loops outperforms its counterpart at two loops with the on-shell scheme. By studying the behavior of the counter-term we find that the on-shell scheme extends the regime from $k\simeq 0.3\,h\,$Mpc$^{-1}$ up to $k \simeq 0.4\,h\,$Mpc$^{-1}$, without overfitting the data. This is consistent with our results to three loop orders. \vskip 4pt

The reader will immediately notice that the power spectrum using our renormalization scheme to two loops in Fig.~\ref{fig:residuals} outperforms the on-shell counter-part. This is not all that surprising. When the a generalized scheme is used, the low $k$ data fits the combination $c^2_{s(1)} + c^2_{s(2)}$, but leaves the `other direction' unconstrained.  Hence, each one of the individual $c^2_{s(\ell)}$ is allowed to run as we move toward shorter distances, $ k \gg k_{\rm ren}$. (This is even more prominent up to three loops, see Fig.~\ref{fig:cs3L}.) This possibility, in particular for $c^2_{s(2)}$ was not included in the on-shell prescription of \cite{Foreman:2015lca}. As a result, the $c^2_s{(1)}$ in the best fit obtained in \cite{Foreman:2015lca} has a flat region ending around $k \simeq 0.3\, h$ Mpc$^{-1}$ (see the top panel in their Fig.~7). This is an indication of a non-trivial running rather than overfitting, as we see in Fig.~\ref{fig:cs2L}, where the total value for the sound speed remains essentially unaltered with our renormalization procedure. In other words, the on-shell scheme produced an earlier failure of the perturbative expansion due to missing running effects. 

\bibliographystyle{utphys}
\bibliography{RefsLSS}

\providecommand{\href}[2]{#2}\begingroup\raggedright\begin{thebibliography}{10}

\bibitem{Abbott:2005bi}
{\bfseries DES} Collaboration, T.~Abbott {\em et~al.}, ``{The dark energy
  survey},''
\href{http://arxiv.org/abs/astro-ph/0510346}{{\ttfamily arXiv:astro-ph/0510346
  [astro-ph]}}.

\bibitem{Ivezic:2008fe}
{\bfseries LSST} Collaboration, Z.~Ivezic {\em et~al.}, ``{LSST: from Science
  Drivers to Reference Design and Anticipated Data Products},''
\href{http://arxiv.org/abs/0805.2366}{{\ttfamily arXiv:0805.2366 [astro-ph]}}.

\bibitem{Laureijs:2011gra}
{\bfseries EUCLID} Collaboration, R.~Laureijs {\em et~al.}, ``{Euclid
  Definition Study Report},''
\href{http://arxiv.org/abs/1110.3193}{{\ttfamily arXiv:1110.3193
  [astro-ph.CO]}}.

\bibitem{Levi:2013gra}
{\bfseries DESI} Collaboration, M.~Levi {\em et~al.}, ``{The DESI Experiment, a
  whitepaper for Snowmass 2013},''
\href{http://arxiv.org/abs/1308.0847}{{\ttfamily arXiv:1308.0847
  [astro-ph.CO]}}.

\bibitem{Aghanim:2018eyx}
{\bfseries Planck} Collaboration, N.~Aghanim {\em et~al.}, ``{Planck 2018
  results. VI. Cosmological parameters},''
\href{http://arxiv.org/abs/1807.06209}{{\ttfamily arXiv:1807.06209
  [astro-ph.CO]}}.

\bibitem{Akrami:2019izv}
{\bfseries Planck} Collaboration, Y.~Akrami {\em et~al.}, ``{Planck 2018
  results. IX. Constraints on primordial non-Gaussianity},''
\href{http://arxiv.org/abs/1905.05697}{{\ttfamily arXiv:1905.05697
  [astro-ph.CO]}}.

\bibitem{threshold1}
D.~Baumann, D.~Green, and R.~A. Porto, ``{B-modes and the Nature of
  Inflation},'' {\em JCAP} {\bfseries 1501} no.~01, (2015) 016,
\href{http://arxiv.org/abs/1407.2621}{{\ttfamily arXiv:1407.2621 [hep-th]}}.

\bibitem{threshold2}
D.~Baumann, D.~Green, H.~Lee, and R.~A. Porto, ``{Signs of Analyticity in
  Single-Field Inflation},'' {\em Phys. Rev.} {\bfseries D93} no.~2, (2016)
  023523,
\href{http://arxiv.org/abs/1502.07304}{{\ttfamily arXiv:1502.07304 [hep-th]}}.

\bibitem{Baumann:2010tm}
D.~Baumann, A.~Nicolis, L.~Senatore, and M.~Zaldarriaga, ``{Cosmological
  Non-Linearities as an Effective Fluid},''
  \href{http://dx.doi.org/10.1088/1475-7516/2012/07/051}{{\em JCAP} {\bfseries
  1207} (2012) 051},
\href{http://arxiv.org/abs/1004.2488}{{\ttfamily arXiv:1004.2488
  [astro-ph.CO]}}.

\bibitem{Carrasco:2012cv}
J.~J.~M. Carrasco, M.~P. Hertzberg, and L.~Senatore, ``{The Effective Field
  Theory of Cosmological Large Scale Structures},'' {\em JHEP} {\bfseries 09}
  (2012) 082,
\href{http://arxiv.org/abs/1206.2926}{{\ttfamily arXiv:1206.2926
  [astro-ph.CO]}}.

\bibitem{Carrasco:2013mua}
J.~J.~M. Carrasco, S.~Foreman, D.~Green, and L.~Senatore, ``{The Effective
  Field Theory of Large Scale Structures at Two Loops},'' {\em JCAP} {\bfseries
  1407} (2014) 057,
\href{http://arxiv.org/abs/1310.0464}{{\ttfamily arXiv:1310.0464
  [astro-ph.CO]}}.

\bibitem{Carrasco:2013sva}
J.~J.~M. Carrasco, S.~Foreman, D.~Green, and L.~Senatore, ``{The 2-loop matter
  power spectrum and the IR-safe integrand},''
  \href{http://dx.doi.org/10.1088/1475-7516/2014/07/056}{{\em JCAP} {\bfseries
  1407} (2014) 056},
\href{http://arxiv.org/abs/1304.4946}{{\ttfamily arXiv:1304.4946
  [astro-ph.CO]}}.

\bibitem{Angulo:2014tfa}
R.~E. Angulo, S.~Foreman, M.~Schmittfull, and L.~Senatore, ``{The One-Loop
  Matter Bispectrum in the Effective Field Theory of Large Scale Structures},''
  \href{http://dx.doi.org/10.1088/1475-7516/2015/10/039}{{\em JCAP} {\bfseries
  1510} no.~10, (2015) 039},
\href{http://arxiv.org/abs/1406.4143}{{\ttfamily arXiv:1406.4143
  [astro-ph.CO]}}.

\bibitem{Foreman:2015lca}
S.~Foreman, H.~Perrier, and L.~Senatore, ``{Precision Comparison of the Power
  Spectrum in the EFTofLSS with Simulations},''
  \href{http://dx.doi.org/10.1088/1475-7516/2016/05/027}{{\em JCAP} {\bfseries
  1605} no.~05, (2016) 027},
\href{http://arxiv.org/abs/1507.05326}{{\ttfamily arXiv:1507.05326
  [astro-ph.CO]}}.

\bibitem{Baldauf:2015aha}
T.~Baldauf, L.~Mercolli, and M.~Zaldarriaga, ``{Effective field theory of large
  scale structure at two loops: The apparent scale dependence of the speed of
  sound},'' \href{http://dx.doi.org/10.1103/PhysRevD.92.123007}{{\em Phys.
  Rev.} {\bfseries D92} no.~12, (2015) 123007},
\href{http://arxiv.org/abs/1507.02256}{{\ttfamily arXiv:1507.02256
  [astro-ph.CO]}}.

\bibitem{Baldauf:2015zga}
T.~Baldauf, E.~Schaan, and M.~Zaldarriaga, ``{On the reach of perturbative
  methods for dark matter density fields},''
  \href{http://dx.doi.org/10.1088/1475-7516/2016/03/007}{{\em JCAP} {\bfseries
  1603} (2016) 007},
\href{http://arxiv.org/abs/1507.02255}{{\ttfamily arXiv:1507.02255
  [astro-ph.CO]}}.

\bibitem{Baldauf:2015tla}
T.~Baldauf, E.~Schaan, and M.~Zaldarriaga, ``{On the reach of perturbative
  descriptions for dark matter displacement fields},''
  \href{http://dx.doi.org/10.1088/1475-7516/2016/03/017}{{\em JCAP} {\bfseries
  1603} no.~03, (2016) 017},
\href{http://arxiv.org/abs/1505.07098}{{\ttfamily arXiv:1505.07098
  [astro-ph.CO]}}.

\bibitem{error}
T.~Baldauf, M.~Mirbabayi, M.~Simonovic, and M.~Zaldarriaga, ``{LSS constraints
  with controlled theoretical uncertainties},''
\href{http://arxiv.org/abs/1602.00674}{{\ttfamily arXiv:1602.00674
  [astro-ph.CO]}}.

\bibitem{left}
R.~A. Porto, L.~Senatore, and M.~Zaldarriaga, ``{The Lagrangian-space Effective
  Field Theory of Large Scale Structures},''
  \href{http://dx.doi.org/10.1088/1475-7516/2014/05/022}{{\em JCAP} {\bfseries
  1405} (2014) 022},
\href{http://arxiv.org/abs/1311.2168}{{\ttfamily arXiv:1311.2168
  [astro-ph.CO]}}.

\bibitem{left2}
M.~McQuinn and M.~White, ``{Cosmological perturbation theory in 1+1
  dimensions},'' \href{http://dx.doi.org/10.1088/1475-7516/2016/01/043}{{\em
  JCAP} {\bfseries 1601} no.~01, (2016) 043},
\href{http://arxiv.org/abs/1502.07389}{{\ttfamily arXiv:1502.07389
  [astro-ph.CO]}}.

\bibitem{left3}
Z.~Vlah, M.~White, and A.~Aviles, ``{A Lagrangian effective field theory},''
  \href{http://dx.doi.org/10.1088/1475-7516/2015/09/014}{{\em JCAP} {\bfseries
  1509} no.~09, (2015) 014},
\href{http://arxiv.org/abs/1506.05264}{{\ttfamily arXiv:1506.05264
  [astro-ph.CO]}}.

\bibitem{Zaldarriaga:2015jrj}
M.~Zaldarriaga and M.~Mirbabayi, ``{Lagrangian Formulation of the
  Eulerian-EFT},''
\href{http://arxiv.org/abs/1511.01889}{{\ttfamily arXiv:1511.01889
  [astro-ph.CO]}}.

\bibitem{review}
R.~A. Porto, ``{The effective field theorist's approach to gravitational
  dynamics},'' \href{http://dx.doi.org/10.1016/j.physrep.2016.04.003}{{\em
  Phys. Rept.} {\bfseries 633} (2016) 1--104},
\href{http://arxiv.org/abs/1601.04914}{{\ttfamily arXiv:1601.04914 [hep-th]}}.

\bibitem{Bernardeau:2001qr}
F.~Bernardeau, S.~Colombi, E.~Gaztanaga, and R.~Scoccimarro, ``{Large scale
  structure of the universe and cosmological perturbation theory},'' {\em Phys.
  Rept.} {\bfseries 367} (2002) 1--248,
\href{http://arxiv.org/abs/astro-ph/0112551}{{\ttfamily arXiv:astro-ph/0112551
  [astro-ph]}}.

\bibitem{Leofin}
M.~Lewandowski and L.~Senatore, ``{An analytic implementation of the
  IR-resummation for the BAO peak},''
\href{http://arxiv.org/abs/1810.11855}{{\ttfamily arXiv:1810.11855
  [astro-ph.CO]}}.

\bibitem{Blas:2013aba}
D.~Blas, M.~Garny, and T.~Konstandin, ``{Cosmological perturbation theory at
  three-loop order},'' {\em JCAP} {\bfseries 1401} no.~01, (2014) 010,
\href{http://arxiv.org/abs/1309.3308}{{\ttfamily arXiv:1309.3308
  [astro-ph.CO]}}.

\bibitem{Sahni:1995rr}
V.~Sahni and S.~Shandarin, ``{Behavior of Lagrangian approximations in
  spherical voids},'' \href{http://dx.doi.org/10.1093/mnras/282.2.641}{{\em
  Mon. Not. Roy. Astron. Soc.} {\bfseries 282} (1996) 641},
\href{http://arxiv.org/abs/astro-ph/9510142}{{\ttfamily arXiv:astro-ph/9510142
  [astro-ph]}}.

\bibitem{Pajer:2017ulp}
E.~Pajer and D.~van~der Woude, ``{Divergence of Perturbation Theory in Large
  Scale Structures},'' {\em JCAP} {\bfseries 1805} no.~05, (2018) 039,
\href{http://arxiv.org/abs/1710.01736}{{\ttfamily arXiv:1710.01736
  [astro-ph.CO]}}.

\bibitem{Senatore:2014via}
L.~Senatore and M.~Zaldarriaga, ``{The IR-resummed Effective Field Theory of
  Large Scale Structures},''
  \href{http://dx.doi.org/10.1088/1475-7516/2015/02/013}{{\em JCAP} {\bfseries
  1502} no.~02, (2015) 013},
\href{http://arxiv.org/abs/1404.5954}{{\ttfamily arXiv:1404.5954
  [astro-ph.CO]}}.

\bibitem{Cataneo:2016suz}
M.~Cataneo, S.~Foreman, and L.~Senatore, ``{Efficient exploration of cosmology
  dependence in the EFT of LSS},''
  \href{http://dx.doi.org/10.1088/1475-7516/2017/04/026}{{\em JCAP} {\bfseries
  1704} no.~04, (2017) 026},
\href{http://arxiv.org/abs/1606.03633}{{\ttfamily arXiv:1606.03633
  [astro-ph.CO]}}.

\bibitem{Senatore:2017pbn}
L.~Senatore and G.~Trevisan, ``{On the IR-Resummation in the EFTofLSS},''
  \href{http://dx.doi.org/10.1088/1475-7516/2018/05/019}{{\em JCAP} {\bfseries
  1805} no.~05, (2018) 019},
\href{http://arxiv.org/abs/1710.02178}{{\ttfamily arXiv:1710.02178
  [astro-ph.CO]}}.

\bibitem{horizon}
J.~Kim, C.~Park, G.~Rossi, S.~M. Lee, and J.~R. Gott, III, ``{The New Horizon
  Run Cosmological N-Body Simulations},''
  \href{http://dx.doi.org/10.5303/JKAS.2011.44.6.217}{{\em J. Korean Astron.
  Soc.} {\bfseries 44} (2011) 217--234},
\href{http://arxiv.org/abs/1112.1754}{{\ttfamily arXiv:1112.1754
  [astro-ph.CO]}}.

\bibitem{Blas:2013bpa}
D.~Blas, M.~Garny, and T.~Konstandin, ``{On the non-linear scale of
  cosmological perturbation theory},'' {\em JCAP} {\bfseries 1309} (2013) 024,
\href{http://arxiv.org/abs/1304.1546}{{\ttfamily arXiv:1304.1546
  [astro-ph.CO]}}.

\bibitem{Pietroni:2018ebj}
M.~Pietroni, ``{Structure formation beyond shell-crossing: nonperturbative
  expansions and late-time attractors},''
  \href{http://dx.doi.org/10.1088/1475-7516/2018/06/028}{{\em JCAP} {\bfseries
  1806} no.~06, (2018) 028},
\href{http://arxiv.org/abs/1804.09140}{{\ttfamily arXiv:1804.09140
  [astro-ph.CO]}}.

\bibitem{Foreman}
S.~Foreman {\em unpublished} .

\bibitem{Ben-Dayan:2014hsa}
I.~Ben-Dayan, T.~Konstandin, R.~A. Porto, and L.~Sagunski, ``{On Soft Limits of
  Large-Scale Structure Correlation Functions},''
  \href{http://dx.doi.org/10.1088/1475-7516/2015/02/026}{{\em JCAP} {\bfseries
  1502} no.~02, (2015) 026},
\href{http://arxiv.org/abs/1411.3225}{{\ttfamily arXiv:1411.3225
  [astro-ph.CO]}}.

\bibitem{Garny:2015oya}
M.~Garny, T.~Konstandin, R.~A. Porto, and L.~Sagunski, ``{On the Soft Limit of
  the Large Scale Structure Power Spectrum: UV Dependence},''
  \href{http://dx.doi.org/10.1088/1475-7516/2015/11/032}{{\em JCAP} {\bfseries
  1511} no.~11, (2015) 032},
\href{http://arxiv.org/abs/1508.06306}{{\ttfamily arXiv:1508.06306
  [astro-ph.CO]}}.

\bibitem{Assassi:2014fva}
V.~Assassi, D.~Baumann, D.~Green, and M.~Zaldarriaga, ``{Renormalized Halo
  Bias},'' \href{http://dx.doi.org/10.1088/1475-7516/2014/08/056}{{\em JCAP}
  {\bfseries 1408} (2014) 056},
\href{http://arxiv.org/abs/1402.5916}{{\ttfamily arXiv:1402.5916
  [astro-ph.CO]}}.

\end{thebibliography}\endgroup

\end{document}